\documentclass[a4paper,12pt]{article}
\ifdefined\pdfoutput\pdfoutput=1\fi

\usepackage[left=1.2cm,right=1.2cm,top=2cm,bottom=1.5cm]{geometry}
\usepackage[T1]{fontenc}
\usepackage[utf8]{inputenc}
\usepackage[dvipsnames]{xcolor}
\usepackage[font=small,format=plain,labelfont=bf,textfont=normal,
singlelinecheck=false]{caption}
\usepackage{subcaption}
\usepackage{amsfonts,amsmath,amssymb,mathrsfs}
\usepackage{setspace}
\usepackage{array,booktabs,multirow,dcolumn}
\usepackage{graphicx}
\usepackage{bm}
\usepackage{verbatim}
\usepackage{textcomp}
\usepackage{tikz}
\usetikzlibrary{calc,patterns,arrows.meta}
\usetikzlibrary{decorations.text,decorations.pathmorphing,decorations.markings}
\usepackage{pgfplots}
\pgfplotsset{compat=1.18}
\usepackage{cite}
\usepackage{etoolbox}
\usepackage[colorlinks=true,linkcolor=black,citecolor=teal,
  urlcolor=MidnightBlue,filecolor=black]{hyperref}
\allowdisplaybreaks[4]
\newcommand{\bea}{\begin{eqnarray}}
\newcommand{\eea}{\end{eqnarray}}
\newcommand{\be}{\begin{equation}}
\newcommand{\ee}{\end{equation}}
\newcommand{\bs}{\begin{subequations}}
\newcommand{\es}{\end{subequations}}
\def\nn{\nonumber}

\newcommand{\beqs}{\begin{eqnarray}}
\newcommand{\eeqs}{\end{eqnarray}}

\numberwithin{equation}{section}

\newcommand{\Rmnum}[1]{\uppercase\expandafter{\romannumeral #1\relax}}

\def\c.c.{\mathrm{c.c.}}

\def\Om{\Omega}

\newif\ifrevision
\revisiontrue

\newcommand{\Disc}{\operatorname{Disc}}
\newcommand{\RevO}{\mathcal{O}}
\AtBeginEnvironment{thebibliography}{\color{black}}

\title{\bfseries Carrollian correlators in black hole perturbation theory}
\author{
Jiang Long$^{a,}$\thanks{\href{mailto:longjiang@hust.edu.cn}{\textcolor{black}{longjiang@hust.edu.cn}}} ,
Zhan-Jia Qu$^{a,}$\thanks{\href{mailto:m202470243@hust.edu.cn}{\textcolor{black}{m202470243@hust.edu.cn}}}  \  and \  
Hong-Yang Xiao$^{a,b,}$\thanks{\href{mailto:Hongyang.Xiao@soton.ac.uk}{\textcolor{black}{Hongyang.Xiao@soton.ac.uk}}}\\[4pt]
\small $^{a}$ School of Physics, Huazhong University of Science and Technology,\\
\small Luoyu Road 1037, Wuhan, Hubei 430074, China  \\[4pt]
\small $^{b}$ STAG Research Centre \& Mathematical Sciences, University of Southampton,\\
\small Highfield, SO17 1BJ, Southampton, UK
}
\date{}

\newcommand{\paperabstract}{In this note, we clarify the relationship between the two-point Carrollian correlator and massless scattering in a black hole background. It turns out that there are two types of Carrollian correlators at the null boundaries of each asymptotically flat spacetime.
The correlator from $\mathscr I^-$ to $\mathscr I^+$ should be regularized by subtracting the flat-space analog, and it is the position space version of the reflection amplitude for massless scattering. On the other hand, the correlator from $\mathscr I^-$ to the future horizon $\mathcal H^+$ is absent in flat space, and it is the position space version of the transmission amplitude. The poles of the Carrollian correlators are governed by null geodesics from $\mathscr I^-$ to either $\mathscr I^+$ or $\mathcal H^+$, and these geodesics define two classes of classical equations in Carrollian space. These equations relate the Shapiro time delay to the deflection angle and are determined purely by the boundary data. We find that, for small deflection angles, the time delay for the correlator from $\mathscr I^-$ to $\mathscr I^+/\mathcal H^+$ exhibits logarithmic/quadratic behavior. On the other hand, when the deflection angle is large, the time delay always increases linearly for both correlators. We further establish a relation between the low-frequency bulk scattering coefficients and the leading non-contact contributions at late times. The correlator from $\mathscr I^-$ to $\mathscr I^+$ features an inverse-square tail $\tau^{-2}$ with an explicit angular distribution $K_2$. We have numerically confirmed this angle-dependent late-time tail.}

\begin{document} 
\begin{titlepage}
\makeatletter\let\@fnsymbol\@arabic\makeatother
\maketitle
\thispagestyle{empty}
\begin{abstract}
\noindent\paperabstract
\end{abstract}
\end{titlepage}
\pdfbookmark[1]{Contents}{toc}
\tableofcontents
\clearpage
\raggedbottom % Avoid stretched gaps around displayed equations and floats.

\section{Introduction}
Gravitational waves, ripples in spacetime that are predicted by general relativity and first detected by LIGO about a decade ago \cite{LIGOScientific:2016aoc}, are of great interest to astronomers since they are generated by some of the most energetic astrophysical processes, such as binary mergers, supernovae, and galaxy collisions. Due to their paramount importance in fundamental physics, various analytic methods, e.g., the Post-Newtonian (PN) expansion \cite{Blanchet:2013haa} and the effective one-body (EOB) formalism \cite{Damour:2001tu}, have been developed to meet the needs of an upcoming era of high signal-to-noise ratio gravitational wave observations. In the PN regime, perturbative calculations are organized in terms of orders of velocity, and ideas from particle physics could play an important role. In the effective field theory (EFT) approach \cite{Goldberger:2004jt}, the two body problem in gravity is converted to the computation of Feynman diagrams. Nowadays, advanced techniques from quantum field theory have been integrated into the Post-Minkowskian (PM) expansion of two body problems in general relativity \cite{Bern:2019nnu}.

In a parallel development, motivated by the success of AdS/CFT and asymptotic analysis \cite{Bondi:1962px,Sachs:1962wk,Barnich:2010eb} on gravitational physics near future/past null infinity ($\mathscr I^\pm$), CFT techniques are applied to field theory at null boundary and lead to Carrollian holography \cite{Bagchi:2025vri}. Carrollian holography is a scenario of flat holography that maps the boundary Carrollian correlator\footnote{ The term ``Carrollian'' has several closely related uses in the literature.
Algebraically, the Carroll group arises from the ultra-relativistic contraction
$c\to0$ of the Poincar\'e group \cite{LevyLeblond:1965}. Geometrically, a
Carrollian manifold carries a degenerate metric together with a preferred null
direction, with null hypersurfaces providing the canonical realization. In
particular, the conformal Carroll group of a spherical spatial section is
isomorphic to the four-dimensional BMS group \cite{Duval:2014uva}. The future
and past null infinities therefore naturally support Carrollian fields labelled
by a null time and a point on the celestial two-sphere. We will adopt the description in terms of Carrollian fields on null infinity.
 }to Carrollian amplitude \cite{Donnay:2022aba,Donnay:2022wvx,Bagchi:2022emh,Mason:2023mti}, a position space scattering amplitude that can be computed perturbatively using the bulk-to-boundary propagator, bulk-to-bulk propagator as well as interaction vertices \cite{Liu:2024nfc}. Most of the recent developments are restricted to theoretical exploration \cite{Li:2024kbo,Liu:2024llk,nguyen2024carrollian,kulkarni2025carrollian,Long:2026cpq}, and its connection to black hole physics is still under investigation. In the Carrollian framework, near-horizon black hole physics may be understood in the context of Carrollian hydrodynamics \cite{Penna:2015gza,Ciambelli:2018wre,Ciambelli:2018xat,Donnay:2019jiz,petkou2022relativistic,freidel2023carrollian,armas2024carrollian}, which provides a macroscopic description based on Carrollian symmetry. One can also find the relationship between helicity flux and topological invariants on null hypersurfaces \cite{Long:2025fbb,Liu:2024rvz}. Furthermore, the observational aspects have been explored for radiative systems \cite{Long:2024yvj,Heng:2025kmr}. Unfortunately, these developments are relatively independent of the scattering amplitude approach based on the Carrollian correlator. %{\xhy Two-point function\cite{bagchi2022scattering}, three-point function can be fixed by Ward identities\cite{bagchi2023ads,salzer2023embedding,bagchi2024holography}, higher-point function and scattering amplitude \cite{nguyen2025operator,nguyen2024carrollian}, Carrollian OPEs\cite{dutta2024stress,saha2023carrollian,nguyen2025operator}}

In principle, the Carrollian correlator should encode the crucial information of gravitational problems in position space, i.e. time domain physics and angle-dependent distributions of observables. In a recent paper \cite{Long:2025bfi}, we propose to study thermal Carrollian correlators at null infinity and expect that they reveal a holographic description of black holes in asymptotically flat spacetimes. For technical reasons, we focus on the thermal Minkowski spacetime in that work. In this work, we will clarify the relationship between the Carrollian correlator and black hole perturbation theory. The latter is important for understanding Hawking radiation, black hole entropy, as well as astrophysical problems related to the detection of gravitational waves. To be more precise, we focus on the Green's functions that are fundamental building blocks of black hole perturbation theory as well as of the Carrollian correlator.  
 By extrapolating the Green's function to $\mathscr I^{\pm}$ and $\mathcal H^+$, we obtain two kinds of two-point Carrollian correlators in a black hole background.  The regularized $\mathscr I^-$ to $\mathscr I^+$ correlator is obtained by subtracting its flat-space analogue, and it represents the time-domain form of the reflection scattering amplitude. It encodes the prompt response, the ringdown, and the angular dependence of the late-time tail. The $\mathscr I^-$ to $\mathcal H^+$ correlator is determined by the transmission amplitude and thus encodes absorption. Interestingly, the leading non-contact late-time tail of the $\mathscr I^-$ to $\mathscr I^+$ correlator decays as $1/\tau^2$ with an angular-dependent factor $K_2$. On the other hand, the leading non-contact late-time tail of the $\mathscr I^-$ to $\mathcal H^+$ correlator decays as $1/s^2$, which is uniform in angle. Both the late-time behaviours have been checked by numerical calculations.   Physically, we expect that the poles of the two-point Carrollian correlators correspond to the null geodesics that emanate from $\mathscr I^-$ and arrive at $\mathscr I^+$ or $\mathcal H^+$. Thus we utilize the null geodesics to define classical equations at null boundaries. These equations impose constraints on the initial and final data for classical dynamics in the bulk. We analyze the asymptotic behaviour of the classical equations in detail.

The structure of this paper is as follows. In section \ref{rev}, we derive the two-point Carrollian correlators in black hole perturbation theory. In section \ref{alg}, we use the null geodesics in the bulk to define classical equations at null boundaries. We conclude with a summary and a discussion of future directions in the final section. Technical details of the integration and the proof of the properties of the classical equations are collected in three appendices.

\section{Carrollian correlator in black hole background}\label{rev}
In this section, we will first review the two-point Carrollian correlator in flat spacetime. Notations and conventions are introduced and the infrared problems in the flat space Carrollian correlator are discussed in the meantime. Then we delve into the two-point Carrollian correlator in black hole perturbation theory.  A regularized Carrollian correlator is obtained by subtracting the flat space analog. We will also show that the Fourier transform of this regularized two-point Carrollian correlator is directly related to the scattering and absorption amplitude.
\subsection{Carrollian correlator in flat spacetime}
We focus on the  four-dimensional Minkowski spacetime $\mathbb{R}^{1,3}$ whose metric in Cartesian coordinates $x^\mu=(t,\bm x)$ is
\begin{align}
    ds^2=\eta_{\mu\nu}dx^\mu dx^\nu=-dt^2+dx^idx^i,\quad \mu,\nu=0,1,2,3,
\end{align}
where the Minkowski matrix is $\eta_{\mu\nu}=\text{diag}(-1,+1,+1,+1)$. In  spherical coordinates $(t,r,\theta,\phi)$, the metric is
\begin{align}
    ds^2=-dt^2+dr^2+r^2(d\theta^2+\sin^2{\theta}d\phi^2).
\end{align} To simplify notation, we will also collect the coordinates $(\theta,\phi)$ as $\Omega$.
Given a massless and real scalar field $\Phi(x)$ in the bulk, we impose the fall-off condition
\begin{align}\label{falloof}
    \Phi(x)=\left\{\begin{array}{cc}
         \frac{\Sigma(u,\Om)}{r}+\mathcal{O}(r^{-2}),& \text{near }\mathscr I^+ \\
         \frac{\Sigma^{(-)}(v,\Om)}{r}+\mathcal{O}(r^{-2}),& \text{near }\mathscr I^- 
    \end{array}\right.
\end{align}
where the coefficients $\Sigma(u,\Omega)$ and $\Sigma^{(-)}(v,\Omega)$ are identified as boundary operators at $\mathscr I^{+}$ and $\mathscr I^{-}$, respectively. The retarded coordinates $(u=t-r,\Omega)$ are used in $\mathscr I^+$ and the advanced coordinates $(v=t+r,\Omega)$ are suitable for $\mathscr I^-$. The fundamental field $\Sigma(u,\Om)/\Sigma^{(-)}(v,\Om)$ encodes the propagating degree of freedom of the bulk field and it is the leading order coefficient in the asymptotic expansion. In the language of the boundary Carrollian field theory, it is a primary operator on $\mathscr I^{\pm}$ with dimension $1$ and spin $0$. 
The Feynman propagator in the bulk is defined in the conventional way
\be 
G_{\text F}(x;x')=\langle \text{T}\Phi(x)\Phi(x')\rangle
\ee where $\text T$ denotes the time-ordering operator
\bea 
\text{T}\Phi(x)\Phi(x')=\Theta(x^0-x'^0)\Phi(x)\Phi(x')+\Theta(x'^0-x^0)\Phi(x')\Phi(x).
\eea By extrapolating one of the bulk field to the boundary, we obtain the retarded and advanced bulk-to-boundary propagators as follows:
\bs\begin{align}
   D(u,\Omega;x')&=\langle \Sigma(u,\Omega)\Phi(x')\rangle=\lim_{r\to\mathscr I^+}r\ G_{\text{F}}(x;x'),\\
   D^{(-)}(v',\Omega';x)&=\langle \Phi(x)\Sigma^{(-)}(v',\Omega')\rangle=\lim_{r'\to\mathscr I^-} r'\ G_{\text{F}}(x;x'). 
\end{align}\es The symbol $\lim_{r\to\mathscr I^+}$ means to take the limit by moving $r\to\infty$ and keeping $u$ finite at the same time. Similarly, the symbol $\lim_{r\to\mathscr I^-}$ means to take the limit by moving $r\to\infty$ and keeping $v$ finite at the same time. The complex conjugate of the Feynman propagator is called the anti-Feynman propagator 
\bea 
G_{\bar F}(x;x')=\langle \bar{\text{T}}\Phi(x)\Phi(x')\rangle=\left(G_F(x;x')\right)^*
\eea where $\bar{\text T}$ denotes the anti-time ordering operator
\bea 
\bar{\text{T}}\Phi(x)\Phi(x')=\Theta(x^0-x'^0)\Phi(x')\Phi(x)+\Theta(x'^0-x^0)\Phi(x)\Phi(x').
\eea Thus we can also obtain the complex conjugate of the retarded and advanced bulk-to-boundary propagators
\bs\begin{align}
    D^*(u,\Omega;x')&=\langle \Phi(x')\Sigma(u,\Omega)\rangle=\lim_{r\to\mathscr I^+} r\ G_{\bar F}(x;x'),\\
    D^{(-)*}(v',\Omega';x)&=\langle \Sigma^{(-)}(v',\Omega')\Phi(x)\rangle=\lim_{r'\to \mathscr I^-}r'\ G_{\bar F}(x;x').
\end{align}\es 
From the Feynman propagator
\be 
G_F(x;x')=\frac{1}{4\pi^2}\frac{1}{(x-x')^2+i\epsilon},
\ee it is straightforward to obtain the retarded and advanced bulk-to-boundary correlators 
\bs\begin{align}
    D(u,\Omega;x')&=-\frac{1}{8\pi^2(u+n\cdot x'-i\epsilon)},\\
    D^{(-)}(v',\Omega';x)&=\frac{1}{8\pi^2(v'-\bar n'\cdot x+i\epsilon)}.
\end{align}\es The null vector $n^\mu$ is determined by the $\Omega$ 
\be 
n^\mu=(1,\sin\theta\cos\phi,\sin\theta\sin\phi,\cos\theta)=(1,n^i)
\ee while 
\be 
\bar n'^\mu=(-1,\sin\theta'\cos\phi',\sin\theta'\sin\phi',\cos\theta')=(-1,n'^i).
\ee The unit vector $n^i$ is the normal vector of the $S^2$ by embedding it into $\mathbb R^3$.
The boundary-to-boundary propagator is formally obtained by extrapolating the remaining field to the boundary 
\bea 
B(u,\Omega;v',\Omega')=\langle\Sigma(u,\Omega)\Sigma^{(-)}(v',\Omega')\rangle=\lim_{r'\to\mathscr I^-}r'\ D(u,\Omega;x').\label{bulktoboundary}
\eea Note that one should get the same result via the alternative formula
\bea 
B^{(-)}(u,\Omega;v',\Omega')=\langle\Sigma(u,\Omega)\Sigma^{(-)}(v',\Omega')\rangle=\lim_{r\to\mathscr I^+}r\ D^{(-)}(v',\Omega';x).\label{bulktoboundarym}
\eea This has been evaluated in \cite{Long:2025bfi} and we have obtained the electric and magnetic branch. However, the parameter in the $i\epsilon$ prescription of the magnetic branch depends on the infrared  radius slightly. To clarify this point, we work in the integral representation of the bulk-to-boundary propagator 
\bea 
D(u,\Omega;x')&=&\frac{i}{8\pi^2}\int_0^\infty  d\omega e^{-i\omega(u+n\cdot x'-i\epsilon)}
\eea Note that $n\cdot x'=-v'+r' n\cdot \bar n'$, we use the identity
\bea 
e^{-i\omega r'n\cdot \bar n'}&=&e^{-i\omega r'-i\omega r'\bm n\cdot\bm n'}=e^{-i\omega r'} 4\pi \sum_{\ell,m}(-i)^\ell j_\ell(\omega r')Y_{\ell,m}(\Omega)Y_{\ell,m}^*(\Omega')
\eea and then the large $r'$ expansion of the exponential function is 
\bea 
e^{-i\omega r'n\cdot\bar n'}&=& \frac{4\pi}{2i\omega r'}\sum_{\ell,m}(-1)^\ell Y_{\ell,m}(\Omega)Y_{\ell,m}^*(\Omega')-\frac{4\pi}{2i\omega r'}e^{-2i\omega r'}\sum_{\ell,m} Y_{\ell,m}(\Omega)Y_{\ell,m}^*(\Omega')+\mathcal {O}(r'^{-2}).
\eea 
The boundary-to-boundary propagator \eqref{bulktoboundary} is 
\bea 
B(u,\Omega;v',\Omega')&=&-\frac{1}{4\pi}\int_0^\infty \frac{d\omega}{\omega}e^{-i\omega(u-v')}\sum_{\ell,m}(-1)^\ell Y_{\ell,m}(\Omega)Y^*_{\ell,m}(\Omega')\nn\\&&+\frac{1}{4\pi}\int_0^\infty \frac{d\omega}{\omega} e^{-i\omega(u-v'+2r')}\sum_{\ell,m}Y_{\ell,m}(\Omega)Y_{\ell,m}^*(\Omega').\label{B}
\eea Note that the second line depends on the infrared radius $r'\to\infty$ explicitly. One may ignore this term to obtain the two-point Carrollian correlator in the electric branch. However, a much more careful treatment of this term would give us a magnetic branch \cite{Long:2025bfi}. In this work, we will leave it in this form for later convenience. Rather interesting, we can also use the formula \eqref{bulktoboundarym} to obtain the two-point correlator 
\bea 
B^{(-)}(u,\Omega;v',\Omega')&=&-\frac{1}{4\pi}\int_0^\infty \frac{d\omega}{\omega}e^{-i\omega(u-v')}\sum_{\ell,m}(-1)^\ell Y_{\ell,m}(\Omega)Y^*_{\ell,m}(\Omega')\nn\\&&+\frac{1}{4\pi}\int_0^\infty \frac{d\omega}{\omega} e^{-i\omega(u-v'+2r)}\sum_{\ell,m}Y_{\ell,m}(\Omega)Y_{\ell,m}^*(\Omega').\label{Bm}
\eea By definition, the two-point correlator $B^{(-)}$ is formally the same as $B$. However, the result depends on the description since the infrared radius in \eqref{Bm} is $r$ while it is $r'$ in \eqref{B}. One may identify $r=r'$ to get the same result. It is a bit confusing that why the two different radius should be identical. In the following, we will find that the paradoxical term should be subtracted in black hole perturbation theory  and there is no need to identify $r$ and $r'$.

\subsection{Carrollian correlator in Schwarzschild black hole background}
In this subsection, we will initiate a definition of Carrollian correlator in an asymptotically flat spacetime. For the sake of clarity, we will focus on the two-point Carrollian correlator of massless scalar theory in Schwarzschild background. The Penrose diagram of two-sided  Schwarzschild black hole is shown in Figure \ref{propagators}, we will focus on the physically relevant region \text{AFS}$_1$. Considering a $1\to 1$ scattering problem in  Schwarzschild background, the incoming wave is from $\mathscr I^-$ and into the future. Part of the wave is scattered by the black hole and becomes the outgoing wave at $\mathscr I^+$. The remaining wave is absorbed by the black hole and becomes the ingoing wave at the future event horizon $\mathcal H^+$. There is no wave coming from the past event horizon $\mathcal H^-$. Thus we will discuss  Carrollian correlator from $\mathscr I^-$ to $\mathscr I^+$ or $\mathcal H^+$. The two kinds of correlator are also shown in Figure \ref{propagators}.

\begin{figure}
    \centering
    \usetikzlibrary{decorations.text}
    \begin{tikzpicture} [scale=0.8]
        \draw[dashed,thick] (-3,-3) node[below]{\footnotesize $i^-$}  -- (3,3) node[above]{\footnotesize $i^+$};
        \draw[draw,thick] (3,3) -- (6,0) node[right]{\footnotesize $i^0$};
        \node at (4.8,1.8) {\footnotesize $\mathscr{I}^+$};
        \draw[dashed,thick] (-3,3) node[above] {\footnotesize $i^+$} -- (3,-3) node[below]{\footnotesize $i^-$};
        \draw[draw, thick](3,-3) -- (6,0);
        \draw[draw, thick](-3,-3) -- (-6,0) node[left]{\footnotesize $i^0$};
        \draw[draw,thick](-6,0) -- (-3,3);
        \node at(-4.8,1.8) {\footnotesize $\mathscr {I}^+$};
        \node at (-4.8,-1.8){\footnotesize $\mathscr {I}^-$};
        \node at (4.8,-1.8) {\footnotesize $\mathscr {I}^-$};
        \node at (3,0){\text{AFS}$_1$};
        \node at (-3,0){\text{AFS}$_2$};
       \draw[decorate, decoration={snake, amplitude=0.4mm, segment length=1mm}] (-3,3) -- (3,3);
       \draw[draw,thick](-3,3) -- (3,3);
       \draw[decorate, decoration={snake, amplitude=0.4mm, segment length=1mm}] (-3,-3) -- (3,-3);
       \draw[draw,thick](-3,-3) -- (3,-3);
       \draw[decorate, decoration={snake, amplitude=0.8mm,segment length=2mm}] (4.5,1.5)  .. controls (3.7,1) and (3.7,-1) .. (4.5,-1.5);
       \draw[decorate, decoration={snake, amplitude=0.8mm,segment length=2mm}] (3.5,-2.5)  -- (0.8,0.8);
       %\draw[decorate, decoration={snake, amplitude=0.8mm,segment length=2mm}] (4.8,1.2)  .. controls (4.6,0.4) and (5.4,0.2) .. (5.6,0.4);
       %\draw[decorate, decoration={snake, amplitude=0.8mm,segment length=2mm}] (3.4,2.6)  .. controls (1.5,1.5) and (-1.5,1.5) .. (-3.4,2.6);
       %\draw[decorate, decoration={snake, amplitude=0.8mm,segment length=2mm}] (3.5,-2.5)  .. controls (1,1) and (-1.5,1.5) .. (-4.8,1.2);
    \end{tikzpicture}
    \caption{Penrose diagram of two-sided Schwarzschild black hole and Carrollian propagators in an asymptotically flat region.}
    \label{propagators}
\end{figure}
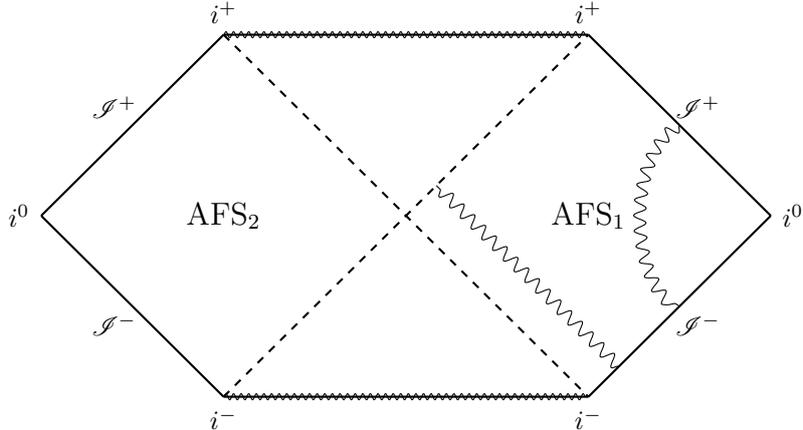

 \begin{figure}
    \centering
    \usetikzlibrary{decorations.text}
    \begin{tikzpicture} [scale=0.8]
        \draw[draw,thick] (-3,0) -- (0,3) ;
        \draw[draw,thick] (0,3) -- (3,0) ;
        \node at (1.8,1.8) {\footnotesize $\mathscr I^+$};
        \node at (1.8,-1.8) {\footnotesize $\mathscr I^-$};
        \node at (-1.8,1.8) {\footnotesize $\mathcal{H}^+$};
        \node at (-1.8,-1.8) {\footnotesize $\mathcal{H}^-$};
        \draw[draw,thick] (-3,0)  -- (0,-3) ;
        \draw[draw, thick](0,-3) -- (3,0);
        \draw[->,thick] (1.5,-1.5) -- (-1.5,1.5);
        \draw[->,thick] (1.7,-1.3) -- (-1.3,1.7);
        \draw[->,thick] (1.3,-1.7) -- (-1.7,1.3);
        \draw[->,thick] (-0.5,0.5) -- (1,2);
        \draw[->,thick] (-1,0.6) -- (0.7,2.3);
        \draw[->,thick] (0,0.4) -- (1.3,1.7);
        \node at (0,-4) {\footnotesize in};
      
       %\draw[decorate, decoration={snake, amplitude=0.8mm,segment length=2mm}] (4.8,1.2)  .. controls (4.6,0.4) and (5.4,0.2) .. (5.6,0.4);
       %\draw[decorate, decoration={snake, amplitude=0.8mm,segment length=2mm}] (3.4,2.6)  .. controls (1.5,1.5) and (-1.5,1.5) .. (-3.4,2.6);
       %\draw[decorate, decoration={snake, amplitude=0.8mm,segment length=2mm}] (3.5,-2.5)  .. controls (1,1) and (-1.5,1.5) .. (-4.8,1.2);
    \end{tikzpicture}\hspace{1cm}
    \begin{tikzpicture} [scale=0.8]
        \draw[draw,thick] (-3,0) -- (0,3) ;
        \draw[draw,thick] (0,3) -- (3,0) ;
        \node at (1.8,1.8) {\footnotesize $\mathscr I^+$};
        \node at (1.8,-1.8) {\footnotesize $\mathscr I^-$};
        \node at (-1.8,1.8) {\footnotesize $\mathcal{H}^+$};
        \node at (-1.8,-1.8) {\footnotesize $\mathcal{H}^-$};
        \draw[draw,thick] (-3,0)  -- (0,-3) ;
        \draw[draw, thick](0,-3) -- (3,0);
        \draw[->,thick] (-1.5,-1.5) -- (1.5,1.5);
        \draw[->,thick] (-1.7,-1.3) -- (1.3,1.7);
        \draw[->,thick] (-1.3,-1.7) -- (1.7,1.3);
        \draw[->,thick] (1.5,-1.5) -- (-0.2,0.2);
        \draw[->,thick] (1.8,-1.2) -- (0.1,0.5);
        \draw[->,thick] (1.2,-1.8) -- (-0.3,-0.3);
        \node at (0,-4) {\footnotesize out};
       %\draw[decorate, decoration={snake, amplitude=0.8mm,segment length=2mm}] (4.8,1.2)  .. controls (4.6,0.4) and (5.4,0.2) .. (5.6,0.4);
       %\draw[decorate, decoration={snake, amplitude=0.8mm,segment length=2mm}] (3.4,2.6)  .. controls (1.5,1.5) and (-1.5,1.5) .. (-3.4,2.6);
       %\draw[decorate, decoration={snake, amplitude=0.8mm,segment length=2mm}] (3.5,-2.5)  .. controls (1,1) and (-1.5,1.5) .. (-4.8,1.2);
    \end{tikzpicture}\vspace{1cm}
    
    \begin{tikzpicture} [scale=0.8]
        \draw[draw,thick] (-3,0) -- (0,3) ;
        \draw[draw,thick] (0,3) -- (3,0) ;
        \node at (1.8,1.8) {\footnotesize $\mathscr I^+$};
        \node at (1.8,-1.8) {\footnotesize $\mathscr I^-$};
        \node at (-1.8,1.8) {\footnotesize $\mathcal{H}^+$};
        \node at (-1.8,-1.8) {\footnotesize $\mathcal{H}^-$};
        \draw[draw,thick] (-3,0)  -- (0,-3) ;
        \draw[draw, thick](0,-3) -- (3,0);
        \draw[->,thick] (-1.5,-1.5) -- (1.5,1.5);
        \draw[->,thick] (-1.7,-1.3) -- (1.3,1.7);
        \draw[->,thick] (-1.3,-1.7) -- (1.7,1.3);
        \draw[->,thick] (0.1,-0.3) -- (-1.6,1.4);
        \draw[->,thick] (-0.1,0.3) -- (-1.4,1.6);
        \draw[->,thick] (0.3,0.3) -- (-1.2,1.8);
        \node at (0,-4) {\footnotesize up};
      
       %\draw[decorate, decoration={snake, amplitude=0.8mm,segment length=2mm}] (4.8,1.2)  .. controls (4.6,0.4) and (5.4,0.2) .. (5.6,0.4);
       %\draw[decorate, decoration={snake, amplitude=0.8mm,segment length=2mm}] (3.4,2.6)  .. controls (1.5,1.5) and (-1.5,1.5) .. (-3.4,2.6);
       %\draw[decorate, decoration={snake, amplitude=0.8mm,segment length=2mm}] (3.5,-2.5)  .. controls (1,1) and (-1.5,1.5) .. (-4.8,1.2);
    \end{tikzpicture}\hspace{1cm}
    \begin{tikzpicture} [scale=0.8]
        \draw[draw,thick] (-3,0) -- (0,3) ;
        \draw[draw,thick] (0,3) -- (3,0) ;
        \node at (1.8,1.8) {\footnotesize $\mathscr I^+$};
        \node at (1.8,-1.8) {\footnotesize $\mathscr I^-$};
        \node at (-1.8,1.8) {\footnotesize $\mathcal{H}^+$};
        \node at (-1.8,-1.8) {\footnotesize $\mathcal{H}^-$};
        \draw[draw,thick] (-3,0)  -- (0,-3) ;
        \draw[draw, thick](0,-3) -- (3,0);
        \draw[->,thick] (1.5,-1.5) -- (-1.5,1.5);
        \draw[->,thick] (1.7,-1.3) -- (-1.3,1.7);
        \draw[->,thick] (1.3,-1.7) -- (-1.7,1.3);
        \draw[->,thick] (-1.5,-1.5) -- (0,0);
        \draw[->,thick] (-1.2,-1.8) -- (0.1,-0.5);
        \draw[->,thick] (-0.9,-2.1) -- (0.8,-0.4);
        \node at (0,-4) {\footnotesize down};
      
       %\draw[decorate, decoration={snake, amplitude=0.8mm,segment length=2mm}] (4.8,1.2)  .. controls (4.6,0.4) and (5.4,0.2) .. (5.6,0.4);
       %\draw[decorate, decoration={snake, amplitude=0.8mm,segment length=2mm}] (3.4,2.6)  .. controls (1.5,1.5) and (-1.5,1.5) .. (-3.4,2.6);
       %\draw[decorate, decoration={snake, amplitude=0.8mm,segment length=2mm}] (3.5,-2.5)  .. controls (1,1) and (-1.5,1.5) .. (-4.8,1.2);
    \end{tikzpicture}
    
    \caption{\centering An illustration of the boundary conditions for the in, out, up and down modes.}
    \label{inout1}
\end{figure}
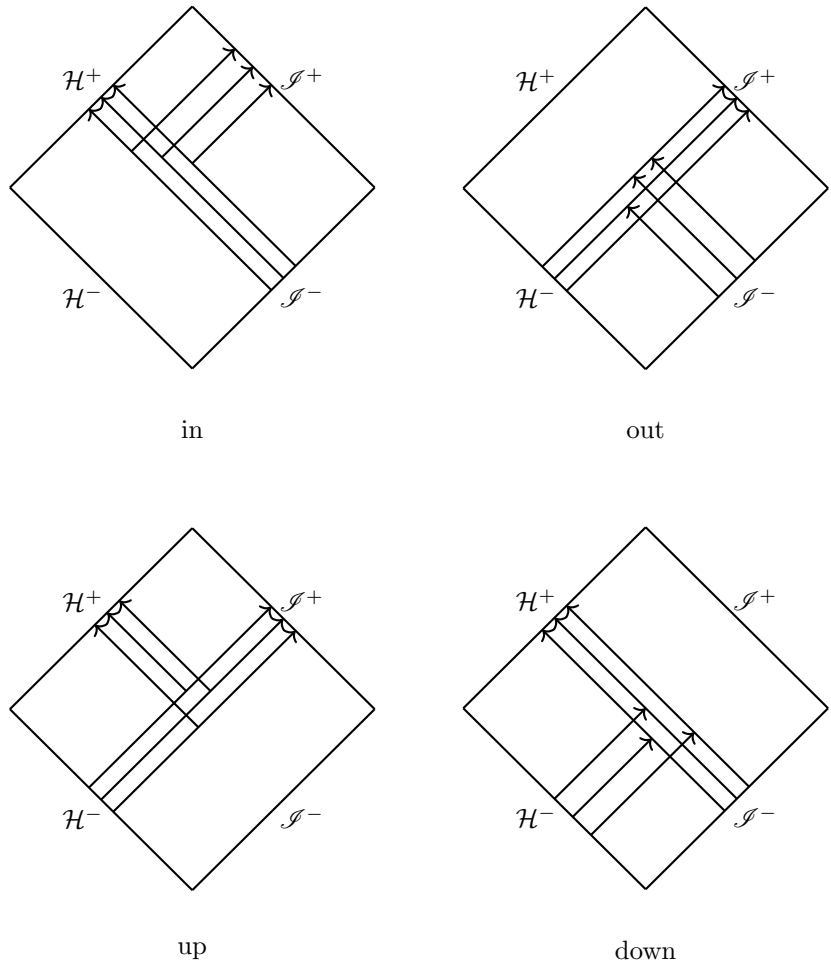

The Klein-Gordon equation in Schwarzschild background is 
\be 
\Box \Phi=0
\ee and one can solve it by separation of variables 
\bea 
\Phi(x)=\int d\omega e^{-i\omega t}\sum_{\ell,m}\frac{1}{r}u_{\omega,\ell}(r)Y_{\ell,m}(\Omega)
\eea where we have inserted a factor $r^{-1}$ in the expansion to adapt to the fall-off behaviour \eqref{falloof}. The function $u_{\omega,\ell}(r)$ satisfies the second order differential equation\footnote{This is the  Regge-Wheeler-Zerilli type equation \cite{Regge:1957td,Zerilli:1970se}. In black hole perturbation theory, the related wave equation for a Weyl scalar in Newman-Penrose formalism \cite{Newman:1961qr} is called the Teukolsky equation \cite{1973ApJ...185..635T}. More general formalism can be found in \cite{Pound:2021qin}.}
\bea 
\frac{d^2}{dr_*^2}u_{\omega,\ell}+\left[\omega^2-V_\ell(r)\right]u_{\omega,\ell}=0\label{uomegal}
\eea where $V_\ell$ is the potential 
\be 
V_\ell(r)=\left(1-\frac{2M}{r}\right)\left(\frac{\ell(\ell+1)}{r^2}+\frac{2M}{r^3}\right).
\ee In this expression, $M$ is the mass of the black hole and $\ell=0,1,2,\cdots$ is the quantum number of angular momentum. The tortoise coordinate $r_*$ is determined by the equation 
\be 
\frac{dr_*}{dr}=\frac{1}{1-\frac{2M}{r}}\quad\Rightarrow\quad r_*(r)=\int_o^r dr'\left(1-\frac{2M}{r'}\right)^{-1}
\ee up to a reference point $o$. The conventional choice is 
\be 
r_*(r)=r+2M\ln\left(\frac{r}{2M}-1\right).\label{rstar1}
\ee However, the other choice is also fine. In general, we have 
\be 
r_*(r)=r-o+2M\ln\frac{r-2M}{o-2M}.\label{rstar2}
\ee To recover the standard $r_*$, the reference point $o=2M\zeta$ should obey the equation 
\be 
\zeta+\ln \left(\zeta-1\right)=0\quad\Rightarrow\quad \zeta\approx 1.2785.
\ee 

The equation \eqref{uomegal} can be solved by imposing suitable boundary condition. In Figure \ref{inout1}, we have shown four kinds of modes according to the boundary conditions\cite{frolov2012black}. The corresponding solutions are denoted as $u_{\omega,\ell}^{\text{in}},u_{\omega,\ell}^{\text{out}},u_{\omega,\ell}^{\text{up}},u_{\omega,\ell}^{\text{down}}$ respectively. As an illustration, the ``in'' mode obeys the boundary condition 
\be 
u_{\omega,\ell}^{\text{in}}\sim\left\{\begin{array}{cc} e^{-i\omega r_*},&r_*\to-\infty\\
A_{\omega,\ell}^{\text{in},\text{up}}e^{i\omega r_*}+A_{\omega,\ell}^{\text{in},\text{down}}e^{-i\omega r_*},&r_*\to\infty.\end{array}\right.
\ee
The coefficients $A_{\omega,\ell}^{\text{in},\text{up}}$ and $A_{\omega,\ell}^{\text{in},\text{down}}$ are determined by solving the equation near the horizon and extend it to null infinities. In black hole scattering problems, the transmission and reflection amplitude are defined via 
\bea 
\mathcal T_{\omega,\ell}=\frac{1}{A_{\omega,\ell}^{\text{in},\text{down}}},\quad \mathcal R_{\omega,\ell}=\frac{A_{\omega,\ell}^{\text{in},\text{up}}}{A_{\omega,\ell}^{\text{in},\text{down}}}\label{refam}
\eea and they satisfy the standard scattering relation 
\begin{align}
    |\mathcal T_{\omega,\ell}|^2+|\mathcal R_{\omega,\ell}|^2=1.
\end{align}
Note that only two of the modes are linearly independent in general. By matching the boundary conditions, we may write 
\be 
u^{\text{in}}_{\omega,\ell}=A_{\omega,\ell}^{\text{in},\text{up}} u^{\text{up}}_{\omega,\ell}+A_{\omega,\ell}^{\text{in},\text{down}}u_{\omega,\ell}^{\text{down}}.
\ee The transmission and reflection amplitude can be obtained by solving the retarded Green's function $G_R(x;x')$
\bea 
\Box G_R(x;x')=\frac{i}{\sqrt{-g}}\delta^{(4)}(x-x')
\eea that satisfies the ingoing boundary condition at $\mathcal H^+$ and outgoing boundary condition at $\mathscr I^+$ together with the condition $G_R(x;x')=0$ for $t<t'$. Expanding the Green's function via the eigenfunctions 
\bea 
G_R(x;x')=-\frac{i}{2\pi r r'}\int_{\mathcal C} d\omega e^{-i\omega(t-t')}\sum_{\ell,m} G_{\omega,\ell}(r;r')Y_{\ell,m}(\Omega)Y_{\ell,m}^*(\Omega'),\label{ret}
\eea the radial function $G_{\omega,\ell}(r;r')$ can be constructed through the ``in'' and ''up'' modes 
\bea 
G_{\omega,\ell}(r;r')=\frac{u_{\omega,\ell}^{\text{in}}(r_<)u_{\omega,\ell}^{\text{up}}(r_>)}{W_{\omega,\ell}}\label{Gomega}
\eea 
where $r_>=\text{max}\{r,r'\}$ and $r_<=\text{min}\{r,r'\}$. The Wronskian $W_{\omega,\ell}$ is defined as
\be 
W_{\omega,\ell}=u^{\text{up}}_{\omega,\ell}\partial_ru^{\text{in}}_{\omega,\ell}-u^{\text{in}}_{\omega,\ell}\partial_ru^{\text{up}}_{\omega,\ell}
\ee and it is independent of $r$ and $r'$
\be 
W_{\omega,\ell}=-2i\omega A_{\omega,\ell}^{\text{in},\text{down}}.
\ee The integral of \eqref{ret} is evaluated by considering a contour $\mathcal C$ that is displaced slightly above the real $\omega$ axis, e.g.,
\be 
G_R(x;x')=-\frac{i}{2\pi r r'}\int_{-\infty+i\epsilon}^{\infty+i\epsilon} d\omega e^{-i\omega(t-t')}\sum_{\ell,m} G_{\omega,\ell}(r;r')Y_{\ell,m}(\Omega)Y_{\ell,m}^*(\Omega').\label{ret2}
\ee As a function of complex frequency, $G_{\omega,\ell}$ has an infinite tower of quasi-normal modes (QNMs) located at the lower half plane \cite{Vishveshwara:1970zz}. These modes are determined by the zeros of the Wronskian 
\be 
W_{\omega,\ell}\Big|_{\omega=\omega_{QNM}}=0.
\ee As a matter of fact, there  is also a branch cut along the imaginary axis \cite{Price:1972pw,Ching:1995tj}. Now we use \eqref{ret2} and \eqref{Gomega} to define the bulk-to-boundary propagators 
\bs\begin{align}
    D_R(u,\Omega;x')&=\lim_{r\to\mathscr I^+}r G_R(x;x'),\\
    D_R^{(-)}(v',\Omega';x)&=\lim_{r'\to\mathscr I^-}r' G_R(x;x'),\\
    D_R^{\mathcal H^+}(v,\Omega;x')&=\lim_{r\to\mathcal H^+}r G_R(x;x'),\\
    D_R^{\mathcal H^-}(u',\Omega';x)&=\lim_{r'\to \mathcal H^-}r' G_R(x;x').
\end{align}\es More precisely, 
\bs\label{dr}\begin{align}
    D_R(u,\Omega;x')&=\frac{1}{4\pi r'}\int_{-\infty+i\epsilon}^{\infty+i\epsilon}\frac{d\omega}{\omega}e^{-i\omega(u-t')}\sum_{\ell,m}\mathcal T_{\omega,\ell}u_{\omega,\ell}^{\text{in}}(r')Y_{\ell,m}(\Omega)Y^*_{\ell',m'}(\Omega'),\\
    D_R^{(-)}(v',\Omega';x)&=\frac{1}{4\pi r}\int_{-\infty+i\epsilon}^{\infty+i\epsilon}\frac{d\omega}{\omega} e^{-i\omega(t-v')}\sum_{\ell,m} \mathcal T_{\omega,\ell}u_{\omega,\ell}^{\text{in}}(r)Y_{\ell,m}(\Omega)Y_{\ell,m}^*(\Omega'),\\
   D_R^{\mathcal H^+}(v,\Omega;x')&=\frac{1}{4\pi r'}\int_{-\infty+i\epsilon}^{\infty+i\epsilon}\frac{d\omega}{\omega}e^{-i\omega(v-t')}\sum_{\ell,m}\mathcal T_{\omega,\ell}u_{\omega,\ell}^{\text{up}}(r')Y_{\ell,m}(\Omega)Y_{\ell,m}^*(\Omega'),\label{bulktohp}\\
   D_R^{\mathcal H^-}(u',\Omega';x)&=0.
\end{align}\es 
The last bulk-to-boundary correlator vanishes since the solution $u^{\text{in}}$ vanishes at $\mathcal H^-$. In all the expressions, we have added a subscript $R$ to distinguish them with the advanced/retarded bulk-to-boundary propagator in previous subsection. Here the bulk-to-boundary correlators are extracted from the retarded Green's function in the bulk. However, the advanced/retarded bulk-to-boundary propagators are extracted from the Feynman propagator in the bulk. Note that the retarded Green's function is related to the Feynman propagator via 
\be 
\begin{aligned}G_R(x;x')&=\theta(x^0-x'^0)\langle [\Phi(x),\Phi(x')]\rangle=\theta(x^0-x'^0)\left(G_F(x;x')-G_{\bar F}(x;x')\right)=2i\theta(x^0-x'^0)\text{Im} G_F(x;x').\end{aligned}
\ee 
Therefore, the flat spacetime analogs of $D_R$ are 
\bs\label{dr0}\begin{align}
D_R^{0}(u,\Omega;x')&=\lim_{r\to\mathscr I^+}r\ G^0_R(x;x')=\frac{i}{8\pi^2}\int_{-\infty}^{\infty} d\omega e^{-i\omega(u+n\cdot x')},\\
D_R^{(-)0}(v',\Omega';x)&=\lim_{r'\to\mathscr I^-}r'\ G^0_R(x;x')=\frac{i}{8\pi^2}\int_{-\infty}^{\infty} d\omega e^{i\omega(v'-\bar n'\cdot x)}.
\end{align}\es Here we add a superscript $0$ to denote the result in flat spacetime. The bulk-to-future horizon $\mathcal H^+$  propagator is absent in the flat spacetime. By extrapolating the remaining field to the null boundary in \eqref{dr}, we may define the following boundary-to-boundary commutators 
\bs\begin{align}
    B_R(u,\Omega;v',\Omega')&=\lim_{r'\to\mathscr I^-} r'\ D_R(u,\Omega;x')=\lim_{r'\to\mathscr I^-}\lim_{r\to\mathscr I^+}r\ r' G_R(x;x'),\\
    B_R^{(-)}(u,\Omega;v',\Omega')&=\lim_{r\to\mathscr I^+}r\ D_R^{(-)}(v',\Omega';x)=\lim_{r\to\mathscr I^+}\lim_{r'\to\mathscr I^-}r\ r'\ G_R(x;x'),\\
    B_R^{(-),\mathcal H^+}(v,\Omega;v',\Omega')&=\lim_{r\to\mathcal H^+}r\ D_R^{(-)}(v',\Omega';x)=\lim_{r\to\mathcal H^+}\lim_{r'\to\mathscr I^-}r\ r'\ G_R(x;x'),\\
    B_R^{\mathcal H^+,(-)}(v,\Omega;v',\Omega')&=\lim_{r'\to\mathscr I^-}r'\ D_R^{\mathcal H^+}(v,\Omega;x')=\lim_{r'\to\mathscr I^-}\lim_{r\to\mathcal H^+}r\ r'\ G_R(x;x').
\end{align}\es Assuming the double limits commute with each other, we may obtain the following identities
\bs\begin{align}
    B_R(u,\Omega;v',\Omega')&=B_R^{(-)}(u,\Omega;v',\Omega'),\label{bru}\\
    B_R^{(-),\mathcal H^+}(v,\Omega;v',\Omega')&= B_R^{\mathcal H^+,(-)}(v,\Omega;v',\Omega').\label{brm}
\end{align}\es To check them, we evaluate the limits explicitly, 
\bs\begin{align}
     B_R(u,\Omega;v',\Omega')&=\frac{1}{4\pi}\int_{-\infty+i\epsilon}^{\infty+i\epsilon}\frac{d\omega}{\omega}\sum_{\ell,m}\left(e^{-i\omega(u-v')}\mathcal R_{\omega,\ell}+e^{-i\omega(u-u')}\right)Y_{\ell,m}(\Omega)Y_{\ell,m}^*(\Omega'),\\
     B_R^{(-)}(u,\Omega;v',\Omega')&=\frac{1}{4\pi}\int_{-\infty+i\epsilon}^{\infty+i\epsilon}\frac{d\omega}{\omega}\sum_{\ell,m}\left(e^{-i\omega(u-v')}\mathcal R_{\omega,\ell}+e^{-i\omega(v-v')}\right)Y_{\ell,m}(\Omega)Y_{\ell,m}^*(\Omega'),\\
     B_R^{(-),\mathcal H^+}(v,\Omega;v',\Omega')&= \frac{1}{4\pi}\int_{-\infty+i\epsilon}^{\infty+i\epsilon}\frac{d\omega}{\omega}e^{-i\omega(v-v')}\sum_{\ell,m}\mathcal T_{\omega,\ell}Y_{\ell,m}(\Omega)Y_{\ell,m}^*(\Omega'),\\
     B_R^{\mathcal H^+,(-)}(v,\Omega;v',\Omega')&=\frac{1}{4\pi}\int_{-\infty+i\epsilon}^{\infty+i\epsilon}\frac{d\omega}{\omega}e^{-i\omega(v-v')}\sum_{\ell,m}\mathcal T_{\omega,\ell}Y_{\ell,m}(\Omega)Y_{\ell,m}^*(\Omega').
\end{align}\es Interestingly, the second identity \eqref{brm} is satisfied while the first identity \eqref{bru} is not valid. The mismatch between $B_R$ and $B_R^{(-)}$ is from the highly oscillating terms that depend on the IR regulators. These terms remind us the similar behaviour of \eqref{B} and \eqref{Bm} in flat spacetime. In fact, we can also define the boundary commutators from \eqref{dr0}
\bs\begin{align}
    B_R^0(u,\Omega;v',\Omega')&=-\frac{1}{4\pi}\int_{-\infty+i\epsilon}^{\infty+i\epsilon}\frac{d\omega}{\omega}\sum_{\ell,m}\left(e^{-i\omega(u-v')}(-1)^\ell-e^{-i\omega(u-u')}\right)Y_{\ell,m}(\Omega)Y^*_{\ell,m}(\Omega') ,\\
    B_R^{(-),0}(u,\Omega;v',\Omega')&=-\frac{1}{4\pi}\int_{-\infty+i\epsilon}^{\infty+i\epsilon}\frac{d\omega}{\omega}\sum_{\ell,m}\left(e^{-i\omega(u-v')}(-1)^\ell-e^{-i\omega(v-u')}\right)Y_{\ell,m}(\Omega)Y^*_{\ell,m}(\Omega').
\end{align}\es Both of them are formally equal to $ \langle [\Sigma(u,\Omega),\Sigma^{(-)}(v',\Omega')]\rangle$ while they actually depend on the IR regulators. Note that the dependence on the IR regulator is exactly the same in Schwarzschild and flat spacetime. We may subtract the flat spacetime correlator from the corresponding one in the Schwarzschild spacetime and get a unique boundary-to-boundary correlator from $\mathscr I^-$ to $\mathscr I^+$
\be 
C(u,\Omega;v',\Omega')=B_R(u,\Omega;v',\Omega')-B_R^0(u,\Omega;v',\Omega')=B_R^{(-)}(u,\Omega;v',\Omega')-B_R^{(-),0}(u,\Omega;v',\Omega').
\ee More precisely, 
\bea  
C(u,\Omega;v',\Omega')&=&\frac{1}{4\pi}\int_{-\infty+i\epsilon}^{\infty+i\epsilon}\frac{d\omega}{\omega}e^{-i\omega(u-v')}\sum_{\ell,m}\left(\mathcal R_{\omega,\ell}+(-1)^\ell\right)Y_{\ell,m}(\Omega)Y^*_{\ell,m}(\Omega')\nn\\&=&-\frac{1}{16\pi^2}\int_{-\infty+i\epsilon}^{\infty+i\epsilon}\frac{d\omega}{\omega}e^{-i\omega(u-v')}\sum_{\ell}(2\ell+1)\left(e^{2i\delta_\ell}-1\right)P_\ell(\cos\gamma^P).\label{Cuv}
\eea  In the second step, we have used the spherical symmetry and the addition theorem 
\be 
\sum_{m=-\ell}^\ell Y_{\ell,m}(\Omega)Y_{\ell,m}^*(\Omega')=\frac{2\ell+1}{4\pi}P_\ell(\cos\gamma)
\ee where $P_\ell(\cos\gamma)$ is the Legendre polynomial of order $\ell$ and $\gamma$ is the angle between the directions $\Omega$ and $\Omega'$. Note that in \eqref{Cuv} the angle is denoted as $\gamma^P$ which means the angle between $\Omega^P$ and $\Omega'$. There is a factor $(-1)^\ell$ in the summation and its effect is to map $\Omega$ to its antipodal point $\Omega^P$. We have also rewritten the reflection amplitude $\mathcal R_{\omega,\ell}$ through a phase shift factor 
\be 
e^{2i\delta_\ell}=(-1)^{\ell+1}\mathcal R_{\omega,\ell}.
\ee Interestingly, the summation of $\ell$ is exactly equal to the scattering amplitude $f(\gamma)$ in the partial wave expansion 
\be 
f(\omega,\gamma)=\frac{1}{2i\omega}\sum_{\ell}(2\ell+1)\left(e^{2i\delta_\ell}-1\right)P_\ell(\cos\gamma)
\ee and then the correlator from $\mathscr I^-$ to $\mathscr I^+$ is the Fourier transform of the scattering amplitude 
\be 
C(u,\Omega;v',\Omega'^P)=-\frac{i}{8\pi^2}\int_{-\infty+i\epsilon}^{\infty+i\epsilon} d\omega e^{-i\omega(u-v')} f(\omega,\gamma).\label{cfour}
\ee Inversely, the scattering amplitude is 
\be 
f(\omega,\gamma)=4\pi i \int_{-\infty}^\infty du e^{i\omega u}C(u,\Omega;0,\Omega'^P).
\ee We have used the time translation symmetry to set $v'=0$ on the right hand side. In particular, the zero frequency limit of the scattering amplitude is proportional to the ``average null'' Carrollian correlator from $\mathscr I^-$ to $\mathscr I^+$
\be 
f(0,\gamma)=4\pi i\int_{-\infty}^\infty du C(u,\Omega;0,\Omega'^P).
\ee The scattering cross section is 
\bea 
\frac{d\sigma_{\text{sca}}}{d\Omega}=|f(\omega,\gamma)|^2=16\pi^2\Big|\int_{-\infty}^\infty du e^{i\omega u}C(u,\Omega;0,\Omega'^P)\Big|^2.
\eea The formula connects the Carrollian correlator to the scattering cross section in black hole perturbation theory.

Now we explore the other Carrollian correlator from $\mathscr I^-$ to $\mathcal H^+$ 
\bea 
\widetilde C(v,\Omega;v',\Omega')&=& B_R^{(-),\mathcal H^+}(v,\Omega;v',\Omega')= B_R^{\mathcal H^+,(-)}(v,\Omega;v',\Omega')\nn\\&=&\frac{i}{8\pi^2}\int_{-\infty+i\epsilon}^{\infty+i\epsilon}d\omega e^{-i\omega(v-v')}\widetilde f(\omega,\gamma)\label{wideC}
\eea where $\widetilde f(\omega,\gamma)$ is  the absorption amplitude 
\be 
\widetilde f(\omega,\gamma)=\frac{1}{2i\omega}\sum_{\ell}(2\ell+1)\mathcal T_{\omega,\ell}P_\ell(\cos\gamma)
\ee which is determined by the transmission amplitude $\mathcal T_{\omega,\ell}$. We find its connection to the Carrollian correlator as
\be 
\widetilde f(\omega,\gamma)=-4\pi i\int_{-\infty}^\infty dv e^{i\omega v}\widetilde C(v,\Omega;0,\Omega').
\ee In the zero frequency limit, this is the ``average null'' Carrollian correlator from $\mathscr I^-$ to $\mathcal H^+$
\be 
\widetilde f(0,\gamma)=-4\pi i \int_{-\infty}^\infty dv \widetilde C(v,\Omega;0,\Omega').
\ee 
The absorption cross section is 
\be 
\frac{d\sigma_{\text{abs}}}{d\Omega}=|\widetilde f(\omega,\gamma)|^2=16\pi^2\Big|\int_{-\infty}^\infty dv e^{i\omega v}\widetilde C(v,\Omega;0,\Omega')\Big|^2.
\ee 
The absorption cross section is proportional to the area of the event horizon for a spherically symmetric black hole in an asymptotically flat spacetime \cite{1974JETP...38....1S,1975CMaPh..44..245G,Page:1976df,Unruh:1976fm,Sanchez:1977si,Das:1996we}. Therefore, we obtain the following universal result for Carrollian correlator
\bea 
\Big|\int_{-\infty}^\infty dv \widetilde C(v,\Omega;0,\Omega')\Big|^2\propto \text{Area of the event horizon}.\label{arealawhor}
\eea 
The left hand side is the square of the ``average null'' Carrollian correlator from $\mathscr I^-$ to $\mathcal H^+$. We conclude that the ``average null'' Carrollian correlator $\widetilde C$ has a geometric meaning and is also related to the entropy of the black hole.
 More properties of the Carrollian correlators are discussed in the following: 
 \begin{enumerate}
     \item The integral representation \eqref{cfour} promotes us to separate $\mathscr I^+$ into two regions. In the region $\mathscr I^+_L$, $u>v'$, we may complete the integral contour from lower half plane. Since $f(\omega,\gamma)$ contains QNMs in the lower half plane and branch cut along the negative imaginary axis \cite{Ching:1995tj}, the contour is chosen as Figure \ref{qnms}. This kind of contour is conventionally used, e.g., \cite{Andersson:1996cm}. In this region, the Carrollian correlator $C(u,\Omega;v',\Omega')$ receives contributions from the residues at the QNMs and the discontinuity while across the branch cut in the negative imaginary axis. On the other hand, in the region $\mathscr I^+_{U}$, $u<v'$, we may complete the integral contour from upper half plane. The contour is chosen as Figure \ref{bcut}. There are no QNMs in the upper half plane and the Carrollian correlator $C(u,\Omega;v',\Omega')$ only receives contribution from the branch cut in the positive imaginary axis. 
     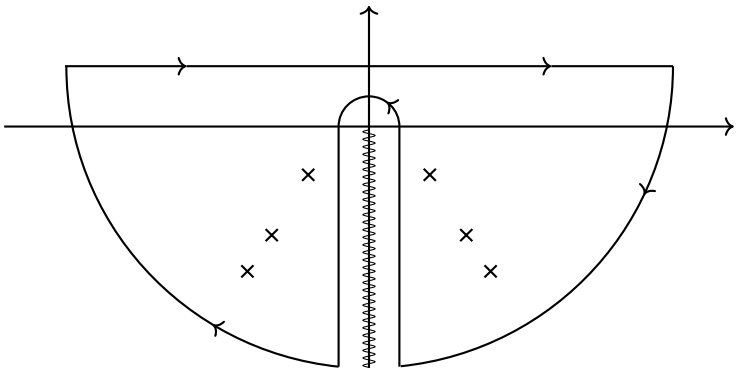
\begin{figure}
    \centering
    \usetikzlibrary{decorations.text}
    \begin{tikzpicture} [scale=0.6]
  % \label{sigmatimepath}
        \draw[->,thick] (-6,0) -- (6,0) ;
        \draw[->,thick] (0,-4) -- (0,2) ;
        \draw[decorate, decoration={snake, amplitude=0.8mm, segment length=1mm}] (0,-4) -- (0,0);
        %\node at (-0.2,0.2) {\footnotesize $0$};
        \draw[->,thick] (-5,1) -- (-3,1);
        \draw[->,thick] (-3,1) -- (3,1);
        \draw[draw,thick] (3,1) -- (5,1);
        \draw[thick, decoration={markings, mark=at position 0.3 with {\arrow{>}}},
          postaction={decorate}]
        (5,1) arc (0:-84:5);
        \draw[draw,thick] (0.5,-3.98) -- (0.5,0);
        \draw[thick, decoration={markings, mark=at position 0.3 with {\arrow{>}}},postaction={decorate}]
        (0.5,0) arc (0:180:0.5);
        \draw[draw,thick] (-0.5,-3.98) -- (-0.5,0);
        \draw[thick, decoration={markings, mark=at position 0.3 with {\arrow{>}}},
          postaction={decorate}]
        (-0.5,-3.98) arc (264:180:5);
        \foreach \x/\y in {1/-0.8, 1.6/-1.8, 2/-2.4, -1/-0.8, -1.6/-1.8, -2/-2.4}
        \draw[thick] (\x-0.1,\y-0.1) -- (\x+0.1,\y+0.1)
                     (\x-0.1,\y+0.1) -- (\x+0.1,\y-0.1);
       
%\caption{The time path $C$.}
    \end{tikzpicture}
    \caption{Contour for region $\mathscr I^+_L$. The QNMs are labeled by small $\times$ and the branch cut is represented by wavy lines.}
    \label{qnms}
\end{figure}

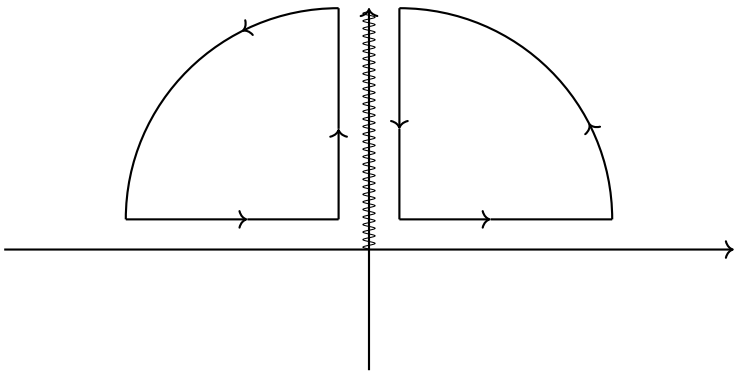
\begin{figure}
    \centering
    \usetikzlibrary{decorations.text}
    \begin{tikzpicture} [scale=0.8]
  % \label{sigmatimepath}
        \draw[->,thick] (-6,0) -- (6,0) ;
        \draw[->,thick] (0,-2) -- (0,4) ;
        \draw[decorate, decoration={snake, amplitude=0.8mm, segment length=1mm}] (0,0) -- (0,4);
        %\node at (-0.2,0.2) {\footnotesize $0$};
        \draw[->,thick] (-4,0.5) -- (-2,0.5);
        \draw[draw,thick] (-2,0.5) -- (-0.5,0.5);
        \draw[->,thick] (-0.5,0.5) -- (-0.5,2);
        \draw[draw,thick] (-0.5,2) -- (-0.5,4);
        \draw[->,thick] (0.5,0.5) -- (2,0.5);
        \draw[draw,thick] (2,0.5) -- (4,0.5);
        \draw[->,thick] (0.5,4) -- (0.5,2);
        \draw[draw,thick] (0.5,2) -- (0.5,0.5);
        \draw[thick, decoration={markings, mark=at position 0.3 with {\arrow{>}}},postaction={decorate}]
        (4,0.5) arc (0:90:3.5);
        \draw[thick, decoration={markings, mark=at position 0.3 with {\arrow{>}}},postaction={decorate}]
        (-0.5,4) arc (90:180:3.5);

%\caption{The time path $C$.}
    \end{tikzpicture}
    \caption{\centering{Contour for region $\mathscr I^+_U$}}
    \label{bcut}
\end{figure}
     
     \item Similarly, the integral representation \eqref{wideC} motivates us to separate $\mathcal H^+$ into another two regions. In the region $\mathcal H^+_U$, $v<v'$, we may complete the integral contour from the upper half plane. Only branch cut may contribute to the integral. However, 
     note that $\mathcal H^+_U$ is located outside of the future light cone of $v'$, the retarded Green's function should be zero due to causality and thus 
     \be 
     \widetilde C(v,\Omega;v',\Omega')=0,\quad v<v'.
     \ee On the other hand, in the region $\mathcal H^+_L$, we can complete the integral contour in the lower half plane. In this case, the contour is still chosen as  in Figure \ref{qnms}, both of the QNMs and the branch cut contribute to the Carrollian correlator $\widetilde C$.
     \item We have shown the regions $\mathscr I^+_{U/L}, \mathcal H^+_{U/L}$ in Figure \ref{divide} and extended them into the bulk. In the bulk, we find three regions I,II,III. Region I is connected to $\mathscr I^+_{L}$ and $\mathcal H^+_L$ and this region is related to the late time behaviour of the retarded Green's function. Region II is connected to $\mathscr I^+_U$ and it is  related to the early time behaviour of the retarded Green's function. Region III is connected to $\mathcal H^+_U$ and $\mathcal H^-$, in which the retarded Green's function vanishes. The conclusion is consistent with the analysis in Schwarzschild-dS spacetime \cite{Arnaudo:2025uos}.
     \begin{figure}
    \centering
    \usetikzlibrary{decorations.text}
    \begin{tikzpicture} [scale=0.8]
        \draw[dashed,thick] (0,0)  -- (3,3) node[above]{\footnotesize $i^+$};
        \draw[draw,thick] (3,3) -- (6,0) node[right]{\footnotesize $i^0$};
        
        \draw[dashed,thick] (0,0) -- (3,-3) node[below]{\footnotesize $i^-$};
        \draw[draw, thick](3,-3) -- (6,0);
       \draw[decorate, decoration={snake, amplitude=0.4mm, segment length=1mm}] (-1,3) -- (3,3);
       \draw[decorate, decoration={snake, amplitude=0.4mm, segment length=1mm}] (-1,-3) -- (3,-3);
       \draw[draw,thick] (1.5,1.5) -- (4.5,-1.5);
       \draw[draw,thick] (3,0) -- (4.5,1.5);
       \node at (3,1.5){\text{I}};
       \node at (4.5,0){\text{II}};
       \node at (2,-0.7){\text{III}};
       \node at (0.5,1) {\footnotesize $\mathcal{H}^+_U$};
       \node at (1.8,2.4) {\footnotesize $\mathcal{H}^+_L$};
       \node at (3.8,2.6) {\footnotesize $\mathscr {I}^+_L$};
       \node at (5.4,1.1) {\footnotesize $\mathscr {I}^+_U$};
       \node at (1.3,-1.7){\footnotesize $\mathcal{H}^-$};
    \end{tikzpicture}
    \caption{Classification of the asymptotically flat region \text{AFS}$_1$ and its boundaries. The  lightrays are emanating from $v'$ in all directions, as well as a third
ray which reflects from the black hole potential. These lightrays divide the asymptotically flat region \text{AFS}$_1$ into three regions I, II, III. Furthermore, the ingoing lightray separates the event horizon $\mathcal H^+$ into two regions $\mathcal H^+_L$ and $\mathcal H^+_U$. The reflected lightray separates the future null infinty $\mathscr I^+$ into another two regions $\mathscr I^+_L$ and $\mathscr I^+_U$.}
    \label{divide}
\end{figure}
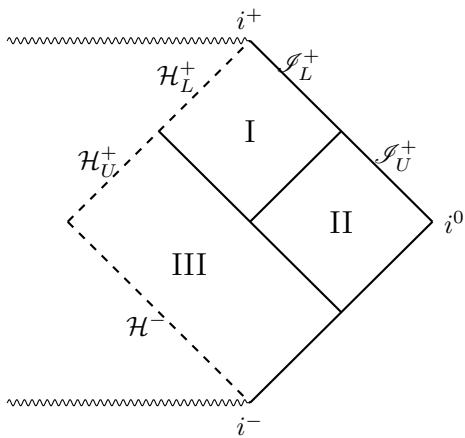
 \end{enumerate}

Before we close this subsection, we should mention that the reflection amplitude, a frequency space representation of the Carrollian amplitude from $\mathscr I^-$ to $\mathscr I^+$, is determined by the formula \eqref{refam}, which closely mirrors the retarded Green's function of an AdS black hole \cite{Son:2002sd}. It would be interesting to explore its relation to the flat space limit of an AdS black hole, following the approach of \cite{Alday:2024yyj}. However, it seems that the flat space limit of the other Carrollian amplitude, from $\mathscr I^-$ to $\mathcal H^+$, is absent in AdS/CFT. This topic certainly deserves further investigation.

% BEGIN INLINED revision_scattering.tex
\subsection{\texorpdfstring{ Late-time tails}{Late-time tails}}
{ In this subsection, we study the late-time tails of the Carrollian correlators. We make the scattering data in \eqref{refam} explicit in the low-frequency
regime. This has been studied in \cite{Mano:1996vt} and reviewed in \cite{Sasaki:2003xr}\footnote{In Schwarzschild black hole, The reflection/transmission amplitude are determined by the confluent Heun equation with boundary conditions of retarded Green's function. The problem is solved by mapping it to the a four-dimensional $\mathcal N=2$ supersymmetric gauge theory with three fundamental hypermultiplets \cite{Aminov:2020yma}. The reflection and transmission amplitudes are obtained by semiclassical Virasoro conformal blocks in the context of
AGT correspondence \cite{Bonelli:2021uvf,Bonelli:2022ten}. We will not use the technologies in these directions.}. }
\subsubsection{\texorpdfstring{ Low-frequency reflection and transmission}{Low-frequency reflection and transmission}}
\label{rev:scattering}
{ 

Throughout this subsection we fix the tortoise convention
\eqref{rstar1}, use $e^{-i\omega t}$, and write
$\mathcal R_\ell=\mathcal R_{\omega,\ell}$ and
$\mathcal T_\ell=\mathcal T_{\omega,\ell}$. The formulas below refer to
positive real $\omega$. The complex frequency continuation requires the { 
branch prescription} in Section~\ref{rev:tails}. Define
\begin{equation}
 \epsilon=2M\omega,\qquad
 \mathcal L=\log(2\epsilon)=\log(4M\omega),\qquad
 H_\ell=\sum_{j=1}^{\ell}\frac1j,\qquad H_0=0.
 \label{rev:epsilon}
\end{equation}
Matching a unit ingoing wave at the horizon to the two asymptotic waves at
infinity yields the fixed-$\ell$ expansion
\begin{align}
 { \mathcal T_\ell
 =(-i)^{\ell+1} B_\ell\epsilon^{\ell+1}
 \left[1+\epsilon\left\{\frac{\pi}{2}
 +i\left(\mathcal L+\gamma_E+\frac12-3H_\ell\right)\right\}
 +\RevO(\epsilon^2\mathcal L^2)\right]},\label{rev:T}
\end{align} with \(B_\ell=\frac{2^{\ell+1}(\ell!)^3}{(2\ell)!(2\ell+1)!}\).
The associated greybody factor is
%\begin{equation}
%{\lon  \Gamma_\ell=|\mathcal T_\ell|^2
% =\left[\frac{(\ell!)^3}{(2\ell)!(2\ell+1)!}\right]^2
% (2\epsilon)^{2\ell+2}
 %\left[1+2\pi M\omega+
 %\RevO\!\left((M\omega)^2\log^2(M\omega)\right)\right].}
% \label{rev:greybody}
%\end{equation}
\begin{equation}
\begin{split}
 \Gamma_\ell
 =&\left[\frac{(\ell !)^{3}}{(2\ell)!(2\ell+1)!}\right]^2
 (2\epsilon)^{2\ell+2}
 \Bigg[1+\pi\epsilon
 +\epsilon^2\left\{-4p_\ell(\mathcal L+\gamma_E)
             +\frac{2\pi^2}{3}+q_\ell\right\}+\RevO\!\left(\epsilon^3\log^3\epsilon\right)\Bigg],
 \label{rev:greybody-next-order}
 \end{split}
\end{equation}
where
\begin{equation}
 p_\ell=
 \frac{15\ell(\ell+1)-11}
 {4(2\ell-1)(2\ell+1)(2\ell+3)}\label{pell}
\end{equation} and \be
 q_\ell=16p_\ell H_{2\ell}-12p_\ell H_{\ell-1}
 +3H_\ell^{(2)}+\frac{-396+539\ell+1766\ell^2-988\ell^3
              -2648\ell^4-960\ell^5}
 {4\ell(2\ell-1)^2(2\ell+1)^2(2\ell+3)^2}
 \label{rev:qell-greybody}
\ee for $\ell\ge 1$. We define 
\be 
 H_n^{(2)}=\sum_{j=1}^{n}\frac1{j^2}.
\ee When $\ell=0$, we find $ q_0=\frac{155}{36}$. 
The coefficient of the putative relative $\epsilon^2\log^2\epsilon$ term cancels in
$\Gamma_\ell$.  The first logarithm at this order is therefore the
single-logarithmic term shown in \eqref{rev:greybody-next-order}.  
\iffalse 
The first
three values of the remaining coefficient are
\begin{equation}
 q_0=\frac{155}{36},\qquad
 q_1=\frac{6853}{900},\qquad
 q_2=\frac{8492}{1225},
\end{equation}
while, for $\ell\geq1$,
\begin{align}
 q_\ell={}&16p_\ell H_{2\ell}-12p_\ell H_{\ell-1}
 +3H_\ell^{(2)} \notag\\
 &+\frac{-396+539\ell+1766\ell^2-988\ell^3
              -2648\ell^4-960\ell^5}
 {4\ell(2\ell-1)^2(2\ell+1)^2(2\ell+3)^2},
 \qquad H_n^{(2)}=\sum_{j=1}^{n}\frac1{j^2}.
 \label{rev:qell-greybody}
\end{align}\fi

This fixed-$\ell$ scaling follows from the low-frequency MST/connection
expansion \cite{Mano:1996vt,Sasaki:2003xr}.
For example, the first three leading greybody factors are
\begin{equation}
 \Gamma_0^{\mathrm{leading}}=4\epsilon^2,\qquad
 \Gamma_1^{\mathrm{leading}}=\frac{1}{9}\epsilon^4,\qquad
 \Gamma_2^{\mathrm{leading}}=\frac{1}{2025}\epsilon^6.
\end{equation}
The standard partial wave absorption cross section consequently satisfies
\begin{equation}
 \sigma_{\mathrm{abs}}(\omega)
 =\frac{\pi}{\omega^2}\sum_{\ell=0}^{\infty}(2\ell+1)\Gamma_\ell.
 \label{rev:area}
\end{equation} In the low frequency limit, the dominant term
\(\sigma_{\mathrm{abs}}=16\pi M^2\) comes from $\ell=0$, in agreement with
the universal low-energy scalar absorption area law \cite{Das:1996we} and
consistent with \eqref{arealawhor}.

For reflection it is convenient to separate phase and norm. With
$c_\ell=\gamma_E-H_\ell-\tfrac12$, the long-range phase and flux relation gives
\begin{equation}
{  \mathcal R_\ell=(-1)^{\ell+1}
 \exp\!\left[2i\epsilon(\mathcal L+c_\ell)
       +i\RevO(\epsilon^2\mathcal L)\right]
 \left[1-\frac{\Gamma_\ell}{2}+\RevO(\Gamma_\ell^2)\right]}.
 \label{rev:Rphase}
\end{equation}
Here the remainder in the phase is real before multiplication by $i$ on the
positive real axis. The first few orders of the reflection amplitude are 
\begin{align}
 \mathcal R_\ell
 =(-1)^{\ell+1}\Big[1+2i\epsilon(\mathcal L+c_\ell)
 +\epsilon^2\big\{-2(\mathcal L+c_\ell)^2
                    +2i\pi p_\ell-2\eta_\ell\big\}\Big]
 +\RevO(\epsilon^3\mathcal L^3)\label{rev:R01}.
\end{align}
Here $p_\ell, \eta_\ell$ are the second-order coefficients. Note that $p_\ell$ is not the
Legendre polynomial $P_\ell(\cos\gamma)$. Instead, it is given by \eqref{pell}. The coefficient $\eta_\ell$ is non-vanishing only for $\ell=0$
\be 
\eta_\ell=\delta_{\ell,0}.
\ee 
For $\ell=0,1$, the explicit expressions match those in  \cite{Correia:2025Analyticity} 
\be 
 (p_0,\eta_0)=\left(\frac{11}{12},1\right),\qquad
 (p_1,\eta_1)=\left(\frac{19}{60},0\right).
\ee 

% END INLINED revision_scattering.tex
% BEGIN INLINED revision_tails.tex
\subsubsection{\texorpdfstring{ Non-contact tails and their angular distributions}{Non-contact tails and their angular distributions}}
\label{rev:tails}
Let $\tau=u-v'>0$, $x=\cos\gamma$, and use the antipodally matched scattering
angle in \eqref{cfour}. We also use  $C(\tau,\gamma)$ below to abbreviate
$C(u,\Omega;v',\Omega'^P)$. Then 
\begin{equation}
 C(\tau,x)=-\frac1{16\pi^2}\int_{\mathbb R+i0}d\omega\,
 e^{-i\omega\tau}\sum_{\ell=0}^{\infty}(2\ell+1)P_\ell(x)F_\ell(\omega).
 \label{rev:F}
\end{equation} with 
\be 
 F_\ell(\omega)=\frac{(-1)^{\ell+1}\mathcal R_\ell(\omega)-1}{\omega}.
\ee 
In the contour decomposition, prompt propagation, QNM residues and the
branch cut contribution are distinct terms\cite{Berti:2009kk}. The branch cut contribution on the negative
imaginary frequency axis is
\begin{equation}
 C_{\mathrm{BC}}(\tau,x)=\frac{i}{16\pi^2}
 \sum_{\ell=0}^{\infty}(2\ell+1)P_\ell(x)
 \int_0^\infty d\sigma\,e^{-\sigma\tau}
 \Disc F_\ell(-i\sigma).
 \label{rev:cutintegral}
\end{equation} The subscript ``BC'' indicates the branch cut contribution.
The discontinuity is the right side value minus the left side value at
$\omega_\pm=\pm0-i\sigma$. In particular,
$\log(\omega^2)_\pm=2\log\sigma\mp i\pi$ and
$\sqrt{\omega^2}_\pm=\mp i\sigma$. Both multivalued structures must be
 continued with the same prescription. We use the branch cut strength
 $q_\ell$ defined by the radial relation  \cite{casals2013analytic,Casals:2015Tail}
 \be 
g_{\ell,+}(-i\sigma)-g_{\ell,-}(-i\sigma)
  =iq_\ell(\sigma)g_\ell(+i\sigma) 
 \ee  where $g_\ell$ is exactly $u_{\omega,\ell}^{\text{up}}$ in our notation. 
 The radial cut relation then gives
 \be 
  \Disc A_\ell^{\rm in,down}=-iq_\ell A_\ell^{\rm in, up},
  \qquad
  \Disc\mathcal R_\ell
  =iq_\ell\mathcal R_{\ell,+}\mathcal R_{\ell,-}.
 \ee 
 Substitution into the definition of $F_\ell$ at $\omega=-i\sigma$ yields
 the identity
 \begin{equation}
 \Disc F_\ell(-i\sigma)
 =(-1)^\ell\frac{q_\ell(\sigma)}{\sigma}
 \mathcal R_{\ell,+}(-i\sigma)\mathcal R_{\ell,-}(-i\sigma).
 \label{rev:cutidentity}
 \end{equation}
 Here $\mathcal R_{\ell,\pm}(-i\sigma)$ is the right/left side value,
 \[
 \mathcal R_{\ell,\pm}(-i\sigma)
  =\lim_{\eta\to0^+}\mathcal R_\ell(\pm\eta-i\sigma).
 \]
 Write $a=2M\sigma$, $L=\log(4M\sigma)$, and
 $y_\ell=\gamma_E-H_\ell$.
 We obtain 
 \be 
  \frac{\mathcal R_{\ell,\pm}}{(-1)^{\ell+1}}
  =1+a\left[2\left(L+\gamma_E-H_\ell-\frac12\right)
             \mp i\pi\right]+\RevO(a^2L^2),
 \ee 
and then the product of $\mathcal R_{\ell,+}$ and $\mathcal R_{\ell,-}$ is 
\be 
 \mathcal R_{\ell,+}\mathcal R_{\ell,-}
=1+a(r_{11}L+r_{10})+a^2(r_{22}L^2+r_{21}L+r_{20})+\RevO(a^3L^3)\label{prodR}
\ee with 
\be 
 r_{11}=4,
 \quad r_{10}=-4H_\ell+4\gamma_E-2,\quad
 r_{22}=8,\quad 
 r_{21}=4r_{10},\quad
 r_{20}=\frac12r_{10}^2+4\delta_{\ell0}.
\ee 
 The expansion of the branch cut strength  is
\be 
  \frac{q_\ell}{a} =2\pi(-1)^\ell\left[1+a(q_{11}L+q_{10})+a^2
(q_{22}L^2+q_{21}L+q_{20})+\RevO(a^3L^3)\right]
\ee with 
\bs\begin{align}
 q_{11}&=-2,
  \quad q_{10}=2H_\ell+A_\ell-2\gamma_E,\quad
 q_{22}=2,
  \quad q_{21}=4(\gamma_E-H_\ell)-2A_\ell,\\
 q_{20}&=2\left(y_\ell-\frac12\right)^2
  +2(1-A_\ell)\left(y_\ell-\frac12\right)-\frac{\pi^2}{6}+\delta_{\ell0}
\end{align}\es where $y_\ell=\gamma_E-H_\ell$.
 Combining the expansion of the cut strength with the product \eqref{prodR} gives
\bs \begin{align}
 \Disc F_\ell(-i\sigma)
 &=4\pi M\left[1+a(2L+2\gamma_E+\kappa_\ell)
                  +a^2\mathcal D^{(2)}_\ell(L)+\cdots\right],
 \label{rev:cutspectrum}\\
 \kappa_\ell&=-2H_\ell+A_\ell-2,
 \label{rev:kappa}\\
 \mathcal D^{(2)}_\ell(L)
 &=2L^2+(4\gamma_E+2\kappa_\ell)L
   +2\gamma_E^2+2\kappa_\ell\gamma_E+\chi_\ell-\frac{\pi^2}{6}
 \label{rev:D2}
\end{align}\es  with 
\bs\begin{align}
 A_\ell&=\frac{16\ell^3+39\ell^2+11\ell-17}
                 {2(2\ell-1)(2\ell+1)(2\ell+3)},\\  \chi_\ell&=2H_\ell^2-2A_\ell H_\ell+4H_\ell
              +\frac32-A_\ell+5\delta_{\ell0}.
 \label{rev:chi-definition}
\end{align}\es 
Terms independent of $\ell$ in \eqref{rev:cutspectrum} give angular contact
distributions to \eqref{rev:cutintegral} through
\be \sum_\ell(2\ell+1)P_\ell(\bm n\cdot\bm n')
=4\pi\delta^{(2)}(\Omega,\Omega').\ee
This term belongs to the contact sector that is from the universal long-range Newtonian (Coulomb-type) dressing of Schwarzschild scattering \cite{Andersson:2000tf,1988sfbh.book.....F}. Since it is supported only at $\Omega=\Omega'$ and may therefore be omitted for the separated angle tail, $0<\gamma\le \pi$.
  We focus on the 
non-contact part of \eqref{rev:cutintegral} below.  With a regulator $\rho^\ell$, one convenient
representative of the first non-contact angular distribution is\footnote{Note that the relevant summation in \eqref{rev:cutintegral} is actually
\be 
\widehat K_2(x)=\lim_{\rho\to 1^-}\sum_{\ell=0}^\infty (2\ell+1)\kappa_\ell \rho^\ell P_\ell(x).
\ee 
This function differs from $K_2(x)$ by a contact term, $K_2(x)-\widehat K_2(x)=4\pi\delta^{(2)}(\Omega,\Omega')$, which we ignore in our analysis.
}
\bs \begin{align}
 K_2(x)&=\lim_{\rho\to1^-}\sum_{\ell=0}^{\infty}
 (2\ell+1)(\kappa_\ell+1)\rho^\ell P_\ell(x),\label{rev:K2abel}\\
 \kappa_\ell+1&=-2H_\ell+\frac{15}{8(2\ell+1)}
 +\frac1{8(2\ell+1)(2\ell-1)(2\ell+3)}.
\end{align}\es 
The sum gives the distribution
\begin{equation}
 K_2(x)=\frac2{1-x}
       +\frac{15}{8\sqrt{2(1-x)}}-\frac{\sqrt{2(1-x)}}{32}.
 \label{rev:K2distribution}
\end{equation}
 For a fixed
$0<\gamma\leq\pi$, this is the ordinary function
\begin{equation}
K_2(\gamma)=\csc^2\frac\gamma2
             +\frac{15}{16}\csc\frac\gamma2
             -\frac1{16}\sin\frac\gamma2.
 \label{rev:K2closed}
\end{equation}
Thus the leading angular dependence contains an inverse-square forward
 singularity and a subleading inverse-angle term
 \be K_2(\gamma)=4/\gamma^2+15/(8\gamma)+1/3+\RevO(\gamma).\ee This is a monotonic positive function of $\gamma$. It blows up at the point 
 $\gamma=0$ and decreases to $\frac{15}{8}$ at $\gamma=\pi$. We have drawn the angular dependence function $K_2(\gamma)$ in Figure \ref{rev:fig-K2-backward}.
 Since the divergence near $\gamma=0$
 compresses the finite large angle structure on a full range  plot, we
 display the backward-angle interval $\gamma\in[\frac{\pi}{2},\pi]$ separately.
 
 \begin{figure}[tbp]
  \centering
  \begin{minipage}[t]{0.48\textwidth}
   \centering
   \begin{tikzpicture}
    \begin{axis}[
      width=0.88\linewidth,height=0.68\linewidth,
      xmin=0,xmax=3.14159265,ymin=1.5,ymax=80,
      xtick={0,1.57079633,3.14159265},
      xticklabels={$0$,$\pi/2$,$\pi$},
      ytick={20,40,60,80},
      xlabel={$\gamma$},ylabel={$K_2$},
      xlabel style={black},ylabel style={black},
      axis line style={black},tick style={black},
      tick label style={black},clip=true]
     \addplot[black,thick,domain=0.02:3.14159265,samples=400,
       restrict y to domain*=1.5:80]
       {1/(sin(deg(x/2))^2)+15/(16*sin(deg(x/2)))-sin(deg(x/2))/16};
     \addplot[black,dashed,domain=0:3.14159265] {15/8};
    \end{axis}
   \end{tikzpicture}
  \end{minipage}\hfill
  \begin{minipage}[t]{0.48\textwidth}
   \centering
   \begin{tikzpicture}
    \begin{axis}[
      width=0.88\linewidth,height=0.68\linewidth,
      xmin=1.57079633,xmax=3.14159265,ymin=1.8,ymax=3.4,
      xtick={1.57079633,3.14159265},
      xticklabels={$\pi/2$,$\pi$},ytick={2,2.5,3},
      xlabel={$\gamma$},ylabel={$K_2$},
      xlabel style={black},ylabel style={black},
      axis line style={black},tick style={black},
      tick label style={black},clip=true]
     \addplot[black,thick,domain=1.57079633:3.14159265,samples=240]
       {1/(sin(deg(x/2))^2)+15/(16*sin(deg(x/2)))-sin(deg(x/2))/16};
     \addplot[black,dashed,domain=1.57079633:3.14159265] {15/8};
    \end{axis}
   \end{tikzpicture}
  \end{minipage}
  \caption{  The leading non-contact angular distribution function $K_2(\gamma)$.  The left panel covers $0<\gamma\leq\pi$ and displays the forward
  divergence; the right panel resolves $\pi/2\leq\gamma\leq\pi$.  The dashed
  line is the backscattering asymptote $K_2=15/8$.}
  \label{rev:fig-K2-backward}
 \end{figure}
At the next order, we may define another function
\begin{align}
 K_\chi(x)
 &=\lim_{\rho\to1^-}\sum_{\ell=0}^{\infty}(2\ell+1)
              (\chi_\ell-\tfrac12)\rho^\ell P_\ell(x)\notag\\
 &=\frac{2}{1-x}\left(\log\frac{1-x}{2}-1\right)
 +\frac{29+x}{8\sqrt{2(1-x)}}
 \left(\log\frac{4\sqrt{2(1-x)}}{2+\sqrt{2(1-x)}}-\frac12\right)+\frac{39+\sqrt{2(1-x)}}{8}.
\label{rev:KchiClosed}
\end{align}
 Using
\be \int_0^\infty\sigma e^{-\sigma\tau}d\sigma=\tau^{-2}\ee  and
\be \int_0^\infty\sigma^2\log(4M\sigma)e^{-\sigma\tau}d\sigma
=2\tau^{-3}[\tfrac32-\gamma_E+\log(4M/\tau)]\ee  yields the non-contact part of the angle-dependent late-time tail 
\begin{align}
 C_{\mathrm{BC}}(\tau,\gamma)
  =\frac{iM^2}{2\pi\tau^2}K_2(\gamma)+\frac{iM^3}{\pi\tau^3}
   \left[4K_2(\gamma)\left(\frac32+\log\frac{4M}{\tau}\right)
                      +2K_\chi(\gamma)\right]+\cdots.
 \label{rev:tail}
\end{align}
The ellipsis
 denotes higher orders. The leading non-contact boundary tail has the same
 low-frequency branch-cut origin as the original Schwarzschild late-time
 power laws \cite{Price:1971fb,Price:1972pw,Gundlach:1993tp}. 
 To the best of our knowledge, the function $K_2(\gamma)$ has not been
 derived previously for this boundary observable. The first surviving logarithmic correction at
 separated angles is $\tau^{-3}\log\tau$. Its form is similar to the
 logarithmic structure of long-range branch-cut tails at timelike infinity
 \cite{Ching:1995tj,Casals:2015Tail}, though in their region the late-time tail is $t^{-2\ell-5}\log t$ for each fixed $\ell$. Interesting,  its coefficient is still fixed by
 $K_2$. The following constant at order $\tau^{-3}$ contains the additional
angular information in $K_\chi$. Related works \cite{CasalsDolanOttewillWardell2013,Zenginoglu:2012xe,Casals:2016qyj} have studied in detail the angle-dependent null-geodesic, caustic, and pole-cut structure of the bulk Schwarzschild Green function. Those results concern the bulk propagator. The quantities \(K_2\) and \(K_\chi\) considered here are not the bulk propagator itself, but rather coefficients of an angle-separated late-time tail.

In Figure \ref{rev:figK2NLOdetail}, we use numerical simulations to demonstrate the time evolution of the Carrollian correlator from \(\mathscr{I}^{-}\) to \(\mathscr{I}^{+}\) for four different angles. The results agree well with the angular distribution function \(K_2(\gamma)\). We also include the analytic result \eqref{rev:tail} and show that it is closer to the numerical simulation than the leading \(K_2\) contribution.

\begin{figure}[tbp]
 \centering
 \pgfplotsset{rev select rows/.style 2 args={x filter/.code={
   \ifnum\coordindex<#1\def\pgfmathresult{nan}\fi
   \ifnum\coordindex>#2\def\pgfmathresult{nan}\fi}}}
 \newcommand{\revKtwoNLOPanel}[3]{%
  \begin{minipage}[t]{0.48\textwidth}\centering
  \begin{tikzpicture}
   \begin{axis}[
    width=0.88\linewidth,height=0.66\linewidth,
    xmin=20,xmax=160,xtick={20,60,100,160},
    xlabel={$\tau$},xlabel style={text=black},
    title={$#1^\circ$},title style={font=\bfseries,text=black},
    axis x line*=bottom,axis y line*=left,
    scaled y ticks=base 10:2,
    ytick scale label code/.code={${\color{black}\times 10^{-2}}$},
    tick label style={font=\scriptsize,text=black},
    label style={font=\small},
    grid=major,grid style={gray!25},unbounded coords=discard]
    \addplot[black,thick,rev select rows={#2}{#3},restrict x to domain=20:160]
     table[x=tau_over_M,y=numerical_ImC_lmax2_sub,col sep=comma]
     {revision_data/four_angle_curves.csv};
    \addplot[black,thick,dashed,rev select rows={#2}{#3},restrict x to domain=20:160]
     table[x=tau_over_M,y=analytic_ImC_K2_full_angle,col sep=comma]
     {revision_data/four_angle_curves.csv};
    \addplot[black,thick,dashdotted,rev select rows={#2}{#3},restrict x to domain=20:160]
     table[x=tau_over_M,y=analytic_ImC_K2_NLO_conditional,col sep=comma]
     {revision_data/four_angle_curves.csv};
   \end{axis}
  \end{tikzpicture}
  \end{minipage}}
 \revKtwoNLOPanel{30}{0}{359}\hfill
 \revKtwoNLOPanel{45}{360}{719}\\[3pt]
 \revKtwoNLOPanel{90}{720}{1079}\hfill
 \revKtwoNLOPanel{120}{1080}{1439}
 \par\smallskip
 {\color{black}%
 \tikz[baseline=-0.5ex]{\draw[black,thick] (0,0)--(0.35,0);}
 Numerical simulation\quad
 \tikz[baseline=-0.5ex]{\draw[black,thick,dashed] (0,0)--(0.35,0);}
 Analytic $K_2$\quad
 \tikz[baseline=-0.5ex]{\draw[black,thick,dashdotted] (0,0)--(0.35,0);}
 Analytic $K_2+K_\chi$}
 \caption{ Analytic late-time details and numerical reconstruction at four angles.
The expanded three-line comparison covers \(20\leq\tau/M\leq160\).
The solid line is the finite-time numerical reconstruction, the dashed line is the leading \(K_2\) prediction, and the dash-dotted line is the next-to-leading-order \(K_2+K_\chi\) prediction of \eqref{rev:tail}.
We set \(M=1\) in the numerical simulation. }
 \label{rev:figK2NLOdetail}
\end{figure}

At higher orders in \(\tau^{-n},\ n\ge 4\), the highest logarithm \(L_\tau^{n-1}\) is a contact term. We have defined \(L_\tau=\log(4M/\tau)\). At separated angles, the highest power is \(L_\tau^{n-2}\), and its coefficient is completely fixed by \(K_2\), while the \(L_\tau^{n-3}\) tower is fixed jointly by \(K_2\) and \(K_\chi\). An analogous phenomenon appears at lower logarithms. Namely, the non-contact late-time tail is given by
\begin{equation}
C_{\text{BC}}(\tau,\gamma)
=
\sum_{n=2}^\infty \sum_{k=0}^{n-2}
C_{n,k}(\gamma)\frac{L_\tau^{n-k-2}}{\tau^n}.
\end{equation}
A resummation leads to
\begin{equation}
C_{\text{BC}}(\tau,\gamma)
=
\sum_{k=0}^\infty \sum_{n=k+2}^\infty
C_{n,k}(\gamma)\frac{L_\tau^{n-k-2}}{\tau^n}
=
\sum_{k=0}^\infty C^{(k)}(\tau,\gamma),
\end{equation}
where \(C^{(0)}\) denotes the leading order, \(C^{(1)}\) the next-to-leading order, and so forth.
With $z=4ML_\tau/\tau$, the first  two orders formally resum to
\bs 
\begin{align}
 C^{(0)}
 &=\frac{iM^2K_2(x)}{2\pi\tau^2(1-z)^2},\\
 C^{(1)}
 &=\frac{iM^3\{[6-4\log(1-z)]K_2(x)+2K_\chi(x)\}}
 {\pi\tau^3(1-z)^3}.
 \label{rev:NCT-resummed}
\end{align}\es 

{ Figure~\ref{rev:figResummedComparison} compares the numerical evaluation
of the resummed expression \eqref{rev:NCT-resummed} with its leading and fixed-order
next-to-leading approximations.}
\begin{figure}[tbp]
 \centering
 \newcommand{\revResummedPanel}[3]{%
  \begin{minipage}[t]{0.48\textwidth}\centering
  \begin{tikzpicture}
   \begin{axis}[
    width=0.88\linewidth,height=0.66\linewidth,
    xmin=5,xmax=80,xtick={5,20,40,60,80},
    xlabel={$\tau$},ylabel={$\operatorname{Im}C$},
    xlabel style={text=black},ylabel style={text=black},
    title={$#1^\circ$},title style={font=\bfseries,text=black},
    axis x line*=bottom,axis y line*=left,
    tick label style={font=\scriptsize,text=black},
    scaled y ticks=base 10:2,
    ytick scale label code/.code={${\color{black}\times 10^{-2}}$},
    label style={font=\small,text=black},
    grid=major,grid style={gray!25},unbounded coords=discard]
    \addplot[black,thick,domain=5:80,samples=220]
     {#2/(2*pi*x^2*(1-4*ln(4/x)/x)^2)
      +((6-4*ln(1-4*ln(4/x)/x))*#2+2*(#3))
       /(pi*x^3*(1-4*ln(4/x)/x)^3)};
    \addplot[black,thick,dashed,domain=5:80,samples=220]
     {#2/(2*pi*x^2)};
    \addplot[black,thick,dashdotted,domain=5:80,samples=220]
     {#2/(2*pi*x^2)+(4*#2*(1.5+ln(4/x))+2*(#3))/(pi*x^3)};
   \end{axis}
  \end{tikzpicture}
  \end{minipage}}
 \revResummedPanel{30}{18.534248888541}{-55.359248782045}\hfill
 \revResummedPanel{45}{9.254314969367}{-16.908169192244}\\[3pt]
 \revResummedPanel{90}{3.281631040901}{1.678095567841}\hfill
 \revResummedPanel{120}{2.361738500327}{3.618622264284}
 \par\smallskip
 {\color{black}%
 \tikz[baseline=-0.5ex]{\draw[black,thick] (0,0)--(0.35,0);}
 Numerical simulation \quad
 \tikz[baseline=-0.5ex]{\draw[black,thick,dashed] (0,0)--(0.35,0);}
 Leading $K_2$\quad
 \tikz[baseline=-0.5ex]{\draw[black,thick,dashdotted] (0,0)--(0.35,0);}
 Fixed-order $K_2+K_\chi$}
 \caption{ Four-angle comparison for $M=1$ over $5\leq\tau\leq80$.
 The solid curve is the numerical evaluation of the resummed expression
 \eqref{rev:NCT-resummed}.
 The dashed and dash-dotted curves are respectively the leading $K_2$ term and
 the fixed-order $K_2+K_\chi$ expression in \eqref{rev:tail}.}
 \label{rev:figResummedComparison}
\end{figure}

The correlator from $\mathscr I^-$ to $\mathcal H^+$ follows from $\mathcal T_\ell/\omega$ in the same
way. Let \be s=v-v'>0,\qquad  
b_\ell=(4M)^{\ell+1}(\ell!)^3/[(2\ell)!(2\ell+1)!].\ee 
The leading non-analytic term of each fixed $\ell$ in \eqref{rev:T} gives
\begin{equation}
 \left.\widetilde C\right|_{
       \text{leading term of each fixed }\ell}
 =\frac{iM}{4\pi}\sum_{\ell=0}^{\infty}(2\ell+1)(-1)^\ell b_\ell
       \frac{\Gamma(\ell+2)}{s^{\ell+2}}P_\ell(\cos\gamma).
 \label{rev:trtower}
\end{equation}
The first three entries are $iM^2/(\pi s^2)$,
$-2iM^3\cos\gamma/(\pi s^3)$ and
$4iM^4P_2(\cos\gamma)/(3\pi s^4)$.
Accordingly, the leading late time tail is  \be \widetilde C= iM^2/(\pi s^2)+\cdots.\ee This leading horizon-channel tail is angle independent and is likewise
 controlled by the low-frequency branch structure underlying the original
 Schwarzschild power-law tails. We have demonstrated this late-time tail using numerical simulation in Figure \ref{rev:figTfourangle}.

\begin{figure}[tbp]
 \centering
 \pgfplotsset{rev select rows/.style 2 args={x filter/.code={
   \ifnum\coordindex<#1\def\pgfmathresult{nan}\fi
   \ifnum\coordindex>#2\def\pgfmathresult{nan}\fi}}}
 \begin{tikzpicture}
  \begin{axis}[
   width=0.88\textwidth,height=0.49\textwidth,
   xmin=5,xmax=80,xtick={5,20,40,60,80},
   xlabel={$s$},ylabel={$\operatorname{Im}\widetilde C$},
   scaled y ticks=base 10:2,
   ytick scale label code/.code={${\color{black}\times 10^{-2}}$},
   axis x line*=bottom,axis y line*=left,
   tick label style={font=\small,text=black},
   label style={font=\small,text=black},
   legend style={font=\scriptsize,text=black,at={(0.5,1.20)},anchor=south,legend columns=3},
   grid=major,grid style={gray!25},unbounded coords=discard]
   \addplot[black,thick,domain=5:80,samples=180]
    {6.274987640349284/(2*pi^2*x^2)
     -6.287303103615529*cos(30)/(pi^2*x^3)
     +2*6.286628931942705*(3*cos(30)^2-1)/(6*pi^2*x^4)};
   \addlegendentry{Numerical $30^\circ$}
   \addplot[black,thick,densely dashed,domain=5:80,samples=180]
    {6.274987640349284/(2*pi^2*x^2)
     -6.287303103615529*cos(45)/(pi^2*x^3)
     +2*6.286628931942705*(3*cos(45)^2-1)/(6*pi^2*x^4)};
   \addlegendentry{Numerical $45^\circ$}
   \addplot[black,thick,dashdotted,domain=5:80,samples=180]
    {6.274987640349284/(2*pi^2*x^2)
     -6.287303103615529*cos(90)/(pi^2*x^3)
     +2*6.286628931942705*(3*cos(90)^2-1)/(6*pi^2*x^4)};
   \addlegendentry{Numerical $90^\circ$}
   \addplot[black,thick,densely dotted,domain=5:80,samples=180]
    {6.274987640349284/(2*pi^2*x^2)
     -6.287303103615529*cos(120)/(pi^2*x^3)
     +2*6.286628931942705*(3*cos(120)^2-1)/(6*pi^2*x^4)};
   \addlegendentry{Numerical $120^\circ$}
   \addplot[black,very thick,loosely dashed,domain=5:80,samples=180]{1/(pi*x^2)};
   \addlegendentry{Analytic leading $1/(\pi s^2)$}
  \end{axis}
 \end{tikzpicture}
 \caption{  Late-time tail of the Carrollian correlator from $\mathscr I^-$ to $\mathcal H^+$. The long-dashed line shows the analytic leading-order result \(1/(\pi s^2)\).
 Four curves are distinguished by
 solid, dashed, dash-dotted and dotted line styles show the numerical reconstructions at four different angles.
 The single dashed curve is the angle-independent leading term.}
 \label{rev:figTfourangle}
\end{figure}}

% END INLINED revision_tails.tex
\subsection{Eikonal approximation}
\begin{figure}
	    \centering
	    \begin{tikzpicture}[scale=1.3]
	      \fill[gray!50] (0,0) circle (0.3);
	      \fill (-3,0.5) circle (1.5pt);
	      \node at (-3,0.2) {$A$};
	      \fill (3,0.5) circle (1.5pt);
	      \node at (3,0.2) {$B$};
	      \node at (0,-0.5) {\footnotesize $M$};
	      \draw[draw, thick] (-3,0.5)  .. controls (-1.5,1.2) and (1.5,1.2) .. (3,0.5);
	      \node at (-1.5,0.6) { $r'$};
	      \node at (1.5,0.6) { $r$};
	      \draw[dashed, thick] (0,0.3) -- (0,1);
	      \node at (0.2,0.7) {\footnotesize $b$};
	     
	    \end{tikzpicture}
	    \caption{Shapiro delay. A light is deflected by the massive object with mass $M$. The distance between $A/B$ and $M$ is $r'/r$ and the turning point to $M$ is approximately equal to the impact parameter $b$.}
	    \label{shapiro}
	\end{figure}
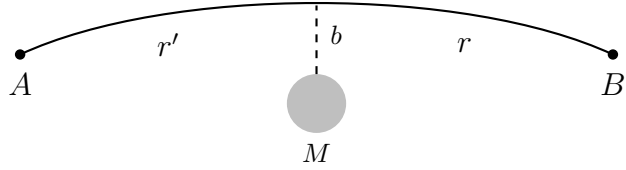

A particularly tractable regime is the eikonal limit, where the energies are large and the momentum transfer is small. This eikonal approximation is known as a method to sum over an infinite tower of Feynman diagrams in QFTs to describe the emergence of classical physics \cite{Abarbanel:1969ek,Levy:1969cr,Cheng:1969eh,Wallace:1973iu}.
In gravitational physics, there is a nice correspondence between the ultra-relativistic $2\to 2$ scalar scattering and $1\to 1$ scattering in a shock wave spacetime \cite{Aichelburg:1970dh, Dray:1984ha} and thus resummation of Feynman diagrams could reproduce the effect of classical geometries \cite{tHooft:1987vrq,Amati:1987wq}. This has been  extended to the $1\to 1$ scattering
amplitude in any stationary spacetime. More precisely, the $1\to 1$ massless scattering amplitude $\mathcal A_2$ in a Schwarzschild spacetime is  \cite{Adamo:2021rfq}
\be 
\mathcal A_2=\frac{\pi}{2M}\delta(p^0-p'^0)\mathcal A_{\text{eik}}(p,p')
\ee where $p^\mu, p'^\mu$ are the incoming and outgoing momenta of the scalar and $M$ is the mass of the black hole. The so-called eikonal amplitude is \bea 
\mathcal A_{\text{eik}}=8\pi  \frac{\alpha(s)^2}{t}\frac{\Gamma(-i \alpha(s))}{\Gamma(i\alpha(s))}\left(\frac{4\mu^2}{-t}\right)^{-i\alpha(s)}
\eea where $s$ and $t$ are Mandelstam invariants and $\mu$ is an IR regulator. We denote the momentum of the background as $P^\mu$ with 
\be 
P^\mu=M(1,0,0,0)
\ee and the incoming and outgoing momenta are 
\bea 
p^\mu=\omega n^\mu=\omega(1,n^i),\quad p'^\mu=\omega' n'^\mu=\omega'(1,n'^i).
\eea  Here $\omega,\omega'$ are energy of the incoming and outgoing massless scalars, respectively. Thus the Mandelstam invariants are 
\bea 
s=-(p+P)^2=M^2+2M\omega,\quad t=-(p-p')^2=2\omega \omega' n\cdot n'=-4\omega \omega'\sin^2\frac{\gamma}{2}.
\eea Thus the function $\alpha(s)$ is  \bea 
\alpha(s)&\equiv &s-M^2=2\omega M \eea  and the eikonal amplitude $\mathcal A_{\text{eik}}$ is 
\bea 
\mathcal A_{\text{eik}}=-\frac{8\pi M^2\omega}{\omega'\sin^2\frac{\gamma}{2}}\frac{\Gamma(-2iM\omega)}{\Gamma(2iM\omega)}\left(\frac{\mu^2}{\omega\omega'\sin^2\frac{\gamma}{2}}\right)^{-2iM\omega}.
\eea The Carrollian amplitude is \footnote{The Carrollian amplitude considered here is slightly generalized because the background is a Schwarzschild black hole. From the perspective of $2\to 2$ scattering, two of the momenta are massive and describe the background, while the other two are massless. The Fourier transform in this Carrollian amplitude applies only to the massless particles, making it a partial Carrollian amplitude \cite{Liu:2025oom}. Further discussion on the celestial analog can be found in \cite{2024JHEP...10..192A}.}
\bea 
B(u,\Omega;v',\Omega')&=&\left(\frac{i}{8\pi^2}\right)^2\int_0^\infty d\omega \int_0^\infty d\omega' e^{-i\omega' u+i\omega v'}\mathcal A_2\nn\\&=&\frac{M}{16\pi^2\sin^2\frac{\gamma}{2}}\int_0^\infty d\omega e^{-i\omega(u-v')} \frac{\Gamma(-2iM\omega)}{\Gamma(2iM\omega)}\left(\frac{\mu^2}{\omega^2\sin^2\frac{\gamma}{2}}\right)^{-2iM\omega}.
\eea To relate it to the Carrollian correlator $C(u,\Omega;v',\Omega')$ from $\mathscr I^-$ to $\mathscr I^+$, we should extract the imaginary part 
\bea 
C(u,\Omega;v',\Omega')&=&B(u,\Omega;v',\Omega')-B^*(u,\Omega;v',\Omega')\nn\\&=&\frac{M}{16\pi^2\sin^2\frac{\gamma}{2}}\int_{-\infty+i\epsilon}^{\infty+i\epsilon}d\omega e^{-i\omega(u-v')}\frac{\Gamma(-2iM\omega)}{\Gamma(2iM\omega)}\left(\frac{\mu^2}{\omega^2\sin^2\frac{\gamma}{2}}\right)^{-2iM\omega}\nn\\&=&\frac{1}{32\pi^2\sin^2\frac{\gamma}{2}}\int_{-\infty+i\epsilon}^{\infty+i\epsilon}dx e^{-ix z} \frac{\Gamma(-ix)}{\Gamma(ix)}\left(x^2\right)^{ix}
\eea where we have defined 
\be 
z=\frac{u-v'}{2M}+2\ln\frac{2M\mu}{|\sin\frac{\gamma}{2}|}.
\ee We have inserted $i\epsilon$ in reference to the previous section. There is no analytic result for the integral \footnote{However, the integral can be evaluated order by order by expanding the last term $x^{-2ix}$. And the first term corresponds to the contribution of the Newton potential. Please find more details in appendix \ref{intI}.}. However, one finds that the Carrollian correlator $C(u,\Omega;v',\Omega')$ is a function of $z$ up to a factor that is determined by the deflection angle 
\be 
C(u,\Omega;v',\Omega')=\frac{1}{32\pi^2\sin^2\frac{\gamma}{2}}\mathcal I(z)
\ee where
\be 
\mathcal I(z)=\int_{-\infty+i\epsilon}^{\infty+i\epsilon}dx e^{-ix z} \frac{\Gamma(-ix)}{\Gamma(ix)}\left(x^2\right)^{ix}.
\ee 
As a function of a complex variable $z$, The poles and branch cuts of $\mathcal I(z)$ should reflect the physical properties of the Carrollian correlator $C(u,\Omega;v',\Omega')$. Indeed, when $z>0$, we may choose the contour in the lower half plane to evaluate the integral. On the other hand, when $z<0$, we may choose the contour in the upper half plane to evaluate the integral. Therefore, $z=0$ determines a constraining equation between the incoming and outgoing data
\be 
\frac{u-v'}{2M}+2\ln\frac{2M\mu}{|\sin\frac{\gamma}{2}|}=0.\label{cri1}
\ee 
Physically, a classical massless particle coming from $(v',\Omega')\in \mathscr I^-$ propagates along a null geodesic and arrives at $(u,\Omega)\in \mathscr I^+$. This already determines a classical equation between the initial and final data via the null geodesic. To check this point, we notice that the deflection angle is small in the eikonal approximation and thus the equation is approximately 
\be 
u-v'+4M\ln \frac{4M\mu}{\gamma}=0. \label{uphir}
\ee The deflection angle of the light is 
\be 
\gamma=\frac{4M}{b}
\ee where $b$ is the impact parameter and thus 
\be 
u-v'+4M \ln \mu b=0.
\ee We introduce two IR regulators $r\gg 2M$ and $r'\gg 2M$ and thus 
\be 
(t-r_*(r))-(t'+r_*(r'))+4M\ln \mu b=0\quad\Rightarrow\quad  t-t'-(r+r')\approx 2M\ln \frac{r r'}{4M^2\mu^2 b^2}.\label{delay}
\ee In flat spacetime, $r+r'$ is the duration for a light that is emitted from the initial point and received in the final point. Thus the left hand side  is the Shapiro time delay during the process 
\be 
\left(\Delta t\right)_{\text{Shapiro}}\approx 2M\ln \frac{r r'}{4M^2\mu^2 b^2}.\label{shp}
\ee 
Figure \ref{shapiro} is a schematic diagram for the Shapiro delay. A light is emitted from A and reflected from B  and finally received by A, the Shapiro delay is \cite{1972gcpa.book.....W}
\be 
\left(\Delta t\right)_{A\to B\to A}=4M\left(1+\ln\frac{4rr'}{b^2}\right).\label{shap2}
\ee 
%The constant $4M$ may be neglected when the IR regulators $r,r'$ are extremly large.
To match the logarithm between \eqref{shp} and \eqref{shap2}, we should identify the IR regulator $\mu$ from the scattering amplitude with\footnote{In the following section, we will find a more precise relation between $\mu$ and the reference point in the definition of tortoise coordinate.}
\be 
1+4\ln 2+2\ln M\mu=0.
\ee 
Note that \eqref{shap2} should divide a factor 2 to match with \eqref{shp} since the time delay in \eqref{shp} is for a single travel.

\section{Classical equations at null boundaries}\label{alg}
In the previous section, we  clarified the relationship between the Carrollian correlators and the reflection and transmission amplitudes in black hole $1\to 1$ scattering. The reflection and transmission amplitudes are defined via the partial wave expansion and the poles in these amplitudes are called QNMs. On the other hand, the Carrollian correlators are the dual descriptions of the same process. Physically, there should be poles for the Carrollian correlators in the position space. According to the propagation theorem of singularity \cite{duistermaat1994fourier,hormander2009analysis}, the poles should be determined by the null geodesics coming from $\mathscr I^-$\footnote{Roughly speaking, this states that singularities propagate along null geodesics. This property can also be found by the Hadamard
form for the retarded Green function \cite{Poisson:2011nh}.  We should mention that Hadamard form is valid only in convex geodesic domains. More discussion on the singularity structure of the Green's function can be found in  \cite{Harte:2012uw,Zenginoglu:2012xe}.}. More precisely, the geodesics from $\mathscr I^-$ to $\mathscr I^+$  define a classical equation 
\be 
\mathcal F(u,\Omega;v',\Omega')=0.\label{F}
\ee Similarly, the poles of the Carrollian correlator $\widetilde C(v,\Omega;v',\Omega')$  should be governed by the null geodesics from $\mathscr I^-$ to $\mathcal H^+$ and thus define another  equation 
\be 
\widetilde{\mathcal F}(v,\Omega;v',\Omega')=0.\label{WF}
\ee Depending on the impact parameter, 
the equation \eqref{F} exists for $b>b_c$ while \eqref{WF} exists for $b<b_c$. The critical value $b_c=3\sqrt{3}M$ is determined by the impact parameter of the photon sphere. In the following, we will derive the two equations and discuss their properties. 
\subsection{\texorpdfstring{From $\mathscr I^-$ to $\mathscr I^+$}{From past null infinity to future null infinity}}
\begin{figure}
    \centering
    \usetikzlibrary{decorations.text}
    \begin{tikzpicture} [scale=0.8]
        \draw[dashed,thick] (0,0)  -- (3,3) node[above]{\footnotesize $i^+$};
        \draw[draw,thick] (3,3) -- (6,0) node[right]{\footnotesize $i^0$};
        
        \draw[dashed,thick] (0,0) -- (3,-3) node[below]{\footnotesize $i^-$};
        \draw[draw, thick](3,-3) -- (6,0);
       \draw[decorate, decoration={snake, amplitude=0.4mm, segment length=1mm}] (-1,3) -- (3,3);
       \draw[decorate, decoration={snake, amplitude=0.4mm, segment length=1mm}] (-1,-3) -- (3,-3);
       \draw[->,thick] (4,-2) -- (3,-1);
       \draw[draw, thick] (3,-1) -- (2,0);
       \node at (1.7,0) {\footnotesize $r_0$};
       \draw[->, thick] (2,0) -- (3,1);
       \draw[draw,thick] (3,1) -- (4,2);
       \node at (4.4,-2.4) {\footnotesize $(v',\Om')$};
       \node at (4.5,2.2) {\footnotesize $(u,\Om)$};

    \end{tikzpicture}
    \caption{\centering Deflection of light in Schwarzschild spacetime.}
    \label{incoming}
\end{figure}
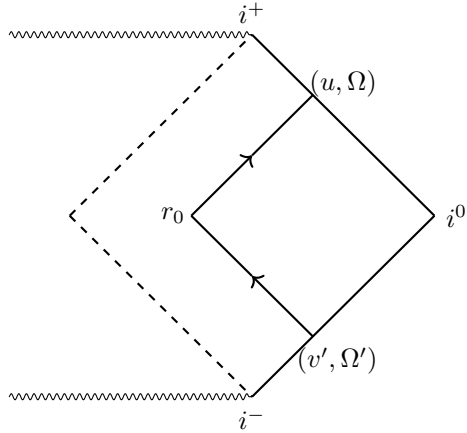

The null geodesics  are shown in Figure \ref{incoming}. 
It starts from $(v',\Omega')$, passes through the turning point $(t_0,r_0,\Omega_0)$ and arrives at $(u,\Omega)$. For the first part of the journey, we use the ingoing Eddington-Finkelstein  coordinates $(v,r,\Omega)$ such that the metric is 
\be ds^2=-A dv^2+2dv dr+r^2d\Omega.
\ee We fix the geodesics in the equatorial plane $\theta=\theta_0=\theta'=\frac{\pi}{2}$ using spherical symmetry. Due to the time translation invariance and the rotational symmetry of the metric, we may assume the Hamilton-Jacobi action as
\be 
S=-E v+R(r;E,L)+L\phi\label{saction}
\ee where $E$ and $L$ are the energy and angular momentum of the light. The action $S$ obeys the eikonal equation
\be 
g^{\mu\nu}\partial_\mu S\partial_\nu S=0.\label{geodesics}
\ee  A constant $S$ defines a hypersurface in the spacetime whose co-normal vector is the  momentum 1-form 
\be 
p_\mu=\frac{\partial S}{\partial x^\mu}. 
\ee To be more precise, 
\be 
p_0=-E,\quad p_\phi=L,\quad p_r=\frac{\partial  R(r;E,L)}{\partial r}.
\ee Substituting into the equation \eqref{geodesics}, we find 
\be 
p_r=\frac{E}{A}\left(1-\sqrt{1-A\frac{b^2}{r^2}}\right).
\ee Since $S$ is a constant along the geodesics, we find 
\bea 
v=\frac{\partial R}{\partial E}\Big|_L,\quad \phi=-\frac{\partial R}{\partial L}\Big|_E.
\eea Thus, we find the relation of the coordinates along the geodesics
\bs\begin{align}
v_0-v'&=v(r_0)-v(r\to\infty)=\int_{\infty}^{r_0} dr \frac{\partial p_r}{\partial E}\Big|_L=\int_{r_0}^\infty dr \frac{1-\sqrt{1-\frac{Ab^2}{r^2}}}{A \sqrt{1-\frac{Ab^2}{r^2}}},\\
\phi_0-\phi'&=\phi(r_0)-\phi(r\to\infty)=-\int_{\infty}^{r_0} dr \frac{\partial p_r}{\partial L}\Big|_E=\int_{r_0}^\infty dr\frac{1}{r\sqrt{\frac{r^2}{b^2}-A}}.
\end{align}\es One can derive the same result using Killing vectors and solve the geodesic equations. The turning point $r_0$ is determined by the largest positive root of the equation 
\be 
r_0^2=A(r_0) b^2.\label{defr0}
\ee The photon sphere corresponds to $r_0=r_{ps}=3M$ and $b=b_c=3\sqrt{3}M$. For $b>b_c$, we find 
\bs\label{vphirelation}\begin{align}
    v_0-v'&=\int_{r_0}^\infty dr \frac{1-\sqrt{1-\frac{A(r)r_0^2}{r^2A(r_0)}}}{A(r)\sqrt{1-\frac{A(r)r_0^2}{r^2A(r_0)}}},\label{vvp}\\
    \phi_0-\phi'&=\int_{r_0}^\infty dr \frac{1}{r\sqrt{\frac{r^2}{r_0^2} A(r_0)-A(r)}}.\label{phip}
\end{align}\es The equations \eqref{vphirelation} define an  equation between $v_0-v'$ and $\phi_0-\phi'$ since both of them are parameterized by $r_0>3M$.

\begin{figure}
    \centering
    \includegraphics[width=5in]{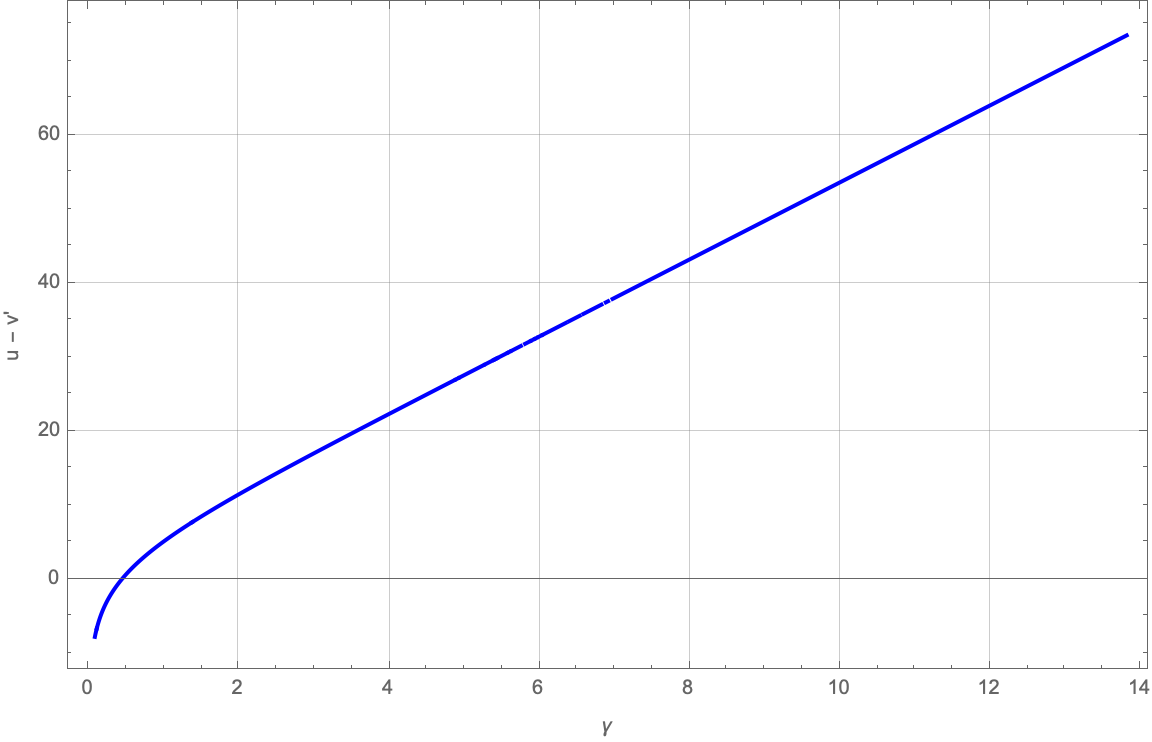}
    \caption{The equation $u-v'=h(\gamma)$ that connects the deflection angle and the time delay for $b>b_c$. We have set $M=1$ in the diagram.}
    \label{defl}
\end{figure}

For the outgoing trip, we switch to the outgoing Eddington-Finkelstein coordinates $(u,r,\Omega)$ and find a set of similar equations
\bs\label{uphirelation}\begin{align}
    u-u_0&=\int_{r_0}^\infty dr \frac{1-\sqrt{1-\frac{A(r)r_0^2}{r^2A(r_0)}}}{A(r)\sqrt{1-\frac{A(r)r_0^2}{r^2A(r_0)}}},\label{uu0}\\
    \phi-\phi_0&=\int_{r_0}^\infty dr \frac{1}{r\sqrt{\frac{r^2}{r_0^2} A(r_0)-A(r)}}.\label{phi0}
\end{align}\es The coordinates at the turning point should be eliminated. Therefore, we should glue the two  sets of equations \eqref{vphirelation} and \eqref{uphirelation} at $r_0$ to get 
\bs
\label{eqnF}
\begin{align}
    u-v'&=2\int_{r_0}^\infty dr \frac{1-\sqrt{1-\frac{A(r)r_0^2}{r^2A(r_0)}}}{A(r)\sqrt{1-\frac{A(r)r_0^2}{r^2A(r_0)}}}-2r_*(r_0),\\
    \phi-\phi'&=2\int_{r_0}^\infty dr \frac{1}{r\sqrt{\frac{r^2}{r_0^2} A(r_0)-A(r)}}.\label{wind}
\end{align}
\es We have used the relation 
\be 
u_0-v_0=-2r_*(r_0)
\ee at the turning point.  We may eliminate $r_0$
and get the equation 
\be 
\frac{u-v'}{M}=h(\gamma)\label{algeeqn}
\ee where $\gamma=\phi-\phi'-\pi$ is the deflection angle. The above equation is the implicit form of the  equation \eqref{F}. Figure \ref{defl} is the curve for the equation that connects the deflection angle and the  time delay \footnote{Strictly speaking, there is a small region that $u-v'$ is negative and it is not suitable to use the word ``delay''. Also, the equation \eqref{algeeqn} is not exactly the same meaning as Shapiro delay. In Shapiro delay, the time delay $t-t'$ is related to the impact parameter. However, in our equation, the  time delay $u-v'$ is related to the deflection angle. It should be understood as a consequence of the combination of the light deflection and Shapiro delay.}. 
Properties are discussed in the following: 
\begin{enumerate}
    \item Note that the deflection angle $\gamma$ can exceed $\pi$ in the Figure. In principle, the azimuthal angle is periodic which leads to the identification
    \be 
    \gamma\sim \gamma+2\pi.
    \ee However, the integration \eqref{wind} ignores this fact and it has the advantage to encode the winding number for the light to wrap around the black hole. For example, when $b\to b_c$, an incoming light should wrap around the black hole multiple times before it escapes from the gravitational potential. Due to this reason, we have extended the domain of deflection angle. The winding number is the integer part of $\frac{\gamma}{2\pi}$
    \be 
    w=\lfloor \frac{\gamma}{2\pi}\rfloor.
    \ee Now the time delay $u-v'$ is a  monotonically  increasing function of $\gamma$.
     \item When the deflection angle $\gamma\to 0$, the impact parameter is large and we find the logarithm behavior 
    \be 
    \frac{u-v'}{M}=4\ln \gamma+2+4\zeta+4\ln\left(\zeta-1\right),\quad \gamma\to 0.
    \ee This result matches with \eqref{uphir} by connecting the IR regulator $\mu$ with the choice of the reference point $o$ via 
    \be 
    1+2\zeta+2\ln(\zeta-1)+2\ln 4M\mu=0.\label{corres}
    \ee With the standard reference point, we find 
    \be 
    \frac{u-v'}{M}=4\ln\gamma+2+\mathcal{O}(\gamma).
    \ee We have proved the logarithmic behavior in appendix \ref{aA}. Note that this region corresponds to the  weak field limit, the difference $u-v'<0$ and the outgoing light arrives at  $\mathscr I^+_U$. The critical point $u-v'=0$ depends on the reference point. For the standard choice of $o$, we find the critical deflection angle 
 \be 
    \gamma_0\approx 0.4352.
    \ee Note that this deflection angle is outside the region where $\gamma\to 0$.
    The corresponding impact parameter and the turning point are 
    \be 
    b_0=12.3455,\quad r_0=11.1877.
    \ee 
    \item When $\gamma\gg 1$, the time delay $u-v'$ is a linear function of $\gamma$. This is in the strong field region. In appendix \ref{aA}, we  have proved that 
    \be 
    \frac{u-v'}{M}=3\sqrt{3}\gamma+c_0+\mathcal{O}(\gamma^{-1})  \label{linser}
    \ee in the large $\gamma$ limit and the constant $c_0$ can be found in \eqref{A44}. The slope of the curve $3\sqrt{3}$ is actually equal to the impact parameter $b_c$. Physically, when $b\to b_c$, the light wraps around the photon sphere and the main contribution of the time delay is from the winding around the photon sphere. The angular frequency of the photon around the photon sphere is 
    \be 
    \Omega_c=\frac{d\phi}{dt}\Big|_{r=r_{ps}}=\frac{1}{b_c}.
    \ee Therefore, the time delay is approximately equal to the product of winding number and $b_c$. This interprets the linear function \eqref{linser} in the large $\gamma$ limit.
   Given a deflection angle $\gamma\in(0,\pi)$, then from the Figure one finds that $\gamma+2\pi w,\ w\ge 1$ is already large and the linear approximation is valid. Consider a light ray emitting near $\mathscr I^-$ with fixed $(v',\Omega')$  and a detector that is located near $\mathscr I^+$ at a fixed angle $\Omega$, as shown in Figure \ref{windingnumber}, the light ray may wrap around the photon sphere many times to arrive at the detector.
   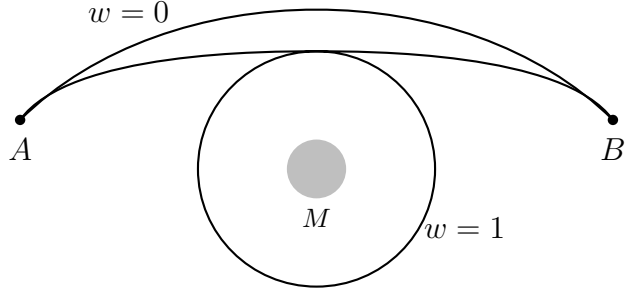
\begin{figure}
    \centering
    \begin{tikzpicture}[scale=1.3]
      % 中心质量 M
      \fill[gray!50] (0,0) circle (0.3);
      \node at (0,-0.5) {\footnotesize $M$};
      
      % 左端点 A
      \fill (-3,0.5) circle (1.5pt);
      \node at (-3,0.2) {$A$};
      % 右端点 B
      \fill (3,0.5) circle (1.5pt);
      \node at (3,0.2) {$B$};
      
      % w = 0 曲线：不缠绕，光滑拱形，远离中心（向上拱起）
      \draw[thick] (-3,0.5) .. controls (-1.5,2.0) and (1.5,2.0) .. (3,0.5);
      \node at (-1.9,1.6) {$w=0$};
      
      % w = 1 曲线：缠绕 M 一圈，半径 1.2，起始段与末段均向上拱起（与 w=0 同向）
      \draw[thick] 
        (-3,0.5) .. controls (-2.5,1.2) and (-0.5,1.2) .. (0,1.2)  % A → 圆弧起点，终点切线水平向右
        arc (90:450:1.2)                                            % 逆时针绕 M 一整圈，起点/终点切线水平向右
        .. controls (0.5,1.2) and (2.5,1.2) .. (3,0.5);            % 圆弧终点 → B，起点切线水平向右
      \node at (1.5,-0.6) {$w=1$};
    \end{tikzpicture}\caption{Light rays with different winding numbers ($w=0,1$) emitted from A and received at B. }\label{windingnumber}
\end{figure}
   Then the observer will detect signals at equal time intervals \footnote{In fact, there are light rays that orbit the black hole in the opposite direction, whose corresponding deflection angle is $2\pi-\gamma$. For an observer at a fixed angle near $\mathscr I^+$, the signals are received alternately, first from one direction and then the other.
}
   \be 
   h(\gamma+2\pi (w+1))-h(\gamma+2\pi w)\approx 2\pi b_c
   \ee for any positive winding number $w\ge 1$. The exception case is $w=0$ where $h(\gamma)$ may not be a linear function of $\gamma$ once the deflection angle is small.
   
    \item Comparison with the results in the literature. In \cite{bozza2002gravitational}, the logarithmic divergence for the deflection angle $\gamma= -\ln\left(\frac{b}{b_c}-1\right)$ has been derived in the strong field region. Then our results show a similar logarithmic divergence for the time delay in the same region 
    \be 
    u-v'=-3\sqrt{3}M \ln \left(\frac{b}{b_c}-1\right).
    \ee This is consistent with \cite{bozza2004time}.
  %  \item Zero point.
   \item Relationship between time delay and deflection angle of neighboring light rays.
For neighboring light rays emitted from the same position $(v',\Omega')$ with slightly different impact parameters, their deflection angles and time delays are also slightly different. The relation is governed by the derivative
\begin{equation}
\frac{d(u-v')}{d\gamma}=M h'(\gamma).
\end{equation}
The asymptotic behavior of $h'(\gamma)$ is
\begin{equation}
h'(\gamma)=\left\{\begin{array}{cc}\frac{4}{\gamma}+\cdots,&\gamma\to 0,\\
3\sqrt{3}+\cdots,&\gamma\to\infty.\end{array}\right.
\end{equation}
The strong-field limit result is consistent with that in \cite{2014AmJPh..82..564M}.
    \item We may use an analytic function $g(\gamma)$ to fit the curve $h(\gamma)$
    \bea 
    g(\gamma)=\alpha_1\ln\left(e^{\alpha_2\gamma}-1\right)+\alpha_3\ln \left(e^{\alpha_4\gamma}-1\right)+\alpha_5
    \eea where $\alpha_i,\ i=1,2,\cdots,5$ are constants. By assuming $\alpha_2,\alpha_4>0$ and  matching with $h(\gamma)$ in the small and large $\gamma$ limit, we find four equations for $\alpha'_i$s
    \bs\begin{align}
        \alpha_1\ln \alpha_2+\alpha_3\ln \alpha_4+\alpha_5&=2,\\
        \alpha_1+\alpha_3&=4,\\
        \alpha_1\alpha_2+\alpha_3\alpha_4&=3\sqrt{3},\\
        \alpha_5&=c_0.
    \end{align}\es This does not fix all the constants uniquely; we can adjust the values to fit the curve $h(\gamma)$. Figure \ref{fit} shows a function $g(\gamma)$ that fits $h(\gamma)$. The function $g(\gamma)$ matches $h(\gamma)$ nicely over the whole region. However, it should be noted that in the intermediate region (e.g., $0.2<\gamma<0.8$), the differences between the two functions are smoothed out due to the large scale of the Figure. In Figure \ref{fitsa}, we zoom into the intermediate region and see that small differences indeed exist.
    \begin{figure}
        \centering
        \includegraphics[width=4in]{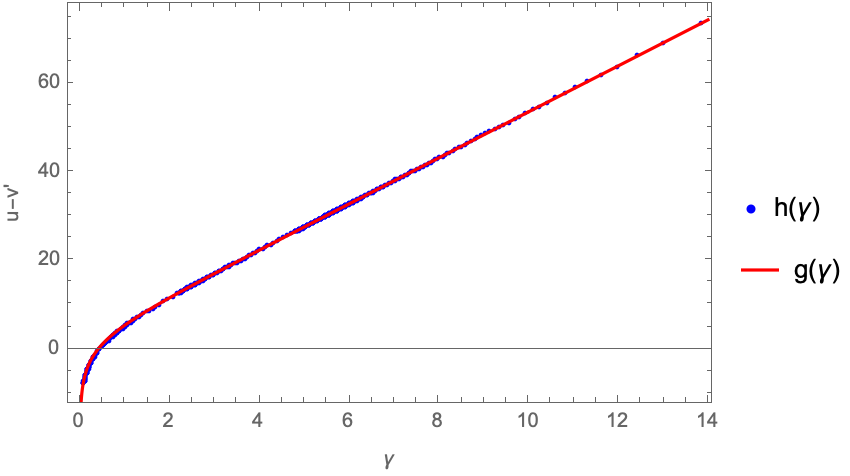}
        \caption{We fit the curve $h(\gamma)$ using the function $g(\gamma)$ by setting $\alpha_1=0.738,\ \alpha_2=0.615,\ \alpha_3=3.262,\ \alpha_4=1.454,\ \alpha_5=1.674$.}.
        \label{fit}
    \end{figure}
     \begin{figure}
        \centering
        \includegraphics[width=4in]{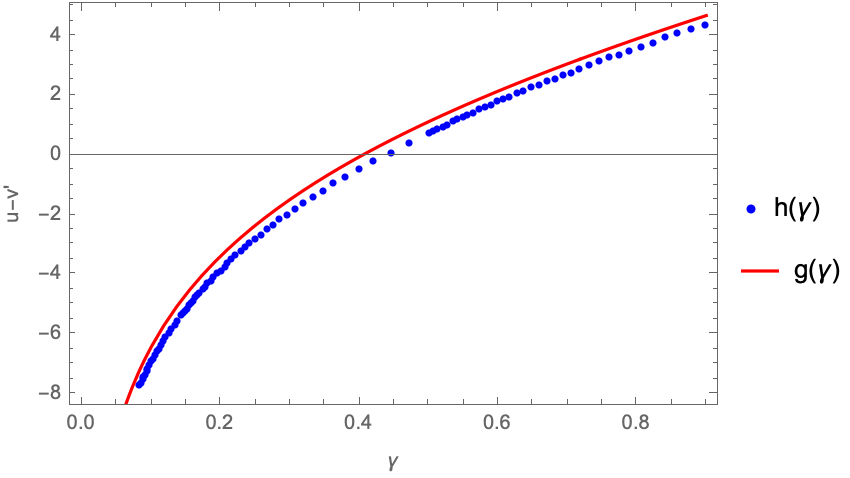}
        \caption{\centering The functions $h(\gamma)$ and  $g(\gamma)$ in the intermediate region $0.1<\gamma<0.9$.}.
        \label{fitsa}
    \end{figure}
   % the function $u-v'=h(\gamma)$ should be a multivalued function  We have 
   \item The previous properties can be extended to general spherically symmetric spacetimes. Further details can be found in appendix \ref{aA}.
\end{enumerate}

\subsection{\texorpdfstring{From $\mathscr I^-$ to $\mathcal H^+$}{From past null infinity to the future horizon}}
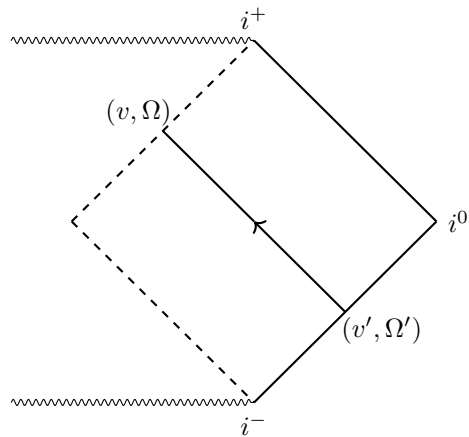
\begin{figure}
    \centering
    \usetikzlibrary{decorations.text}
    \begin{tikzpicture} [scale=0.8]
        \draw[dashed,thick] (0,0)  -- (3,3) node[above]{\footnotesize $i^+$};
        \draw[draw,thick] (3,3) -- (6,0) node[right]{\footnotesize $i^0$};
        
        \draw[dashed,thick] (0,0) -- (3,-3) node[below]{\footnotesize $i^-$};
        \draw[draw, thick](3,-3) -- (6,0);
       \draw[decorate, decoration={snake, amplitude=0.4mm, segment length=1mm}] (-1,3) -- (3,3);
       \draw[decorate, decoration={snake, amplitude=0.4mm, segment length=1mm}] (-1,-3) -- (3,-3);
       \draw[->,thick] (4.5,-1.5) -- (3,0);
       \draw[draw, thick] (3,0) -- (1.5,1.5);
       \node at (5.1,-1.8) {\footnotesize $(v',\Om')$};
       \node at (1.1,1.8) {\footnotesize $(v,\Om)$};

    \end{tikzpicture}
    \caption{\centering The geodesics that falls into the Schwarzschild black hole.}
    \label{falln}
\end{figure}

\begin{figure}
    \centering
    \includegraphics[width=4in]{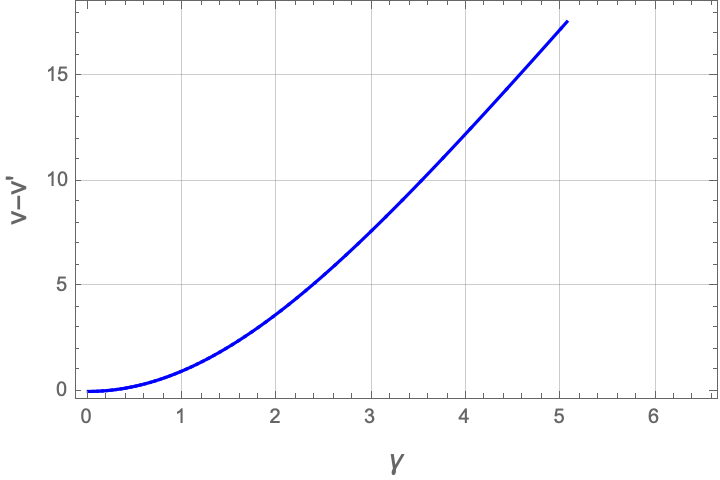}
    \caption{The equation $v-v'=H(\gamma)$ that connects the deflection angle and the time delay for $b<b_c$. We have set $M=1$ in the diagram.}
    \label{htilde}
\end{figure}

The discussion in the previous subsection should be modified when $0\le b<b_c$. In this case, a light ray from $\mathscr I^-$ will cross the potential barrier and arrive at $\mathcal H^+$. This is shown in Figure \ref{falln}.
The time delay is $v-v'$ and the deflection angle is $\gamma=\phi-\phi'$\footnote{Note that the definition of the deflection angle is not the same as the previous subsection. We will not subtract $\pi$ from $\phi-\phi'$ which ensures that $\gamma$ is always greater than zero.}
\bs\label{uvb}\begin{align}
    v-v'&=\int_{r_h}^{\infty}dr \frac{1-\sqrt{1-\frac{A(r)b^2}{r^2}}}{A(r)\sqrt{1-\frac{A(r)b^2}{r^2}}},\\
    \phi-\phi'&=\int_{r_h}^\infty dr\frac{1}{r\sqrt{\frac{r^2}{b^2}-A(r)}} 
\end{align}\es where $r_h=2M$ is the Schwarzschild radius. The functions \eqref{uvb} are parameterized by the impact parameter $0\le b<b_c$ and we write the function \eqref{WF} as follows:
\be 
v-v'=H(\gamma).
\ee Figure \ref{htilde} is the function $H(\gamma)$. Similar to $h(\gamma)$, it is also a monotonically increasing function of $\gamma$ but with a completely different behaviour.
\begin{enumerate}
    \item The time delay $v-v'$ is always non-negative
    \be 
    v-v'\ge 0.
    \ee This is consistent with the Carrollian correlator from $\mathscr I^-$ to $\mathcal H^+$ since it is non-vanishing only for $v>v'$
    \be 
    \widetilde C(v,\Omega;v',\Omega')\propto \Theta(v-v').
    \ee When $v<v'$, no light can propagate to $\mathcal H^+$ as this scenario breaks causality.
    \item When the impact parameter $b\to 0$, the light plunges into the black hole directly. This corresponds to the small deflection angle limit $\gamma\to 0$. One finds a hyperbolic behaviour
    \be 
    \frac{v-v'}{M}=\gamma^2+\mathcal{O}(\gamma^4).
    \ee This is proven in appendix \ref{aA2}.
    \item When the impact parameter $b\to b_c^-$, the light  circles around the photon sphere many times before it falls into the black hole. This corresponds to the limit $\gamma\to\infty$, and we find a linear behaviour 
    \be 
    \frac{v-v'}{M}=3\sqrt{3}\gamma+\widetilde{c}_0+\mathcal{O}(\gamma^{-1}). \label{linearfunction}
    \ee This is discussed in appendix \ref{aA2}.
    \item We can use a rational function 
    \be 
    G(\gamma)=\frac{a \gamma^2+3 \sqrt{3} \gamma^3}{\frac{2 a \gamma}{3 \sqrt{3}}+a+\gamma^2},\quad a=-\widetilde{c}_0=8.878
    \ee to fit the function $H(\gamma)$. One can check that $G(\gamma)$ and $H(\gamma)$ have the same asymptotic expansion for $\gamma\to\infty$ and $\gamma\to 0$. Figure \ref{fitH} shows that the difference of these two functions is small. 
    \begin{figure}
        \centering
        \includegraphics[width=4in]{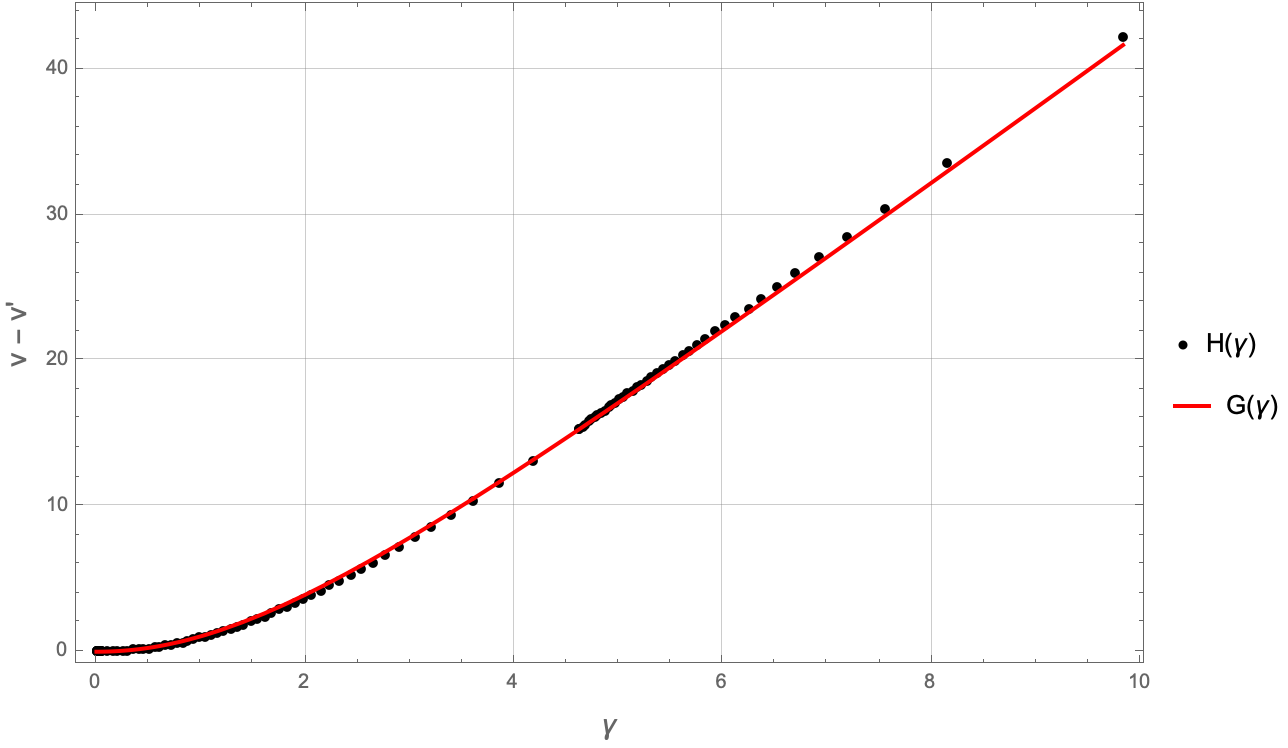}
        \caption{\centering We fit the curve $H(\gamma)$ using the rational function $G(\gamma)$.}
        \label{fitH}
    \end{figure}
\end{enumerate}

\section{Conclusion and discussion}
In this work, we have defined two null-boundary two-point Carrollian correlators in Schwarzschild black hole perturbation theory. The regularized correlator from $\mathscr I^-$ to $\mathscr I^+$ is the Fourier transform of the scattering amplitude in a partial wave expansion and is free of IR divergence. By contrast, the correlator from $\mathscr I^-$ to $\mathcal H^+$ has not been discussed in previous works. These two correlators are determined by the reflection and transmission amplitudes, respectively. 

{ For separated angles, the leading late-time tail of $\mathscr I^-\to \mathscr I^+$ Carrollian correlator is
\be C\sim \frac{iM^2K_2(\gamma)}{2\pi\tau^2}\ee  with the closed angular distribution function $K_2$ given by 
\eqref{rev:K2closed}. At order $\tau^{-3}$, the non-contact $\log^2\tau$
coefficient vanishes and the $\log\tau$ coefficient is fixed by $K_2$.
The constant term contains the function $K_\chi$. On the other hand, the leading late-time tail of $\mathscr I^-\to \mathcal H^+$ Carrollian correlator obeys a similar power law 
\be 
\widetilde C\sim \frac{iM^2}{\pi s^2}. 
\ee There is no angular dependence in this leading term. We have checked the late-time tails using numerical simulation.}

We have also studied the poles of the correlators that are determined by the light rays connecting the initial and final null boundaries. These poles are encoded in two equations that determine the relationship between Shapiro time delay and deflection angle. We analyze the asymptotic behavior of the equations in the limits of large and small deflection angles. The corresponding functions are monotonically increasing curves and can be fitted by simple analytic functions. Furthermore, the linear growth at large deflection angles is controlled by the photon sphere and thus reflects strong-field information.

Many topics deserve further study:

% END INLINED revision_conclusion.tex
 \begin{itemize}
\item \textbf{Black holes in flat holography.}
 % {\lon In this work, we have defined two-point Carrollian correlators in black holes. The derivation follows from the extrapolation dictionary and matches  black hole perturbation theory perfectly. However, due to the lack of knowledge of Carrollian conformal field theory on the boundary, it is not easy to identify the dual description of the asymptotically flat black holes. Nevertheless, our work is motivated by analogy of the flat holography and AdS/CFT. In AdS/CFT, an AdS black hole is dual to a certain thermal conformal field theory. The two point function on an AdS black hole background can describe the thermal correlators in the dual CFT and correspond to the linear-response theory (D.T.Son).
  %In AdS$_3$/CFT$_2$, there are also evidence that the black hole two-point function is captured by 
 % the heavy-heavy-light-light four-point correlation functions in the dual CFT, where the heavy state generates the black hole and the light state corresponds to the perturbations. Therefore, one may conjecture that the dual of an asymptotic black hole is given by a non-perturbative state or ensemble in the dual Carrollian CFT. The explicit description needs more understanding on Carrollian CFTs.
%  }
   This work is largely motivated by the established AdS/CFT correspondence: an
AdS--Schwarzschild geometry represents a thermal boundary state
\cite{Witten:1998zw}, retarded two-point functions encode linear response
\cite{Son:2002sd}, and heavy--heavy--light--light correlators in
AdS$_3$/CFT$_2$ describe light probes in a heavy background and exhibit
black-hole thermality \cite{Fitzpatrick:2015zha}. This suggests that a black
hole in an asymptotically flat spacetime may represent a thermal boundary
Carrollian state. In this work, we do not consider the thermal behaviour of
the correlator\footnote{Including the Hawking thermal factors would require a
separate choice of the Hartle--Hawking state. This is not considered here.}.
Instead, we study the retarded two-point function and find that Carrollian
correlators exhibit behaviour similar to that of the recent source--response
prescription based on holographic renormalization \cite{Ammon:2025flat}.
This requires a complementary flat-space version of the source--response
dictionary from holographic renormalization and a GKPW-like prescription,
with the renormalized radial canonical momentum playing the role of the
boundary response. Moreover, it would be better to identify black-hole microstates
or to construct a complete microscopic dual Carrollian CFT that reproduces the Carrollian correlator in the future.
     \item \textbf{Poles and reconstruction of Carrollian correlator.}
It is well known that the null geodesics in Schwarzschild and even in Kerr spacetime are governed by elliptic functions \cite{Gralla:2019ceu}. Applications of elliptic functions in black hole physics can be found in \cite{Gibbons:2011rh}. In this work, the poles of the Carrollian correlators are parameterized by the impact parameter or the turning point. One should eliminate these parameters to obtain an analytic expression for the poles. It would be interesting to explore whether it is possible to reconstruct the Carrollian correlator from the knowledge of the poles. Such an approach may provide an alternative way to understand the QNMs \cite{1984PhRvD..30..295F} and the branch cuts of the Green's function in momentum space. QNMs have been studied for decades using various methods \cite{1972ApJ...172L..95G,Blome:1981azp,Schutz:1985km,Leaver:1985ax,Leaver:1986gd,Mano:1996vt,Sasaki:2003xr} and are also important for black hole spectroscopy \cite{Berti:2025hly}. Note that the geometric interpretations of the QNMs have already been discussed in \cite{Dolan:2011fh,PhysRevD.86.104006}. It would be interesting to explore this topic from a Carrollian perspective.
     
      \item \textbf{Ambiguity.}  We have used Bondi coordinates to derive the Carrollian correlators and classical equations that characterize the relationship between the time delay and the deflection angle of light. An ambiguity arises from the choice of the reference point $o$ corresponding to the definition of the tortoise coordinates. In \eqref{corres}, we could relate the reference point to the IR regulator $\mu$ in the scattering amplitude. The Bondi coordinate system is motivated by the need to adapt to asymptotic inertial observers, e.g., GW detectors. This is crucial because gravitational waveforms are always studied in a specific coordinate system. However, it should be emphasized that the choice of coordinates is one of the most delicate and, sometimes, confusing aspects of general relativity. Besides the previous ambiguity, there is an additional ambiguity from supertranslations. A supertranslation corresponds to an angle-dependent choice of the origin of the time coordinate. In terms of Bondi coordinates, this is the nontrivial transformation that preserves the asymptotic structure near $\mathscr I^+$,
\begin{equation}
\tilde u = u + T(\Omega) + \mathcal{O}(r^{-1}), \quad \tilde r = r + \mathcal{O}(1), \quad \tilde \theta^A = \theta^A + \mathcal{O}(r^{-1}),
\end{equation}
where $T(\Omega)$ is a smooth function on $S^2$ that parameterizes the supertranslation.\footnote{Choosing a function $T(\Omega)$ means fixing a BMS frame. For each solution of general relativity, there is a supertranslated solution \cite{Hawking:2016msc,Compere:2016jwb,Compere:2016hzt}. This BMS frame-fixing problem is also important in numerical relativity \cite{Mitman:2024uss,Mitman:2021xkq,MaganaZertuche:2021syq}.} Note that there is a corresponding supertranslation near $\mathscr I^-$,
\begin{equation}
\tilde v = v + T^{(-)}(\Omega) + \mathcal{O}(r^{-1}), \quad \tilde r = r + \mathcal{O}(1), \quad \tilde \theta^A = \theta^A + \mathcal{O}(r^{-1}).
\end{equation}
It is claimed that the functions $T^{(-)}$ and $T$ are related via an antipodal map \cite{Strominger:2013jfa} such that there is a unique BMS transformation in the whole spacetime. Now the poles of the Carrollian correlator $u-v' = h(\gamma)$ are sensitive to supertranslations.\footnote{Note that this is in contrast with the one in flat spacetime, where the pole equation reduces to $u=v'$, $\Omega=\Omega'^{\text{P}}$, and this equation is invariant upon imposing the antipodal map.} This sensitivity suggests that the poles of the Carrollian correlator may be used to determine the BMS frame.
    \item \textbf{Kerr black hole.} Kerr spacetime is important for understanding various phenomena in astronomy \cite{Chandrasekhar:1985kt}. Interestingly, the method presented in this work can be extended directly to Kerr spacetime. Several topics in this direction are worth exploring. First, in a rotating background, geodesics are not always confined to the equatorial plane. The constraint equations that describe the poles of the Carrollian correlator should include the polar angle. Moreover, there are various orbits in Kerr spacetime, depending on the initial data \cite{Compere:2021bkk}. It would be interesting to explore how to understand these aspects from a Carrollian approach. Second, in Kerr spacetime, the scattering of radiation produces waves with an amplitude larger than that of the incident wave under certain conditions. This phenomenon is called superradiance \cite{1971JETPL..14..180Z,Bekenstein:1998nt,Brito:2015oca} and is absent in Schwarzschild spacetime. Note that in the spherically symmetric case, the reflection amplitude is related to the Carrollian correlator via
\begin{equation}
(-1)^{\ell+1}\mathcal R_{\omega,\ell}=1-4\pi\int_{-\infty}^\infty du\omega e^{i\omega u}\int_{0}^\pi \sin\gamma d\gamma C(u,\Omega;0,\Omega'^{\text P}) P_\ell(\cos\gamma).
\end{equation}
The absence of superradiance imposes constraints on the Carrollian correlator:
\begin{equation}
|\mathcal R_{\omega,\ell}|^2\le 1.
\end{equation}
For a rotating spacetime, this inequality can be violated, and the constraints on the Carrollian correlator should be relaxed. It would be interesting to explore this point in the future. 
    \item \textbf{Non-linear corrections.} In this work, we have studied the Carrollian correlator in black hole perturbation theory at the linearized level. From the perspective of scattering amplitudes, there are various non-linear corrections that can be written down systematically via Feynman rules, as shown in Figure \ref{correction}. Note that one should extract the classical aspects from the loop corrections. It would be interesting to explore their relation to second-order perturbation theory \cite{Gleiser:1995gx,Campanelli:1998jv} and the self-force problem \cite{DeWitt:1960fc,Poisson:2011nh}.
    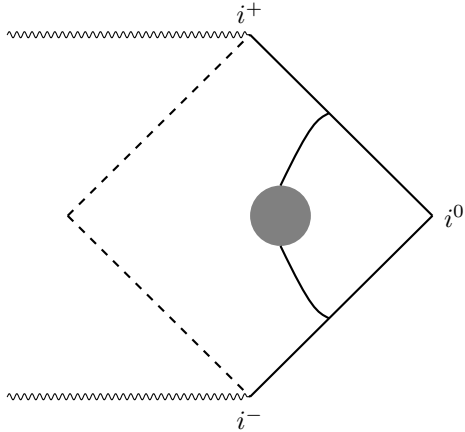
\begin{figure}
    \centering
    \usetikzlibrary{decorations.text}
    \begin{tikzpicture} [scale=0.8]
        \draw[dashed,thick] (0,0)  -- (3,3) node[above]{\footnotesize $i^+$};
        \draw[draw,thick] (3,3) -- (6,0) node[right]{\footnotesize $i^0$};
        
        \draw[dashed,thick] (0,0) -- (3,-3) node[below]{\footnotesize $i^-$};
        \draw[draw, thick](3,-3) -- (6,0);
       \draw[decorate, decoration={snake, amplitude=0.4mm, segment length=1mm}] (-1,3) -- (3,3);
       \draw[decorate, decoration={snake, amplitude=0.4mm, segment length=1mm}] (-1,-3) -- (3,-3);
       \fill[gray!100] (3.5,0) circle (0.5);
       \draw[thick] (4.3,-1.7) .. controls (4.1,-1.6) and (4,-1.55) .. (3.5,-0.5);
       \draw[thick] (3.5,0.5) .. controls (4,1.55) and (4.1,1.6) .. (4.3,1.7);

       %\node at (4.4,-2.4) {\footnotesize $(v',\Om')$};
       %\node at (4.5,2.2) {\footnotesize $(u,\Om)$};

    \end{tikzpicture}
    \caption{\centering Higher-order corrections.}
    \label{correction}
\end{figure}
 \end{itemize}

 \vspace{3pt}
{\bf Acknowledgments.} 
The work of J.L. is supported by NSFC Grant No. 12575074.

\appendix

%\section{Carrollian null-boundary data and the Fourier transform}
%\label{app:carroll-background}

%For the massless scalar used in this work, the bulk falloff is given by \eqref{falloof}.
%The leading coefficients $\Sigma$ and $\Sigma^{(-)}$ are spin-zero 
%operators of Carrollian weight $\Delta=1$.  Their correlator is ``Carrollian'' because
%its insertions live on these null boundaries. 

%\section{\lon The summation in $K_2$ and $K_\chi$}\label{rev:appendix}
\iffalse For completeness, the Abel sum \eqref{rev:Kchi} can be evaluated by
ordinary one-dimensional quadratures for $x<1$. Define
\begin{align}
 d&=\sqrt{2(1-x)},\qquad D(t,x)=1-2xt+t^2,\\
 U(t,x)&=D(t,x)^{-1/2}
 \log\!\left[\frac{1-xt+\sqrt{D(t,x)}}{2D(t,x)}\right],\\
 W(t,x)&=\frac{1-t^2}{D(t,x)^{3/2}},\\
 V(x)&=\frac14\int_0^1(t^{-2}-t^2)U(t^2,x)\,dt,\\
 \mathcal H_2(x)&=-\int_0^1
 \frac{-2\log(1-t)+\log t}{1-t}\,W(t,x)\,dt.
\end{align}
Then
\begin{equation}
 K_\chi(x)=2\mathcal H_2(x)-\frac2{1-x}
 -\frac{15}{4}U(1,x)-\frac14V(x)
 -\frac{15}{8d}+\frac{d}{32}+5.
 \label{rev:KchiIntegral}
\end{equation}
This identity sums the candidate \eqref{rev:chi-definition}; it does not remove its
physical qualification. The values used in the plots are
\begin{center}
\begin{tabular}{rrr}
\toprule
$\gamma$ & $K_2(\gamma)$ & candidate $K_\chi(\gamma)$\\
\midrule
$30^\circ$ & $18.53424889$ & $-55.35924878$\\
$45^\circ$ & $9.25431497$ & $-16.90816919$\\
$90^\circ$ & $3.28163104$ & $1.67809557$\\
$120^\circ$ & $2.36173850$ & $3.61862226$\\
\bottomrule
\end{tabular}
\end{center}\fi
\section{Integration}\label{intI}
In this appendix, we will explore the relation between the  integral $\mathcal I(z)$ and the effect of Newton potential. We expand $x^{2ix}$ as 
\be 
x^{2ix}=\sum_{k=0}^\infty \frac{1}{k!}(2ix \ln x)^k
\ee and then we can write $\mathcal I(z)$ as a formal expansion
\be 
\mathcal I(z)=\sum_{k=0}^\infty \mathcal I_k(z)
\ee where the $\mathcal I_0$ is 
\bea 
\mathcal I_0(z)=\int_{-\infty+i\epsilon}^{\infty+i\epsilon} dx e^{-iz x}\frac{\Gamma(-ix)}{\Gamma(ix)}.
\eea This is non-vanishing only for $z>0$
\bea 
\mathcal I_0(z)&=&\Theta(z)\int_{-\infty+i\epsilon}^{\infty+i\epsilon} dx e^{-iz x}\frac{\Gamma(-ix)}{\Gamma(ix)}\nn\\&=&-2\pi i \Theta(z)\sum_{n=1}^\infty \text{Res}_{x=-in}\left(e^{-iz x}\frac{\Gamma(-ix)}{\Gamma(ix)}\right)\nn\\&=&{ -2 \pi \Theta(z) e^{-\frac{z}{2}} J_1\left(2 e^{-\frac{z}{2}}\right)}
\eea We have used residue theorem to find the integration. The Bessel function $J_1(2e^{-z/2})$ is real and non-negative for $z>0$ . Thus there is no pole for $\mathcal I_0(z)$ in the region $z>0$.
Now we show that the same structure appears in the Carrollian correlator contributed by Newton potential. 

In the geometric optics limit, one can use the WKB approximation to derive the phase shift factor \cite{Andersson:2000tf,Decanini:2011xi}. The starting point is the  massless scattering problem in Schwarzschild spacetime 
\be 
\frac{d^2}{dr_*^2}G_{\ell,\omega}(r)+Q G_{\ell,\omega}(r)=0
\ee where 
$
Q=\omega^2-V_\ell(r).
$ The lowest order WKB solution of this equation is
\bea 
G_{\ell,\omega}(r)=Q^{-1/4}e^{i\int^r dr_* \sqrt{Q}}\propto e^{iS_r}
\eea where the phase $S_r$ satisfies the Hamilton-Jacobi equation for a light ray with energy $E=\omega$ and angular momentum $L\approx \ell+\frac{1}{2}$.
\bea 
\left(\frac{dS_r}{dr}\right)^2=\frac{1}{\Delta^2}(E^2 r^4-L^2\Delta)\quad\Rightarrow\quad \frac{dS_r}{dr}=\pm\frac{E}{A(r)}\sqrt{1-\frac{A(r)b^2}{r^2}}
\eea where $\Delta=r^2-2Mr=r^2 A(r)$ and the impact parameter $b=\frac{|L|}{E}$. To match the solution with the boundary condition at $r_*\to\infty$, the phase shift factor should be identified as \cite{Glampedakis:2001cx}
\bea 
\delta_\ell&\approx &\int_{r_0}^\infty dr \frac{1}{A(r)}(Q^{1/2}-E)-Er_*(r_0)+\frac{\pi}{2}L.
\eea We would expect that for the large $L$ partial waves, it 
feels only the far-zone Newtonian field. In calculating the partial-wave sum for the scattering amplitude, it is convenient to split it into two terms
\bea 
f(\omega,\gamma)=f_d(\omega,\gamma)+f_n(\omega,\gamma)
\eea where $f_d$ represents the part of the scattering amplitude that carries the information of the main diffraction effects\footnote{Please find the associated glory effects in \cite{Anninos:1992ih,Nambu:2015aea}.} while $f_n$ denotes the  amplitude coming from the Newton potential. The phase shift from the Newton potential has been solved explicitly by \cite{1988sfbh.book.....F} 
\be 
e^{2i\delta_\ell}=\frac{\Gamma(\ell+1-2iM\omega)}{\Gamma(\ell+1+2iM\omega)} 
\ee  and  thus we have
\bea 
f_n(\omega,\gamma)=M\frac{\Gamma(1-2iM\omega)}{\Gamma(1+2iM\omega)}\left(\sin^2\frac{\gamma}{2}\right)^{-1+2iM\omega}.
\eea We can also decompose the boundary-to-boundary correlator into a  Newtonian and a diffraction part
\bea 
C(u,\Omega;v',\Omega')=C_n(u,\Omega;v',\Omega')+C_d(u,\Omega;v',\Omega')
\eea where contribution from the Newton potential  is
\bea 
C_n(u,\Omega;v',\Omega'^P)&\propto& \int_{-\infty+i\epsilon}^{\infty+i\epsilon} d\omega f_n(\omega,\gamma)e^{-i\omega(u-v')}\nn\\&=& M\int_{-\infty+i\epsilon}^{\infty+i\epsilon} d\omega e^{-i\omega(u-v')}\frac{\Gamma(1-2iM\omega)}{\Gamma(1+2iM\omega)}\left(\sin^2\frac{\gamma}{2}\right)^{-1+2iM\omega}\nn\\&=&\frac{1}{2\sin^2\frac{\gamma}{2}}\int_{-\infty+i\epsilon}^{\infty+i\epsilon} dx e^{-ix\frac{u-v'}{2M}}\frac{\Gamma(-ix)}{\Gamma(ix)}e^{ix\ln\sin^2\frac{\gamma}{2}}\nn\\&=&\frac{1}{2\sin^2\frac{\gamma}{2}}\int_{-\infty+i\epsilon}^{\infty+i\epsilon} dx e^{-i(\frac{u-v'}{2M}-2\ln|\sin\frac{\gamma}{2}|)x}\frac{\Gamma(-ix)}{\Gamma(ix)}\nn\\&\propto &\frac{1}{\sin^2\frac{\gamma}{2}}\int_{-\infty+i\epsilon}^{\infty+i\epsilon} dx e^{-iz' x}\frac{\Gamma(-ix)}{\Gamma(ix)}\nn\\&=&\Theta(z')\frac{\mathcal I_0(z')}{\sin^2\frac{\gamma}{2}}
\eea where 
\be 
z'=\frac{u-v'}{2M}-2\ln|\sin\frac{\gamma}{2}|.
\ee 
The critical equation in this case is 
\be 
z'=\frac{u-v'}{2M}-2\ln|\sin\frac{\gamma}{2}|=0\label{cri2}
\ee which is approximately the same as \eqref{cri1} up to an IR regulator. %since the deflection angle is small in both cases.  However, the Bessel function $J_1(2e^{-z'})$ has no poles for any  $z'\in\mathbb R$ and thus point $z'=\text{const.}$ is not the pole of $C_n(u,\Omega;v',\Omega')$. Thus, we conclude that to find the poles of the Carrollian propagator, it is no enough to only consider  the contribution from the Newton potential. Interestingly, there are infinite many zeros in the negative $z'$ axis. Figure \ref{Bessel} shows the Bessel function $J_1(e^{-z'})$
%\begin{figure}
 %   \centering
 %   \includegraphics[width=4in]{Bessel.png}
 %   \caption{The Bessel function $J_1(e^{-z'})$ has infinite many zeros }
%    \label{Bessel}
%\end{figure}

\section{\texorpdfstring{The equation from $\mathscr I^-$ to $\mathscr I^+$}{The equation from past null infinity to future null infinity}}\label{aA}
In this appendix, we will investigate the properties of the function $h(\gamma)$ in \eqref{algeeqn}. When the impact parameter is large, we can prove that $r_0$ is large via \eqref{defr0}. As a matter of fact, 
\be 
b=\frac{r_0}{\sqrt{A(r_0)}}=r_0(1+\mathscr R^{-1}+\mathcal{O}(\mathscr R^{-2})),\quad r_0\to\infty.
\ee We have defined dimensionless parameter 
\be 
\mathscr R=\frac{r_0}{M}.
\ee The equation \eqref{eqnF} can be expanded in large $r_0$ limit
\bs\begin{align}
    u-v'&=-4M\ln \mathscr R+2M\left(1+2\zeta+2\ln 4+2\ln(\zeta-1)\right)+\mathcal{O}(\mathscr{R}^{-1}),\\
    \gamma&=\frac{4}{\mathscr R}+\mathcal{O}(\mathscr R^{-2}).
\end{align}\es As a consequence, we find 
\be 
\frac{u-v'}{M}=4\ln\gamma+2+4\zeta+4\ln\left(\zeta-1\right)+\mathcal{O}(\gamma).
\ee 
The result can be extended to a general spherically symmetric in an asymptotically flat spacetime.  The general line element is described by
\begin{align}
    ds^2=-A(r)dt^2+B(r)dr^2+C(r)(d\theta^2+\sin^2{\theta}d\phi^2)\label{spherical}
\end{align}
in which $A(r),B(r)$ and $C(r)$ satisfy an asymptotically flat condition
\begin{align}
    \lim_{r\rightarrow\infty}A(r)=1,\quad \lim_{r\rightarrow\infty}B(r)=1,\quad \lim_{r\rightarrow\infty}C(r)=r^2.
\end{align} Note that one can always redefine the radius $r$ to set $C=r^2$. The tortoise coordinate is 
\be 
r_*(r)=\int_o^r dr' \sqrt{\frac{B(r')}{A(r')}}.
\ee 

However, we will keep this function in this work. A straightforward derivation leads to 
\bs\label{suv}\begin{align} 
u-v'&=2\int_{r_0}^\infty dr \sqrt{\frac{B}{A}}\left(\frac{1}{\sqrt{1-\frac{Ab^2}{C}}}-1\right)-2r_*(r_0),\\
\phi-\phi'&=2\int_{r_0}^\infty dr \sqrt{\frac{B}{C}}\frac{1}{\sqrt{\frac{C}{Ab^2}-1}}.
\end{align}\es The turning point $r_0$ is related to $b$ via 
\be b^2=\frac{C(r_0)}{A(r_0)}.
\ee When $r_0\to\infty$, the functions $A,B,C$ are expanded as 
\be 
A(r)=1+\frac{a_{-1}}{r}+\cdots,\quad B(r)=1+\frac{b_{-1}}{r}+\cdots,\quad C(r)=r^2+c_1 r+\cdots.
\ee Then we find\footnote{ We have ignored the $\cdots$ terms in the expression. This is valid for Schwarzchild black hole or large $o$.}
\bs\begin{align}
    u-v'&=-(b_{-1}-a_{-1})\ln r_0+c_1-a_{-1}+{ (b_{-1}-a_{-1})\ln 2o+2o}+\cdots,\\
    \phi-\phi'&=\pi+\frac{b_{-1}-a_{-1}}{r_0}+\cdots.
\end{align}\es The turning point is approximately equal to the impact parameter
\be 
b=r_0+\frac{c_{1}-a_{-1}}{2}+\mathcal{O}(r_0^{-1}).
\ee Therefore, the relation between the time delay and the deflection angle is 
\be 
u-v'=(b_{-1}-a_{-1})\ln\gamma+{ (b_{-1}-a_{-1})\ln\frac{2o}{b_{-1}-a_{-1}}+c_{1}-a_{-1}+2o}+\mathcal{O}(\gamma).
\ee The logarithmic behavior only depends on the coefficient $b_{-1}-a_{-1}$ which is the same for both Schwarzschild and  RN black holes. 

In the opposite limit, $\gamma\to\infty$, we should deal with the strong field region. The light deflection in the strong field region has been discussed in \cite{bozza2002gravitational} and we will adopt a similar method as  in \cite{tsukamoto2017deflection} in this appendix. We still use the metric \eqref{spherical}. We assume the photon sphere is located at $r_{ps}$ and thus obeys the equation 
\begin{align}
    D(r)\equiv\frac{C'(r)}{C(r)}-\frac{A'(r)}{A(r)}\Big{|}_{r=r_{ps}}=0.
\end{align} We define a function 
\bea 
K(r)=\frac{C}{Ab^2}-1=\frac{C(r)A(r_0)}{A(r)C(r_0)}-1
\eea that satisfies the condition 
\be
K(r_0)=0.
\ee The equation \eqref{suv} can be written as 
\bs\label{uvc}\begin{align}
    u-v'&=2\int_{r_0}^\infty dr \sqrt{\frac{B}{A}}\frac{1}{\sqrt{ K}\left(\sqrt{K+1}+\sqrt{K}\right)}-2r_*(r_0),\\
    \phi-\phi'&=2\int_{r_0}^\infty dr \sqrt{\frac{B}{CK}}.
\end{align}\es 
When $r_0\to r_{ps}$, the integral may be divergent. Therefore, we change the variable $r=\frac{r_0}{1-x}$ with $0<x<1$. Then 
\bea 
K(r)=K(\frac{r_0}{1-x})=K_1(r_0)x+K_2(r_0)x^2+\cdots
\eea where 
\bs\begin{align}
    K_1(r_0)&=r_0 D_0,\\
    K_2(r_0)&=\frac{1}{2} r_0 \left(D_0^2 r_0+2 D_0+r_0D_0' \right)
\end{align}\es  where $D_0=D(r_0),\ D'_0=\frac{d}{dr_0}D(r_0)$. When $r_0\to r_{ps}$, we have 
\be 
D_0=D(r_{ps})+D'(r_{ps})(r_0-r_{ps})+\cdots=D'_{ps}(r_0-r_{ps})+\cdots
\ee with \be D'_{ps}=\frac{dD(r_0)}{dr_0}\Big|_{r_0=r_{ps}}.\ee  
Expanding $K_1(r_0)$ and $K_2(r_0)$ around $r_{ps}$, we find 
\bea 
K_1(r_0)=r_{ps}D'_{ps}(r_0-r_{ps})+\cdots,\quad K_2(r_0)=\frac{1}{2}r_{ps}^2 D'_{ps}+\cdots.
\eea 

The integrals \eqref{uvc} are improper integrals and they diverge logarithmically
\bs\label{log}\begin{align} 
u-v'&\sim \ln (r_0-r_{ps}),\\
\phi-\phi'&\sim \ln (r_0-r_{ps}).
\end{align}\es 
We use the limit comparison test (LCT) for improper integrals to find the ratio 
\be 
\lim_{r_0\to r_{ps}}\frac{u-v'}{\gamma}=\lim_{r\to r_{ps}}\frac{\sqrt{\frac{B}{A}}\frac{1}{\sqrt{ K}\left(\sqrt{K+1}+\sqrt{K}\right)}}{\sqrt{\frac{B}{CK}}}=\sqrt{\frac{C(r_{ps})}{A(r_{ps})}}=b_c
\ee where $b_c$ is the  critical impact parameter. Thus, we prove the linear relation 
\be 
u-v'=b_c\gamma+\cdots.\label{uvgamma}
\ee To determine the constant, we should compute \eqref{log} more precisely
\bs\label{GFv}\begin{align}
    u-v'&=2r_0\int_0^1 \frac{dx}{\sqrt{G(r_0,x)}}-2r_*(r_0),\\
    \phi-\phi'&=2r_0\int_0^1 \frac{dx}{\sqrt{F(r_0,x)}}
\end{align}\es where 
\bs\begin{align}
    G(r_0,x)&=(1-x)^4 \frac{A}{B}K \left(\sqrt{K}+\sqrt{K+1}\right)^2,\\
    F(r_0,x)&=(1-x)^4 \frac{CK}{B}.
\end{align}\es Note that the functions are evaluated at $r=\frac{r_0}{1-x}$. Now it is straightforward to find the asymptotic  expansion
\bs\begin{align}
    G(r_0,x)&=g_1 x+g_{\frac{3}{2}}x^{\frac{3}{2}}+g_2 x^2+\cdots,\\
    F(r_0,x)&=f_1 x+f_2 x^2+\cdots
\end{align}\es where 
\bs\label{coefg}\begin{align}
    f_1&=\frac{K_1(r_0)C(r_0)}{B(r_0)}=\frac{r_{ps}D'_{ps}C_{ps}}{B_{ps}}(r_0-r_{ps})+\cdots,\\
    f_2&=\frac{K_2(r_0)C(r_0)}{B(r_0)}+K_1(r_0)\times (\cdots)=\frac{r_{ps}^2D'_{ps}C_{ps}}{2B_{ps}}+\cdots,\\
    g_1&=\frac{K_1(r_0)A(r_0)}{B(r_0)}=\frac{r_{ps}D'_{ps}A_{ps}}{B_{ps}}(r_0-r_{ps})+\cdots,\\
    g_{\frac{3}{2}}&=\frac{2K_1^{\frac{3}{2}}(r_0) A(r_0)}{B(r_0)}=\cdots,\\
    g_2&=\frac{K_2(r_0)A(r_0)}{B(r_0)}+K_1(r_0)\times(\cdots)=\frac{r_{ps}^2D'_{ps}A_{ps}}{2B_{ps}}+\cdots.
\end{align}\es The terms denoted by $\cdots$ are not important for extracting the logarithmic behavior, so we do not write them out explicitly. The subscript ``ps'' denotes the value of the corresponding function at the photon sphere.

We define the divergent part as 
\bs\begin{align}
    \mathcal I_D&=2r_0\int_0^1 \frac{dx}{\sqrt{g_1x+g_{\frac{3}{2}}x^{\frac{3}{2}}+g_2 x^2}}=\frac{4r_0}{\sqrt{g_2}}\ln\left(\frac{ 2\sqrt{g_1g_2}-g_{\frac{3}{2}}}{2\sqrt{g_2(g_1+g_2+g_{\frac{3}{2}})}-2g_2-g_{\frac{3}{2}}}\right),\\
    \mathcal J_D&=2r_0\int_0^1 \frac{dx}{\sqrt{f_1 x+f_2 x^2}}=\frac{4r_0}{\sqrt{f_2}}\ln \left(\frac{\sqrt{f_1}}{\sqrt{f_1+f_2}-\sqrt{f_2}}\right)
\end{align}\es 
and separate the integrals \eqref{GFv} into divergent and convergent parts 
\bs\begin{align}
    u-v'&=\mathcal I_D+\mathcal I_C,\\
    \gamma&=\mathcal J_D+\mathcal J_C
\end{align}\es where 

\bs\begin{align}
    \mathcal I_C&=2r_0\int_0^1 dx\left(\frac{1}{\sqrt{G(r_0,x)}}-\frac{1}{\sqrt{g_1x+g_{\frac{3}{2}}x^{\frac{3}{2}}+g_2 x^2}}\right)-2r_*(r_0),\\
    \mathcal J_C&=2r_0\int_0^1 dx \left(\frac{1}{\sqrt{F(r_0,x)}}-\frac{1}{\sqrt{f_1 x+f_2 x^2}}\right)-\pi.
\end{align}\es Utilizing \eqref{coefg}, we find

\bs\begin{align}
    \mathcal I_D&=\frac{2r_{ps}}{\sqrt{g_2(r_{ps})}}\ln \frac{2r_{ps}}{(r_0-r_{ps})}+\cdots,\\
    \mathcal J_D&=\frac{2r_{ps}}{\sqrt{f_2(r_{ps})}}\ln \frac{2r_{ps}}{(r_0-r_{ps})}+\cdots.
\end{align}\es The ratio of the coefficients is 
\be 
\lim_{r_0\to r_{ps}}\frac{u-v'}{\gamma}=\lim_{r_0\to r_{ps}}\frac{\mathcal I_D}{\mathcal J_D}=\sqrt{\frac{f_2(r_{ps})}{g_2(r_{ps})}}=\sqrt{\frac{C_{ps}}{A_{ps}}}=b_c
\ee which is consistent with \eqref{uvgamma}.
Thus, the time delay and the deflection angle in the strong field limit are

\bs\begin{align}
    u-v'&=-\frac{2r_{ps}}{\sqrt{g_2(r_{ps})}} \ln\frac{ r_0-r_{ps}}{2r_{ps}}+\mathcal I_C+\mathcal{O}(r_0-r_{ps}),\\
    \gamma&=-\frac{2r_{ps}}{\sqrt{f_2(r_{ps})}}\ln \frac{r_0-r_{ps}}{2r_{ps}} +\mathcal J_C+\mathcal{O}(r_0-r_{ps}).
\end{align}\es  As a consequence, 
\be 
u-v'=b_c\gamma+\mathcal I_C-b_c\mathcal J_C+\mathcal{O}(r_0-r_{ps}).
\ee 

This is a linear function and we can find the constant 
\be 
c_0=\frac{\mathcal I_C-b_c\mathcal J_C}{M}\Big|_{r_0=r_{ps}}.
\ee 
Note that the logarithmic divergence can be written as 
\bs\begin{align}
    \mathcal I_D&=-\sqrt{\frac{2B_{ps}}{A_{ps}D'_{ps}}}\ln \left(\frac{b_0}{b_c}-1\right)+\sqrt{\frac{2B_{ps}}{A_{ps}D'_{ps}}}\ln \left(D'_{ps}r_{ps}^2\right)+\cdots,\\
    \mathcal J_D&=-\sqrt{\frac{2B_{ps}}{C_{ps}D'_{ps}}}\ln \left(\frac{b_0}{b_c}-1\right)+\sqrt{\frac{2B_{ps}}{C_{ps}D'_{ps}}}\ln \left(D'_{ps} r_{ps}^2\right)+\cdots
\end{align}\es
alternatively using the identity 
\be 
\frac{b_0-b_{c}}{b_c}=\frac{\sqrt{\frac{C(r_0)}{A(r_0)}}-\sqrt{\frac{C_{ps}}{A_{ps}}}}{\sqrt{\frac{C_{ps}}{A_{ps}}}}=\frac{1}{4}D'_{ps}(r_0-r_{ps})^2.
\ee 
For the Schwarzschild black hole, we find 
\bs\begin{align}
    \mathcal I_D&=-3\sqrt{3}M\ln\left(\frac{b_0}{b_c}-1\right)+3\sqrt{3}M\ln 6 ,\\
    \mathcal J_D&=-\ln\left(\frac{b_0}{b_c}-1\right)+\ln 6.
\end{align}\es Furthermore, 
\bs\begin{align}
    G(r_{ps},x)&=\frac{1}{9} x^2 (3-2 x) \left(\sqrt{(3-2 x) x^2}+1\right)^2,\\
    F(r_{ps},x)&=3 M^2 (3-2 x) x^2,\\
    f_2(r_{ps})&=9M^2,\quad g_2(r_{ps})=\frac{1}{3}.
\end{align}\es We find 
\be 
\mathcal J_C\Big|_{r_0=r_{ps}}=2\ln \left(12-6\sqrt{3}\right)-\pi.
\ee Similarly, 
\begin{align}
    \mathcal I_C\Big|_{r_0=r_{ps}}=&2 M \Big[-3 \sqrt{3}+3+\ln \left(5404-3120 \sqrt{3}\right)\nn\\&\qquad -6 \sqrt{3} \ln \left(\sqrt{3}+1\right)+3 \sqrt{3} \ln (12)+2\ln 2\Big]\nn\\&+4M\left(\zeta+\ln(\zeta-1)\right).
\end{align} With the standard choice of the reference point, we find 
\be 
c_0\approx 1.674.\label{A44}
\ee We checked this result numerically.

\section{\texorpdfstring{The equation from $\mathscr I^-$ to $\mathcal H^+$}{The equation from past null infinity to the future horizon}}\label{aA2}
To prove the hyperbolic behaviour in the small $b$ limit, we expand the integrals directly 
\bs\begin{align}
    \frac{v-v'}{M}&=\frac{b^2}{4}+\mathcal{O}(b^4),\\
    \phi-\phi'&=\frac{b}{2}+\mathcal{O}(b^3).
\end{align}\es Thus they obey the law 
\be 
\frac{v-v'}{M}=\gamma^2+\mathcal{O}(\gamma^4).
\ee 
Now in the limit $b\to b_c$, both of the integrals \eqref{uvb} diverge logarithmically
\be 
v-v'\sim \ln\left( 1-\frac{b}{b_c}\right),\quad \phi-\phi'\sim \ln\left( 1-\frac{b}{b_c}\right).
\ee 
We can still use the limit comparison test for improper integrals to find the ratio 
\bea 
\lim_{r\to r_{ps}}\frac{\frac{1-\sqrt{1-\frac{A(r)b^2}{r^2}}}{A(r)\sqrt{1-\frac{A(r)b^2}{r^2}}}}{\frac{1}{r\sqrt{\frac{r^2}{b^2}-A(r)}} }=3\sqrt{3}M.
\eea This fixes the coefficient $3\sqrt{3}$ in \eqref{linearfunction}. The constant $\widetilde c_0$ can be found by taking the limit 
\be 
\widetilde c_0=\lim_{b\to b_c}\left(\frac{v-v'}{M}-3\sqrt{3}(\phi-\phi')\right)\approx -8.878.
\ee

\iffalse

\fi
%\input{output.bbl}
\providecommand{\href}[2]{#2}\begingroup\raggedright\endgroup


\begin{thebibliography}{99}

\bibitem{a}
Author, \emph{Title}, \emph{J. Abbrev.} {\bf vol} (year) pg.

\bibitem{b}
Author, \emph{Title},
arxiv:1234.5678.

\bibitem{c}
Author, \emph{Title},
Publisher (year).


% Please avoid comments such as "For a review'', "For some examples",
% "and references therein" or move them in the text. In general,
% please leave only references in the bibliography and move all
% accessory text in footnotes.

% Also, please have only one work for each \bibitem.


\end{thebibliography}

\begin{thebibliography}{100}

\bibitem{LIGOScientific:2016aoc}
{\bf LIGO Scientific, Virgo} Collaboration, B.~P. Abbott {\em et al.},
  ``{Observation of Gravitational Waves from a Binary Black Hole Merger},''
  {\em Phys. Rev. Lett.} {\bf 116} (2016), no.~6, 061102,
  \href{http://www.arXiv.org/abs/1602.03837}{{\tt 1602.03837}}.

\bibitem{Blanchet:2013haa}
L.~Blanchet, ``{Post-Newtonian Theory for Gravitational Waves},'' {\em Living
  Rev. Rel.} {\bf 17} (2014) 2, \href{http://www.arXiv.org/abs/1310.1528}{{\tt
  1310.1528}}.

\bibitem{Damour:2001tu}
T.~Damour, ``{Coalescence of two spinning black holes: an effective one-body
  approach},'' {\em Phys. Rev. D} {\bf 64} (2001) 124013,
  \href{http://www.arXiv.org/abs/gr-qc/0103018}{{\tt gr-qc/0103018}}.

\bibitem{Goldberger:2004jt}
W.~D. Goldberger and I.~Z. Rothstein, ``{An Effective field theory of gravity
  for extended objects},'' {\em Phys. Rev. D} {\bf 73} (2006) 104029,
  \href{http://www.arXiv.org/abs/hep-th/0409156}{{\tt hep-th/0409156}}.

\bibitem{Bern:2019nnu}
Z.~Bern, C.~Cheung, R.~Roiban, C.-H. Shen, M.~P. Solon, and M.~Zeng,
  ``{Scattering Amplitudes and the Conservative Hamiltonian for Binary Systems
  at Third Post-Minkowskian Order},'' {\em Phys. Rev. Lett.} {\bf 122} (2019),
  no.~20, 201603, \href{http://www.arXiv.org/abs/1901.04424}{{\tt 1901.04424}}.

\bibitem{Bondi:1962px}
H.~Bondi, M.~G.~J. van~der Burg, and A.~W.~K. Metzner, ``{Gravitational waves
  in general relativity. 7. Waves from axisymmetric isolated systems},'' {\em
  Proc. Roy. Soc. Lond. A} {\bf 269} (1962) 21--52.

\bibitem{Sachs:1962wk}
R.~K. Sachs, ``{Gravitational waves in general relativity. 8. Waves in
  asymptotically flat space-times},'' {\em Proc. Roy. Soc. Lond. A} {\bf 270}
  (1962) 103--126.

\bibitem{Barnich:2010eb}
G.~Barnich and C.~Troessaert, ``{Aspects of the BMS/CFT correspondence},'' {\em
  JHEP} {\bf 05} (2010) 062, \href{http://www.arXiv.org/abs/1001.1541}{{\tt
  1001.1541}}.

\bibitem{Bagchi:2025vri}
A.~Bagchi, A.~Banerjee, P.~Dhivakar, S.~Mondal, and A.~Shukla, ``{The
  Carrollian Kaleidoscope},'' \href{http://www.arXiv.org/abs/2506.16164}{{\tt
  2506.16164}}.

\bibitem{LevyLeblond:1965}
J.-M. L{\'e}vy-Leblond, ``{Une nouvelle limite non-relativiste du groupe de
  Poincar{\'e}},'' {\em Ann. Inst. H. Poincar{\'e} Phys. Th{\'e}or.} {\bf 3}
  (1965), no.~1, 1--12.

\bibitem{Duval:2014uva}
C.~Duval, G.~W. Gibbons, and P.~A. Horvathy, ``{Conformal Carroll groups and
  BMS symmetry},'' {\em Class. Quant. Grav.} {\bf 31} (2014) 092001,
  \href{http://www.arXiv.org/abs/1402.5894}{{\tt 1402.5894}}.

\bibitem{Donnay:2022aba}
L.~Donnay, A.~Fiorucci, Y.~Herfray, and R.~Ruzziconi, ``{Carrollian Perspective
  on Celestial Holography},'' {\em Phys. Rev. Lett.} {\bf 129} (2022), no.~7,
  071602, \href{http://www.arXiv.org/abs/2202.04702}{{\tt 2202.04702}}.

\bibitem{Donnay:2022wvx}
L.~Donnay, A.~Fiorucci, Y.~Herfray, and R.~Ruzziconi, ``{Bridging Carrollian
  and celestial holography},'' {\em Phys. Rev. D} {\bf 107} (2023), no.~12,
  126027, \href{http://www.arXiv.org/abs/2212.12553}{{\tt 2212.12553}}.

\bibitem{Bagchi:2022emh}
A.~Bagchi, S.~Banerjee, R.~Basu, and S.~Dutta, ``{Scattering Amplitudes:
  Celestial and Carrollian},'' {\em Phys. Rev. Lett.} {\bf 128} (2022), no.~24,
  241601, \href{http://www.arXiv.org/abs/2202.08438}{{\tt 2202.08438}}.

\bibitem{Mason:2023mti}
L.~Mason, R.~Ruzziconi, and A.~Yelleshpur~Srikant, ``{Carrollian Amplitudes and
  Celestial Symmetries},'' \href{http://www.arXiv.org/abs/2312.10138}{{\tt
  2312.10138}}.

\bibitem{Liu:2024nfc}
W.-B. Liu, J.~Long, and X.-Q. Ye, ``{Feynman rules and loop structure of
  Carrollian amplitudes},'' {\em JHEP} {\bf 05} (2024) 213,
  \href{http://www.arXiv.org/abs/2402.04120}{{\tt 2402.04120}}.

\bibitem{Li:2024kbo}
A.~Li, J.~Long, and J.-L. Yang, ``{Carrollian propagator and amplitude in
  Rindler spacetime},'' {\em JHEP} {\bf 03} (2025) 186,
  \href{http://www.arXiv.org/abs/2410.20372}{{\tt 2410.20372}}.

\bibitem{Liu:2024llk}
W.-B. Liu, J.~Long, H.-Y. Xiao, and J.-L. Yang, ``{On the definition of
  Carrollian amplitudes in general dimensions},'' {\em JHEP} {\bf 11} (2024)
  027, \href{http://www.arXiv.org/abs/2407.20816}{{\tt 2407.20816}}.

\bibitem{nguyen2024carrollian}
K.~Nguyen, ``Carrollian conformal correlators and massless scattering
  amplitudes,'' {\em Journal of High Energy Physics} {\bf 2024} (2024), no.~1,
  76.

\bibitem{kulkarni2025carrollian}
H.~Kulkarni, R.~Ruzziconi, and A.~Y. Srikant, ``On carrollian and celestial
  correlators in general dimensions,'' {\em Journal of High Energy Physics}
  {\bf 2025} (2025), no.~10, 1--36.

\bibitem{Long:2026cpq}
J.~Long and J.-L. Yang, ``{Constraining bulk-to-boundary correlators in the
  theories with Poincar{\'e} symmetry},''
  \href{http://www.arXiv.org/abs/2601.18461}{{\tt 2601.18461}}.

\bibitem{Penna:2015gza}
R.~F. Penna, ``{BMS invariance and the membrane paradigm},'' {\em JHEP} {\bf
  03} (2016) 023, \href{http://www.arXiv.org/abs/1508.06577}{{\tt 1508.06577}}.

\bibitem{Ciambelli:2018wre}
L.~Ciambelli, C.~Marteau, A.~C. Petkou, P.~M. Petropoulos, and K.~Siampos,
  ``{Flat holography and Carrollian fluids},'' {\em JHEP} {\bf 07} (2018) 165,
  \href{http://www.arXiv.org/abs/1802.06809}{{\tt 1802.06809}}.

\bibitem{Ciambelli:2018xat}
L.~Ciambelli, C.~Marteau, A.~C. Petkou, P.~M. Petropoulos, and K.~Siampos,
  ``{Covariant Galilean versus Carrollian hydrodynamics from relativistic
  fluids},'' {\em Class. Quant. Grav.} {\bf 35} (2018), no.~16, 165001,
  \href{http://www.arXiv.org/abs/1802.05286}{{\tt 1802.05286}}.

\bibitem{Donnay:2019jiz}
L.~Donnay and C.~Marteau, ``{Carrollian Physics at the Black Hole Horizon},''
  {\em Class. Quant. Grav.} {\bf 36} (2019), no.~16, 165002,
  \href{http://www.arXiv.org/abs/1903.09654}{{\tt 1903.09654}}.

\bibitem{petkou2022relativistic}
A.~C. Petkou, P.~M. Petropoulos, D.~Rivera-Betancour, and K.~Siampos,
  ``Relativistic fluids, hydrodynamic frames and their galilean versus
  carrollian avatars,'' {\em Journal of High Energy Physics} {\bf 2022} (2022),
  no.~9, 162.

\bibitem{freidel2023carrollian}
L.~Freidel and P.~Jai-akson, ``Carrollian hydrodynamics from symmetries,'' {\em
  Classical and Quantum Gravity} {\bf 40} (2023), no.~5, 055009.

\bibitem{armas2024carrollian}
J.~Armas and E.~Have, ``Carrollian fluids and spontaneous breaking of boost
  symmetry,'' {\em Physical Review Letters} {\bf 132} (2024), no.~16, 161606.

\bibitem{Long:2025fbb}
J.~Long and X.-H. Zhou, ``{Reduction of topological invariants on null
  hypersurfaces},'' {\em JHEP} {\bf 01} (2026) 116,
  \href{http://www.arXiv.org/abs/2509.06073}{{\tt 2509.06073}}.

\bibitem{Liu:2024rvz}
W.-B. Liu, J.~Long, and X.-H. Zhou, ``{Electromagnetic helicity flux operators
  in higher dimensions},'' {\em JHEP} {\bf 04} (2025) 026,
  \href{http://www.arXiv.org/abs/2407.20077}{{\tt 2407.20077}}.

\bibitem{Long:2024yvj}
J.~Long and R.-Z. Yu, ``{Gravitational helicity flux density from two-body
  systems},'' {\em Class. Quant. Grav.} {\bf 42} (2025), no.~4, 045005,
  \href{http://www.arXiv.org/abs/2403.18627}{{\tt 2403.18627}}.

\bibitem{Heng:2025kmr}
Z.-Y. Heng, J.~Long, R.-Z. Yu, and X.-H. Zhou, ``{Electromagnetic helicity flux
  density for radiative systems},'' {\em Class. Quant. Grav.} {\bf 43} (2026),
  no.~1, 015016, \href{http://www.arXiv.org/abs/2507.14966}{{\tt 2507.14966}}.

\bibitem{Long:2025bfi}
J.~Long and H.-Y. Xiao, ``{Thermal correlator at null infinity},'' {\em JHEP}
  {\bf 10} (2025) 127, \href{http://www.arXiv.org/abs/2501.15714}{{\tt
  2501.15714}}.

\bibitem{Regge:1957td}
T.~Regge and J.~A. Wheeler, ``{Stability of a Schwarzschild singularity},''
  {\em Phys. Rev.} {\bf 108} (1957) 1063--1069.

\bibitem{Zerilli:1970se}
F.~J. Zerilli, ``{Effective potential for even parity Regge-Wheeler
  gravitational perturbation equations},'' {\em Phys. Rev. Lett.} {\bf 24}
  (1970) 737--738.

\bibitem{Newman:1961qr}
E.~Newman and R.~Penrose, ``{An Approach to gravitational radiation by a method
  of spin coefficients},'' {\em J. Math. Phys.} {\bf 3} (1962) 566--578.

\bibitem{1973ApJ...185..635T}
S.~A. {Teukolsky}, ``{Perturbations of a Rotating Black Hole. I. Fundamental
  Equations for Gravitational, Electromagnetic, and Neutrino-Field
  Perturbations},'' {\em Astrophysics Journal} {\bf 185} (Oct., 1973) 635--648.

\bibitem{Pound:2021qin}
A.~Pound and B.~Wardell, ``{Black hole perturbation theory and gravitational
  self-force},'' \href{http://www.arXiv.org/abs/2101.04592}{{\tt 2101.04592}}.

\bibitem{frolov2012black}
V.~Frolov and I.~Novikov, {\em Black hole physics: Basic concepts and new
  developments}, vol.~96.
\newblock Springer Science \& Business Media, 2012.

\bibitem{Vishveshwara:1970zz}
C.~V. Vishveshwara, ``{Scattering of Gravitational Radiation by a Schwarzschild
  Black-hole},'' {\em Nature} {\bf 227} (1970) 936--938.

\bibitem{Price:1972pw}
R.~H. Price, ``{Nonspherical Perturbations of Relativistic Gravitational
  Collapse. II. Integer-Spin, Zero-Rest-Mass Fields},'' {\em Phys. Rev. D} {\bf
  5} (1972) 2439--2454.

\bibitem{Ching:1995tj}
E.~S.~C. Ching, P.~T. Leung, W.~M. Suen, and K.~Young, ``{Wave propagation in
  gravitational systems: Late time behavior},'' {\em Phys. Rev. D} {\bf 52}
  (1995) 2118--2132, \href{http://www.arXiv.org/abs/gr-qc/9507035}{{\tt
  gr-qc/9507035}}.

\bibitem{1974JETP...38....1S}
A.~A. {Starobinski{\v{i}}} and S.~M. {Churilov}, ``{Amplification of
  electromagnetic and gravitational waves scattered by a rotating ``black
  hole''},'' {\em Soviet Journal of Experimental and Theoretical Physics} {\bf
  38} (Jan., 1974) 1.

\bibitem{1975CMaPh..44..245G}
G.~W. {Gibbons}, ``{Vacuum polarization and the spontaneous loss of charge by
  black holes},'' {\em Communications in Mathematical Physics} {\bf 44} (Oct.,
  1975) 245--264.

\bibitem{Page:1976df}
D.~N. Page, ``{Particle Emission Rates from a Black Hole: Massless Particles
  from an Uncharged, Nonrotating Hole},'' {\em Phys. Rev. D} {\bf 13} (1976)
  198--206.

\bibitem{Unruh:1976fm}
W.~G. Unruh, ``{Absorption Cross-Section of Small Black Holes},'' {\em Phys.
  Rev. D} {\bf 14} (1976) 3251--3259.

\bibitem{Sanchez:1977si}
N.~G. Sanchez, ``{Absorption and Emission Spectra of a Schwarzschild Black
  Hole},'' {\em Phys. Rev. D} {\bf 18} (1978) 1030.

\bibitem{Das:1996we}
S.~R. Das, G.~W. Gibbons, and S.~D. Mathur, ``{Universality of low-energy
  absorption cross-sections for black holes},'' {\em Phys. Rev. Lett.} {\bf 78}
  (1997) 417--419, \href{http://www.arXiv.org/abs/hep-th/9609052}{{\tt
  hep-th/9609052}}.

\bibitem{Andersson:1996cm}
N.~Andersson, ``{Evolving test fields in a black hole geometry},'' {\em Phys.
  Rev. D} {\bf 55} (1997) 468--479,
  \href{http://www.arXiv.org/abs/gr-qc/9607064}{{\tt gr-qc/9607064}}.

\bibitem{Arnaudo:2025uos}
P.~Arnaudo, J.~Carballo, and B.~Withers, ``{Beyond quasinormal modes: a
  complete mode decomposition of black hole perturbations},''
  \href{http://www.arXiv.org/abs/2510.18956}{{\tt 2510.18956}}.

\bibitem{Son:2002sd}
D.~T. Son and A.~O. Starinets, ``{Minkowski space correlators in AdS / CFT
  correspondence: Recipe and applications},'' {\em JHEP} {\bf 09} (2002) 042,
  \href{http://www.arXiv.org/abs/hep-th/0205051}{{\tt hep-th/0205051}}.

\bibitem{Alday:2024yyj}
L.~F. Alday, M.~Nocchi, R.~Ruzziconi, and A.~Yelleshpur~Srikant, ``{Carrollian
  Amplitudes from Holographic Correlators},''
  \href{http://www.arXiv.org/abs/2406.19343}{{\tt 2406.19343}}.

\bibitem{Mano:1996vt}
S.~Mano, H.~Suzuki, and E.~Takasugi, ``{Analytic solutions of the Teukolsky
  equation and their low frequency expansions},'' {\em Prog. Theor. Phys.} {\bf
  95} (1996) 1079--1096, \href{http://www.arXiv.org/abs/gr-qc/9603020}{{\tt
  gr-qc/9603020}}.

\bibitem{Sasaki:2003xr}
M.~Sasaki and H.~Tagoshi, ``{Analytic black hole perturbation approach to
  gravitational radiation},'' {\em Living Rev. Rel.} {\bf 6} (2003) 6,
  \href{http://www.arXiv.org/abs/gr-qc/0306120}{{\tt gr-qc/0306120}}.

\bibitem{Aminov:2020yma}
G.~Aminov, A.~Grassi, and Y.~Hatsuda, ``{Black Hole Quasinormal Modes and
  Seiberg{\textendash}Witten Theory},'' {\em Annales Henri Poincare} {\bf 23}
  (2022), no.~6, 1951--1977, \href{http://www.arXiv.org/abs/2006.06111}{{\tt
  2006.06111}}.

\bibitem{Bonelli:2021uvf}
G.~Bonelli, C.~Iossa, D.~P. Lichtig, and A.~Tanzini, ``{Exact solution of Kerr
  black hole perturbations via CFT2 and instanton counting: Greybody factor,
  quasinormal modes, and Love numbers},'' {\em Phys. Rev. D} {\bf 105} (2022),
  no.~4, 044047, \href{http://www.arXiv.org/abs/2105.04483}{{\tt 2105.04483}}.

\bibitem{Bonelli:2022ten}
G.~Bonelli, C.~Iossa, D.~Panea~Lichtig, and A.~Tanzini, ``{Irregular Liouville
  Correlators and Connection Formulae for Heun Functions},'' {\em Commun. Math.
  Phys.} {\bf 397} (2023), no.~2, 635--727,
  \href{http://www.arXiv.org/abs/2201.04491}{{\tt 2201.04491}}.

\bibitem{Correia:2025Analyticity}
M.~Correia, T.~Gopalka, G.~Isabella, and A.~M. Wolz, ``Analyticity of the black
  hole s-matrix,'' \href{http://www.arXiv.org/abs/2511.11794}{{\tt
  2511.11794}}.

\bibitem{Berti:2009kk}
E.~Berti, V.~Cardoso, and A.~O. Starinets, ``{Quasinormal modes of black holes
  and black branes},'' {\em Class. Quant. Grav.} {\bf 26} (2009) 163001,
  \href{http://www.arXiv.org/abs/0905.2975}{{\tt 0905.2975}}.

\bibitem{casals2013analytic}
M.~Casals and A.~Ottewill, ``Analytic investigation of the branch cut of the
  green function in schwarzschild space-time,'' {\em Physical Review
  D---Particles, Fields, Gravitation, and Cosmology} {\bf 87} (2013), no.~6,
  064010.

\bibitem{Casals:2015Tail}
M.~Casals and A.~C. Ottewill, ``High-order tail in schwarzschild space-time,''
  {\em Phys. Rev. D} {\bf 92} (2015) 124055,
  \href{http://www.arXiv.org/abs/1509.04702}{{\tt 1509.04702}}.

\bibitem{Andersson:2000tf}
N.~Andersson and B.~P. Jensen, ``{Scattering by black holes. Chapter 0.1},''
  \href{http://www.arXiv.org/abs/gr-qc/0011025}{{\tt gr-qc/0011025}}.

\bibitem{1988sfbh.book.....F}
J.~A.~H. {Futterman}, F.~A. {Handler}, and R.~A. {Matzner}, {\em {Scattering
  from black holes}}.
\newblock Cambridge University Press, 1988.

\bibitem{Price:1971fb}
R.~H. Price, ``{Nonspherical perturbations of relativistic gravitational
  collapse. 1. Scalar and gravitational perturbations},'' {\em Phys. Rev. D}
  {\bf 5} (1972) 2419--2438.

\bibitem{Gundlach:1993tp}
C.~Gundlach, R.~H. Price, and J.~Pullin, ``{Late time behavior of stellar
  collapse and explosions: 1. Linearized perturbations},'' {\em Phys. Rev. D}
  {\bf 49} (1994) 883--889, \href{http://www.arXiv.org/abs/gr-qc/9307009}{{\tt
  gr-qc/9307009}}.

\bibitem{CasalsDolanOttewillWardell2013}
M.~Casals, S.~Dolan, A.~C. Ottewill, and B.~Wardell, ``Self-force and green
  function in schwarzschild spacetime via quasinormal modes and branch cut,''
  {\em Phys. Rev. D} {\bf 88} (2013), no.~4, 044022.

\bibitem{Zenginoglu:2012xe}
A.~Zenginoglu and C.~R. Galley, ``{Caustic echoes from a Schwarzschild black
  hole},'' {\em Phys. Rev. D} {\bf 86} (2012) 064030,
  \href{http://www.arXiv.org/abs/1206.1109}{{\tt 1206.1109}}.

\bibitem{Casals:2016qyj}
M.~Casals and B.~C. Nolan, ``{Global Hadamard form for the Green function in
  Schwarzschild spacetime},'' {\em Phys. Rev. D} {\bf 108} (2023), no.~4,
  044033, \href{http://www.arXiv.org/abs/1606.03075}{{\tt 1606.03075}}.

\bibitem{Abarbanel:1969ek}
H.~D.~I. Abarbanel and C.~Itzykson, ``{Relativistic eikonal expansion},'' {\em
  Phys. Rev. Lett.} {\bf 23} (1969) 53.

\bibitem{Levy:1969cr}
M.~Levy and J.~Sucher, ``{Eikonal approximation in quantum field theory},''
  {\em Phys. Rev.} {\bf 186} (1969) 1656--1670.

\bibitem{Cheng:1969eh}
H.~Cheng and T.~T. Wu, ``{High-energy elastic scattering in quantum
  electrodynamics},'' {\em Phys. Rev. Lett.} {\bf 22} (1969) 666.

\bibitem{Wallace:1973iu}
S.~J. Wallace, ``{Eikonal expansion},'' {\em Annals Phys.} {\bf 78} (1973)
  190--257.

\bibitem{Aichelburg:1970dh}
P.~C. Aichelburg and R.~U. Sexl, ``{On the Gravitational field of a massless
  particle},'' {\em Gen. Rel. Grav.} {\bf 2} (1971) 303--312.

\bibitem{Dray:1984ha}
T.~Dray and G.~'t~Hooft, ``{The Gravitational Shock Wave of a Massless
  Particle},'' {\em Nucl. Phys. B} {\bf 253} (1985) 173--188.

\bibitem{tHooft:1987vrq}
G.~'t~Hooft, ``{Graviton Dominance in Ultrahigh-Energy Scattering},'' {\em
  Phys. Lett. B} {\bf 198} (1987) 61--63.

\bibitem{Amati:1987wq}
D.~Amati, M.~Ciafaloni, and G.~Veneziano, ``{Superstring Collisions at
  Planckian Energies},'' {\em Phys. Lett. B} {\bf 197} (1987) 81.

\bibitem{Adamo:2021rfq}
T.~Adamo, A.~Cristofoli, and P.~Tourkine, ``{Eikonal amplitudes from curved
  backgrounds},'' {\em SciPost Phys.} {\bf 13} (2022), no.~2, 032,
  \href{http://www.arXiv.org/abs/2112.09113}{{\tt 2112.09113}}.

\bibitem{Liu:2025oom}
W.-B. Liu and J.~Long, ``{Extrapolating the massive fields to future timelike
  infinity},'' {\em JHEP} {\bf 12} (2025) 042,
  \href{http://www.arXiv.org/abs/2508.15619}{{\tt 2508.15619}}.

\bibitem{2024JHEP...10..192A}
T.~{Adamo}, W.~{Bu}, P.~{Tourkine}, and B.~{Zhu}, ``{Eikonal amplitudes on the
  celestial sphere},'' {\em Journal of High Energy Physics} {\bf 2024} (Oct.,
  2024) 192, \href{http://www.arXiv.org/abs/2405.15594}{{\tt 2405.15594}}.

\bibitem{1972gcpa.book.....W}
S.~{Weinberg}, {\em {Gravitation and Cosmology: Principles and Applications of
  the General Theory of Relativity}}.
\newblock John Wiley \& Sons, 1972.

\bibitem{duistermaat1994fourier}
J.~J. Duistermaat and L.~H{\"o}rmander, ``Fourier integral operators. ii,'' in
  {\em Mathematics Past and Present Fourier Integral Operators}, pp.~129--215.
\newblock Springer, 1994.

\bibitem{hormander2009analysis}
L.~H{\"o}rmander, {\em The analysis of linear partial differential operators
  IV: Fourier integral operators}.
\newblock Springer, 2009.

\bibitem{Poisson:2011nh}
E.~Poisson, A.~Pound, and I.~Vega, ``{The Motion of point particles in curved
  spacetime},'' {\em Living Rev. Rel.} {\bf 14} (2011) 7,
  \href{http://www.arXiv.org/abs/1102.0529}{{\tt 1102.0529}}.

\bibitem{Harte:2012uw}
A.~I. Harte and T.~D. Drivas, ``{Caustics and wave propagation in curved
  spacetimes},'' {\em Phys. Rev. D} {\bf 85} (2012) 124039,
  \href{http://www.arXiv.org/abs/1202.0540}{{\tt 1202.0540}}.

\bibitem{bozza2002gravitational}
V.~Bozza, ``Gravitational lensing in the strong field limit,'' {\em Physical
  Review D} {\bf 66} (2002), no.~10, 103001.

\bibitem{bozza2004time}
V.~Bozza and L.~Mancini, ``Time delay in black hole gravitational lensing as a
  distance estimator,'' {\em General Relativity and Gravitation} {\bf 36}
  (2004), no.~2, 435--450.

\bibitem{2014AmJPh..82..564M}
G.~{Mu{\~n}oz}, ``{Orbits of massless particles in the Schwarzschild metric:
  Exact solutions},'' {\em American Journal of Physics} {\bf 82} (June, 2014)
  564--573.

\bibitem{Witten:1998zw}
E.~Witten, ``{Anti-de Sitter space, thermal phase transition, and confinement
  in gauge theories},'' {\em Adv. Theor. Math. Phys.} {\bf 2} (1998) 505--532,
  \href{http://www.arXiv.org/abs/hep-th/9803131}{{\tt hep-th/9803131}}.

\bibitem{Fitzpatrick:2015zha}
A.~L. Fitzpatrick, J.~Kaplan, and M.~T. Walters, ``{Virasoro conformal blocks
  and thermality from classical background fields},'' {\em JHEP} {\bf 11}
  (2015) 200, \href{http://www.arXiv.org/abs/1501.05315}{{\tt 1501.05315}}.

\bibitem{Ammon:2025flat}
M.~Ammon, F.~Capone, and C.~Sieling, ``{Flat Holography \& Holographic
  Renormalization: Scalar Field},'' {\em JHEP} {\bf 07} (2026) 124,
  \href{http://www.arXiv.org/abs/2512.14818}{{\tt 2512.14818}}.

\bibitem{Gralla:2019ceu}
S.~E. Gralla and A.~Lupsasca, ``{Null geodesics of the Kerr exterior},'' {\em
  Phys. Rev. D} {\bf 101} (2020), no.~4, 044032,
  \href{http://www.arXiv.org/abs/1910.12881}{{\tt 1910.12881}}.

\bibitem{Gibbons:2011rh}
G.~W. Gibbons and M.~Vyska, ``{The Application of Weierstrass elliptic
  functions to Schwarzschild Null Geodesics},'' {\em Class. Quant. Grav.} {\bf
  29} (2012) 065016, \href{http://www.arXiv.org/abs/1110.6508}{{\tt
  1110.6508}}.

\bibitem{1984PhRvD..30..295F}
V.~{Ferrari} and B.~{Mashhoon}, ``{New approach to the quasinormal modes of a
  black hole},'' {\em Phys.Review D} {\bf 30} (July, 1984) 295--304.

\bibitem{1972ApJ...172L..95G}
C.~J. {Goebel}, ``{Comments on the ``vibrations'' of a Black Hole.},'' {\em
  Astrophysics Journal} {\bf 172} (Mar., 1972) L95.

\bibitem{Blome:1981azp}
H.-J. Blome and B.~Mashhoon, ``{Quasi-normal oscillations of a schwarzschild
  black hole},'' {\em Phys. Lett. A} {\bf 100} (1981), no.~5, 231--234.

\bibitem{Schutz:1985km}
B.~F. Schutz and C.~M. Will, ``{BLACK HOLE NORMAL MODES: A SEMIANALYTIC
  APPROACH},'' {\em Astrophys. J. Lett.} {\bf 291} (1985) L33--L36.

\bibitem{Leaver:1985ax}
E.~W. Leaver, ``{An Analytic representation for the quasi normal modes of Kerr
  black holes},'' {\em Proc. Roy. Soc. Lond. A} {\bf 402} (1985) 285--298.

\bibitem{Leaver:1986gd}
E.~W. Leaver, ``{Spectral decomposition of the perturbation response of the
  Schwarzschild geometry},'' {\em Phys. Rev. D} {\bf 34} (1986) 384--408.

\bibitem{Berti:2025hly}
J.~Abedi {\em et al.}, ``{Black hole spectroscopy: from theory to
  experiment},'' \href{http://www.arXiv.org/abs/2505.23895}{{\tt 2505.23895}}.

\bibitem{Dolan:2011fh}
S.~R. Dolan and A.~C. Ottewill, ``{Wave Propagation and Quasinormal Mode
  Excitation on Schwarzschild Spacetime},'' {\em Phys. Rev. D} {\bf 84} (2011)
  104002, \href{http://www.arXiv.org/abs/1106.4318}{{\tt 1106.4318}}.

\bibitem{PhysRevD.86.104006}
H.~Yang, D.~A. Nichols, F.~Zhang, A.~Zimmerman, Z.~Zhang, and Y.~Chen,
  ``Quasinormal-mode spectrum of kerr black holes and its geometric
  interpretation,'' {\em Phys. Rev. D} {\bf 86} (Nov, 2012) 104006.

\bibitem{Hawking:2016msc}
S.~W. Hawking, M.~J. Perry, and A.~Strominger, ``{Soft Hair on Black Holes},''
  {\em Phys. Rev. Lett.} {\bf 116} (2016), no.~23, 231301,
  \href{http://www.arXiv.org/abs/1601.00921}{{\tt 1601.00921}}.

\bibitem{Compere:2016jwb}
G.~Comp{\`e}re and J.~Long, ``{Vacua of the gravitational field},'' {\em JHEP}
  {\bf 07} (2016) 137, \href{http://www.arXiv.org/abs/1601.04958}{{\tt
  1601.04958}}.

\bibitem{Compere:2016hzt}
G.~Comp{\`e}re and J.~Long, ``{Classical static final state of collapse with
  supertranslation memory},'' {\em Class. Quant. Grav.} {\bf 33} (2016),
  no.~19, 195001, \href{http://www.arXiv.org/abs/1602.05197}{{\tt 1602.05197}}.

\bibitem{Mitman:2024uss}
K.~Mitman {\em et al.}, ``{A review of gravitational memory and BMS frame
  fixing in numerical relativity},'' {\em Class. Quant. Grav.} {\bf 41} (2024),
  no.~22, 223001, \href{http://www.arXiv.org/abs/2405.08868}{{\tt 2405.08868}}.

\bibitem{Mitman:2021xkq}
K.~Mitman {\em et al.}, ``{Fixing the BMS frame of numerical relativity
  waveforms},'' {\em Phys. Rev. D} {\bf 104} (2021), no.~2, 024051,
  \href{http://www.arXiv.org/abs/2105.02300}{{\tt 2105.02300}}.

\bibitem{MaganaZertuche:2021syq}
L.~Maga{\~n}a~Zertuche {\em et al.}, ``{High precision ringdown modeling:
  Multimode fits and BMS frames},'' {\em Phys. Rev. D} {\bf 105} (2022),
  no.~10, 104015, \href{http://www.arXiv.org/abs/2110.15922}{{\tt 2110.15922}}.

\bibitem{Strominger:2013jfa}
A.~Strominger, ``{On BMS Invariance of Gravitational Scattering},'' {\em JHEP}
  {\bf 07} (2014) 152, \href{http://www.arXiv.org/abs/1312.2229}{{\tt
  1312.2229}}.

\bibitem{Chandrasekhar:1985kt}
S.~Chandrasekhar, {\em {The mathematical theory of black holes}}.
\newblock Oxford University Press, 1985.

\bibitem{Compere:2021bkk}
G.~Comp{\`e}re, Y.~Liu, and J.~Long, ``{Classification of radial Kerr geodesic
  motion},'' {\em Phys. Rev. D} {\bf 105} (2022), no.~2, 024075,
  \href{http://www.arXiv.org/abs/2106.03141}{{\tt 2106.03141}}.

\bibitem{1971JETPL..14..180Z}
Y.~B. {Zel'Dovich}, ``{Generation of Waves by a Rotating Body},'' {\em Soviet
  Journal of Experimental and Theoretical Physics Letters} {\bf 14} (Aug.,
  1971) 180.

\bibitem{Bekenstein:1998nt}
J.~D. Bekenstein and M.~Schiffer, ``{The Many faces of superradiance},'' {\em
  Phys. Rev. D} {\bf 58} (1998) 064014,
  \href{http://www.arXiv.org/abs/gr-qc/9803033}{{\tt gr-qc/9803033}}.

\bibitem{Brito:2015oca}
R.~Brito, V.~Cardoso, and P.~Pani, ``{Superradiance}: {New Frontiers in Black
  Hole Physics},'' {\em Lect. Notes Phys.} {\bf 906} (2015) pp.1--237,
  \href{http://www.arXiv.org/abs/1501.06570}{{\tt 1501.06570}}.

\bibitem{Gleiser:1995gx}
R.~J. Gleiser, C.~O. Nicasio, R.~H. Price, and J.~Pullin, ``{Second order
  perturbations of a Schwarzschild black hole},'' {\em Class. Quant. Grav.}
  {\bf 13} (1996) L117--L124,
  \href{http://www.arXiv.org/abs/gr-qc/9510049}{{\tt gr-qc/9510049}}.

\bibitem{Campanelli:1998jv}
M.~Campanelli and C.~O. Lousto, ``{Second order gauge invariant gravitational
  perturbations of a Kerr black hole},'' {\em Phys. Rev. D} {\bf 59} (1999)
  124022, \href{http://www.arXiv.org/abs/gr-qc/9811019}{{\tt gr-qc/9811019}}.

\bibitem{DeWitt:1960fc}
B.~S. DeWitt and R.~W. Brehme, ``{Radiation damping in a gravitational
  field},'' {\em Annals Phys.} {\bf 9} (1960) 220--259.

\bibitem{Decanini:2011xi}
Y.~Decanini, G.~Esposito-Farese, and A.~Folacci, ``{Universality of high-energy
  absorption cross sections for black holes},'' {\em Phys. Rev. D} {\bf 83}
  (2011) 044032, \href{http://www.arXiv.org/abs/1101.0781}{{\tt 1101.0781}}.

\bibitem{Glampedakis:2001cx}
K.~Glampedakis and N.~Andersson, ``{Scattering of scalar waves by rotating
  black holes},'' {\em Class. Quant. Grav.} {\bf 18} (2001) 1939--1966,
  \href{http://www.arXiv.org/abs/gr-qc/0102100}{{\tt gr-qc/0102100}}.

\bibitem{Anninos:1992ih}
P.~Anninos, C.~DeWitt-Morette, R.~A. Matzner, P.~Yioutas, and T.~R. Zhang,
  ``{Orbiting cross-sections: Application to black hole scattering},'' {\em
  Phys. Rev. D} {\bf 46} (1992) 4477--4494.

\bibitem{Nambu:2015aea}
Y.~Nambu and S.~Noda, ``{Wave Optics in Black Hole Spacetimes: Schwarzschild
  Case},'' {\em Class. Quant. Grav.} {\bf 33} (2016) 075011,
  \href{http://www.arXiv.org/abs/1502.05468}{{\tt 1502.05468}}.

\bibitem{tsukamoto2017deflection}
N.~Tsukamoto, ``Deflection angle in the strong deflection limit in a general
  asymptotically flat, static, spherically symmetric spacetime,'' {\em Physical
  Review D} {\bf 95} (2017), no.~6, 064035.

\end{thebibliography}
\end{document}